\documentclass[aps,prl,reprint,amsmath,amssymb,superscriptaddress,longbibliography]{revtex4-2}
\usepackage[colorlinks=true,linkcolor=blue,citecolor=blue,urlcolor=blue]{hyperref}
\usepackage{bm}
\usepackage{graphicx}
\usepackage{multirow}
\usepackage{physics}
\usepackage{extarrows}
\usepackage{array}
\usepackage{appendix}
\usepackage{xr}
\usepackage{bbding}
\usepackage{cases}
\usepackage[normalem]{ulem}
\usepackage{color}
\DeclareMathAlphabet\mathbfcal{OMS}{cmsy}{b}{n}

\begin{document}
\title{Intrinsic second-order nonlinear Hall effect from perturbation theory}
\author{Zhi-Hao Wei}
\email{These authors contributed equally to this work.}
\affiliation{State Key Laboratory of Quantum Functional Materials, Department of Physics, Guangdong Provincial Key Laboratory of Topological Matter, and Guangdong Basic Research Center of Excellence for Quantum Science, Southern University of Science and Technology (SUSTech), Shenzhen 518055, China}

\author{Xiao-Bin Qiang}
\email{These authors contributed equally to this work.}
\affiliation{State Key Laboratory of Quantum Functional Materials, Department of Physics, Guangdong Provincial Key Laboratory of Topological Matter, and Guangdong Basic Research Center of Excellence for Quantum Science, Southern University of Science and Technology (SUSTech), Shenzhen 518055, China}
\affiliation{Division of Physics and Applied Physics, School of Physical and Mathematical Sciences, Nanyang Technological University, 21 Nanyang Link, 637371, Singapore}

\author{Renjie Tong}
\affiliation{State Key Laboratory of Quantum Functional Materials, Department of Physics, Guangdong Provincial Key Laboratory of Topological Matter, and Guangdong Basic Research Center of Excellence for Quantum Science, Southern University of Science and Technology (SUSTech), Shenzhen 518055, China}

\author{Hai-Zhou Lu}
\email{Corresponding author: luhz@sustech.edu.cn}
\affiliation{State Key Laboratory of Quantum Functional Materials, Department of Physics, Guangdong Provincial Key Laboratory of Topological Matter, and Guangdong Basic Research Center of Excellence for Quantum Science, Southern University of Science and Technology (SUSTech), Shenzhen 518055, China}
\affiliation{Quantum Science Center of Guangdong-Hong Kong-Macao Greater Bay Area (Guangdong), Shenzhen 518045, China}

\author{X. C. Xie}
\affiliation{International Center for Quantum Materials, School of Physics, Peking University, Beijing 100871, China}
\affiliation{Interdisciplinary Center for Theoretical Physics and Information
Sciences (ICTPIS), Fudan University, Shanghai 200433, China}
\affiliation{Hefei National Laboratory, Hefei 230088, China}
\date{\today}

\begin{abstract}
Nonlinear transport provides an efficient probe of the electronic structure and symmetry of quantum materials. As a prominent example, the intrinsic nonlinear Hall effect has attracted considerable attention in recent years because it is independent of impurity scattering and directly reflects the quantum geometry of Bloch electrons. Here we establish a simple textbook-style perturbation theory for the intrinsic nonlinear conductivity. Based on the well-defined Bloch-state velocity, our formulation perturbatively determines the field-corrected velocity and distribution function, from which the conductivity follows directly. We further compare our approach with wavepacket-based theories and discuss their consistency. 
\end{abstract}

\maketitle

\textit{\textcolor{blue}{Introduction.}--}The anomalous Hall effect has emerged as an indispensable tool for probing the interplay among symmetry, topology, and magnetism~\cite{Hall1880,Karplus54pr,Haldane88prl,Jungwirth02prl,Fang03science,Yao04prl,Nagaosa10rmp,YuR10science,XueQK13science,Burkov14prl,ChangCZ23rmp}. As its nonlinear counterpart, the nonlinear Hall effect has attracted broad interest in recent years~\cite{Du21natrevphys,Liu24nsr,YanBH25rppp}, as it provides access to richer geometric and symmetry properties of quantum materials. The nonlinear Hall effect has been experimentally observed in a wide range of materials, including transition-metal dichalcogenides~\cite{Ma19nature,Kang19natmat,Tiwari21natcomm,Ma22natcomm,Ye23prl,Huang23nsr}, topological semimetals~\cite{Shvetsov19jetpl,Kumar21natnano,Min23natcomm,Lee24natcomm,Jiang25natcomm,Xi25natelectron}, graphene-based systems~\cite{Ho21natelectron,Sinha22natphys,Huang23prl}, semiconductor and poor-metal thin films ~\cite{Lu24natcomm,Makushko24natelectron,Cheng24natcomm,Hu25natnano}, oxide interfaces~\cite{ZhangHB26natcomm}, and magnets~\cite{Jungwirth18nc,Gao23science,Wang23nature,ZhaoW25natcomm,Mali26natcomm,LiX26natcomm}.
Theoretically, the nonlinear Hall response can arise  from distinct mechanisms, including intrinsic~\cite{Gao14prl,Gao19fop,Gao21prl,YangSY21prl}, Berry-curvature-dipole~\cite{Moore10prl,Sodemann15prl,Low15prb,Du18prl,Watanabe20prb,Resta21arXiv}, Drude~\cite{Hurd72book}, skew-scattering, and side-jump contributions~\cite{Du19natcomm,Xiao19prb,Du21natcomm,Ma23prl,GongZH26SB}. Among these mechanisms, the intrinsic contribution is particularly appealing because of its independence from impurity scattering and purely geometric origin.

\begin{figure}
\centering
\includegraphics[width=0.85\linewidth]{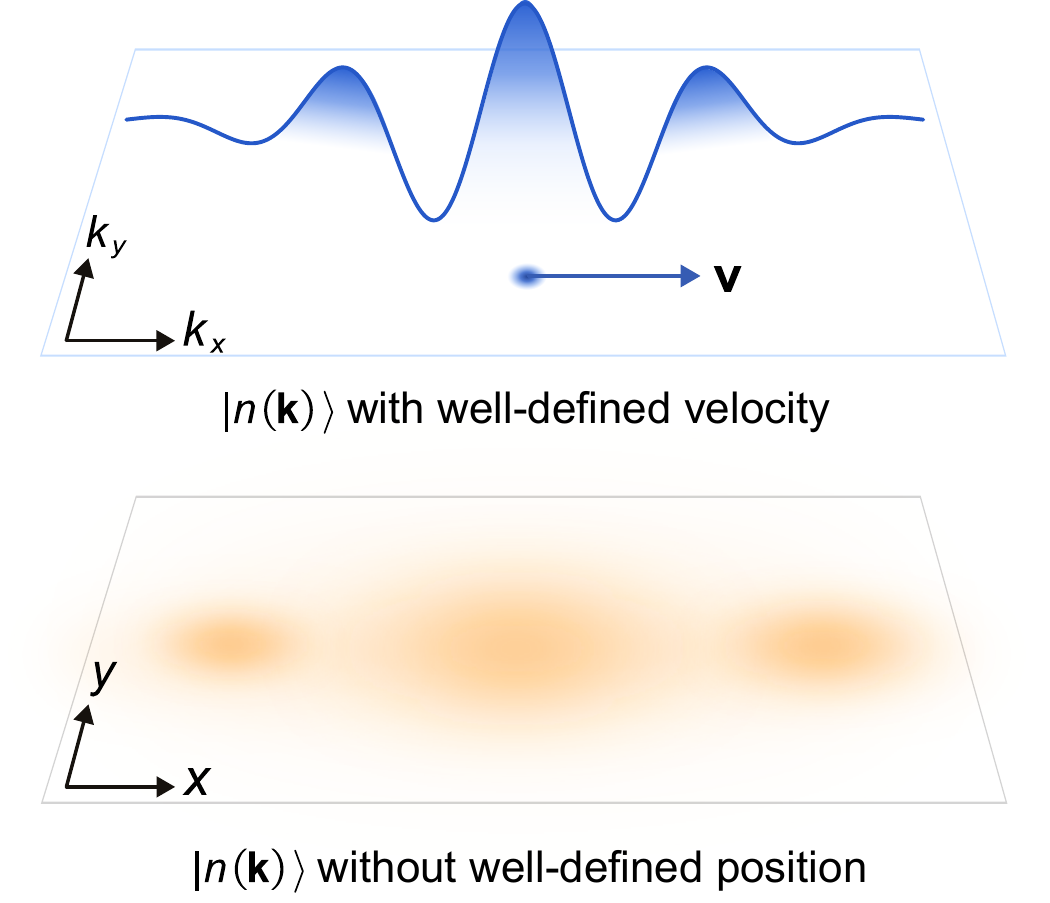}
\caption{Schematic illustration of a Bloch state $|n(\mathbf{k})\rangle$ in momentum space (upper panel) and real space (lower panel). A Bloch state possesses a well-defined velocity $\mathbf{v}$,  which serves as the starting point of our theory. By contrast, its position is not well-defined.  }\label{Fig: demo}
\end{figure}

Nevertheless, the intrinsic nonlinear Hall effect remains theoretically elusive. It has been formulated within a variety of frameworks, including wavepacket dynamics~\cite{Gao14prl,Gao19fop,Gao21prl,YangSY21prl,Xiao25prl}, nonlinear response theory~\cite{Jia24prb}, the Schrieffer–Wolff transformation approach~\cite{Kaplan24prl}, Green's function methods~\cite{Wang25prb,Ulrich26prb,Zhang26arXiv,Junya26prb},  density-matrix/quantum-kinetic approaches~\cite{Das23prb,Wang24prb,Guo26prl,Park26prb,Shakeel26prb}, and many-body theory~\cite{Resta26prb}. Astonishingly, these approaches have not converged on a unified expression for the intrinsic nonlinear conductivity. This inconsistency has led to considerable confusion in both theory and experiment, highlighting the need for a simpler and more standard theoretical framework.

In this Letter, we develop a textbook-style perturbation theory for the intrinsic nonlinear conductivity. We start from the well-defined velocity matrix element in the periodic function representation, and then perturbatively expand the periodic function to second order in the electric field.
With this field-dressed periodic function, the field-corrected velocity is evaluated directly as the expectation value of the velocity operator. Together with the field-corrected distribution function, we subsequently obtain the intrinsic nonlinear conductivity. 
We further compare our approach with wavepacket-based theories~\cite{Gao14prl,Kaplan24prl,Qiang26advs}.
The difference is that wavepacket-based theories start from the ill-defined Bloch-state position, while our approach starts from the well-defined Bloch-state velocity, as illustrated in Fig.~\ref{Fig: demo}. Despite these differences, we show that our approach yields the same intrinsic nonlinear conductivity as wavepacket-based theories once the effective energy is properly identified.
Finally, as a representative example, we calculate the intrinsic nonlinear Hall conductivity in bilayer antiferromagnet CrSBr.

\textit{\textcolor{blue}{Velocity in perturbation theory}--}We consider a Bloch electron in the presence of a uniform electric field $\mathbf{E}$. The Hamiltonian is given by
\begin{equation}
\hat{\mathcal H}
=
\hat{\mathcal H}_0
+
e\mathbf E\cdot \hat{\mathbf r},
\label{Eq: H}
\end{equation}
where $\hat{\mathcal H}_0$ is the unperturbed Hamiltonian, $-e$ is the electron charge, and $\hat{\mathbf r}$ is the position operator. The eigenstate of $\hat{\mathcal H}_0$ is the Bloch state $|\psi_n(\mathbf k)\rangle$, and the corresponding eigenenergy forms the energy band $\varepsilon_n(\mathbf k)$, where $n$ is the band index and $\mathbf k$ is the wavevector. For convenience, we formulate our theory in the periodic function representation. The periodic part of the Bloch state (hereafter referred to as the periodic function) is
$|n(\mathbf k)\rangle=e^{-i\mathbf k\cdot\hat{\mathbf r}}|\psi_n(\mathbf k)\rangle$, which
is an eigenstate of the Bloch Hamiltonian
$\hat{\mathcal H}_0(\mathbf k)=
e^{-i\mathbf k\cdot\hat{\mathbf r}}\hat{\mathcal H}_0
e^{i\mathbf k\cdot\hat{\mathbf r}}$. 

To obtain the velocity of a Bloch electron in an electric field, we first need the velocity matrix element in the periodic function representation. From the Heisenberg equation, the velocity operator reads
$\hat{\mathbf v}(\mathbf k)=\partial_{\mathbf k}\hat{\mathcal H}_0(\mathbf k)/\hbar$, 
where $\hbar$ is the reduced Planck constant and 
$\partial_{\mathbf k}$ denotes $\partial/(\partial{\mathbf k})$ for simplicity. In the periodic function representation, the matrix element of the velocity operator is given by (see Sec.~\ref{supp:bare-velocity} of Supplemental Material~\cite{Supp} for details)
\begin{equation}
\langle n(\mathbf k)|\hat{\mathbf v}(\mathbf k)|m(\mathbf k)\rangle
=
\frac{1}{\hbar}
\partial_{\mathbf k}\varepsilon_n\,\delta_{nm}
+
\frac{i}{\hbar}
(\varepsilon_n-\varepsilon_m)
\mathbfcal A_{nm},
\label{Eq: vmatrix}
\end{equation}
where  
$\mathbfcal A_{nm}=\langle n|i\partial_{\mathbf k}|m\rangle$ is the interband Berry connection~\cite{Berry84proc,Xiao10rmp} [for brevity, we suppress the $\mathbf{k}$ dependence of $|n(\mathbf{k})\rangle$ and $\varepsilon_n(\mathbf{k})$ hereafter].
Equation~\eqref{Eq: vmatrix} shows that the velocity operator admits a well-defined matrix element in the periodic function representation. This is in sharp contrast to the position operator, whose diagonal matrix element is ill-defined~\cite{Blount62ssp}, and forms the basis of our theory.

We then use perturbation theory to construct the field-dressed periodic function 
$|\tilde n\rangle$. 
To obtain the velocity up to second order in $\mathbf{E}$, we expand the periodic function as (see Sec.~\ref{supp:perturbation} of  Supplemental Material~\cite{Supp})
\begin{equation}
|\tilde n\rangle
=
|n\rangle
+
|n^{(1)}\rangle
+
|n^{(2')}\rangle
+
|n^{(2)}\rangle,
\label{Eq: nexpansion}
\end{equation}
where 
$|n^{(1)}\rangle=\sum_{m\neq n}\mathcal C_{mn}|m\rangle$ 
and 
$|n^{(2)}\rangle=\sum_{m\neq n}\mathcal D_{mn}|m\rangle$ 
are the first- and second-order interband corrections, respectively, and
$|n^{(2')}\rangle=\delta_2|n\rangle$ is the second-order normalization correction. 
Here, $\mathcal C_{mn}=e\mathbf E\cdot\mathbfcal A_{mn}/(\varepsilon_n-\varepsilon_m)$, 
$\mathcal D_{mn}=e\mathbf E\cdot
(i\partial_{\mathbf k}\mathcal C_{mn}
+\sum_{l\neq n}\mathcal C_{ln}\mathbfcal A_{ml}
-\mathbfcal A_n\mathcal C_{mn})/(\varepsilon_n-\varepsilon_m)$ with  $\mathbfcal A_n=\langle n|i\partial_{\mathbf k}|n\rangle$ denoting the intraband Berry connection, and
$\delta_2=\sum_{m\neq n}\mathcal C_{nm}\mathcal C_{mn}/2$.

Using the velocity matrix element in Eq.~\eqref{Eq: vmatrix} and the field-dressed periodic function in Eq.~\eqref{Eq: nexpansion}, we directly evaluate the field-corrected velocity as
$\langle \tilde n|
\hat{\mathbf v}
|\tilde n\rangle$.
After some algebra detailed in Sec.~\ref{supp:perturbed-velocity} of  Supplemental Material~\cite{Supp}, the velocity component is given by
\begin{equation}
\tilde v_a
=
\frac{1}{\hbar}\partial_a\varepsilon_n
+
\frac{e}{\hbar}\Omega_n^{ab}E_b
+
\frac{e^2}{\hbar}\Big(
\frac{1}{2}\partial_a\mathcal{G}_n^{bc}
-
\partial_b\mathcal{G}_n^{ac}
\Big)
E_bE_c ,
\label{Eq: v}
\end{equation}
where  $\{a,b,c\}$ are Cartesian indices and repeated indices are implicitly summed over, $\partial_a$ denotes $\partial/(\partial k_a)$ for simplicity,
$\Omega_n^{ab}=-2\,\mathrm{Im}\sum_{m\neq n}\mathcal A_{nm}^a\mathcal A_{mn}^b$ 
is the Berry curvature~\cite{Berry84proc,Xiao10rmp}, and
$\mathcal{G}_n^{ab}=2\,\mathrm{Re}\sum_{m\neq n}\mathcal A_{nm}^a\mathcal A_{mn}^b/(\varepsilon_n-\varepsilon_m)$ 
is the band-normalized quantum metric (also referred to the Berry-connection polarizability)~\cite{Gao14prl,Das23prb,Kaplan24prl}.

\textit{\textcolor{blue}{Field-corrected distribution function}--}Besides modifying the Bloch state and velocity, the electric field also shifts the band energy. 
Accordingly,  the energy argument of the Fermi-Dirac distribution $f_0$, which describes the scattering-independent part of the nonequilibrium distribution, is also modified.
Within perturbation theory (see Sec.~\ref{supp:perturbation} of  Supplemental Material~\cite{Supp}), the field-corrected energy up to second order is given by 
\begin{equation}
\tilde{\mathcal E}_n
=
\varepsilon_n
+
\varepsilon_n^{(1)}
+
\varepsilon_n^{(2)} ,
\label{Eq: energy_pt}
\end{equation}
where 
$\varepsilon_n^{(1)}=eE_a\mathcal A_n^a$ 
and 
$\varepsilon_n^{(2)}=(e^2/2)\mathcal{G}_n^{ab}E_aE_b$ 
are the first- and second-order energy corrections, respectively. 
However, the field-corrected energy $\tilde{\mathcal E}_n$ includes the gauge-dependent potential energy associated with the Berry connection.
Using Eq.~\eqref{Eq: nexpansion}, the field-corrected Berry connection to first order is given by
\begin{equation}
\tilde{\mathbfcal A}_n
=
\langle \tilde n|i\partial_{\mathbf k}|\tilde n\rangle
=
\mathbfcal A_n
+
\mathbfcal A_n^{(1)},
\label{Eq: a_pt}
\end{equation}
where $\mathcal A_n^{(1),a}=e\mathcal{G}_n^{ab}E_b$ denotes the first-order Berry-connection correction. Subtracting the gauge-dependent potential energy $e\mathbf E\cdot\tilde{\mathbfcal A}_n$ from $\tilde{\mathcal E}_n$ yields the effective energy entering the distribution function,
\begin{equation}
\tilde\varepsilon_n
=
\tilde{\mathcal E}_n
-
eE_a\big[
\mathcal A_n^a
+
\mathcal A_n^{(1),a}
\big]
=
\varepsilon_n
+
\bar\varepsilon_n^{(2)},
\label{Eq: ee}
\end{equation}
where
$\bar\varepsilon_n^{(2)}=-(e^2/2)\mathcal{G}_n^{ab}E_aE_b$ is the effective second-order energy correction. Accordingly, the energy argument of $f_0$  is the effective energy $\tilde\varepsilon_n$.

\textit{\textcolor{blue}{Intrinsic nonlinear conductivity}--}We now derive the intrinsic nonlinear conductivity. By definition, the electric current is written as
\begin{equation}
J_a^{\mathrm{int}}
=
-e\int [d\mathbf k]\,
\tilde v_a
f_0(\tilde\varepsilon_n),
\label{Eq: J}
\end{equation}
where $[d\mathbf k]=\sum_n d^d\mathbf k/(2\pi)^d$ with $d$ the dimension. Substituting the field-corrected velocity $\tilde v_a$ [Eq.~\eqref{Eq: v}] and the field-corrected distribution function $f_0(\tilde\varepsilon_n)$ into Eq.~\eqref{Eq: J}, and retaining terms quadratic in the electric field,
the intrinsic nonlinear conductivity is then given by
\begin{equation}
\sigma_{abc}^{\mathrm{int}}
=
e^3
\int [d\mathbf k]\,
\left(
v_a\mathcal{G}_n^{bc}
-
v_b\mathcal{G}_n^{ac}
\right)
f^\prime_0(\varepsilon_n),
\label{Eq: sigma_int}
\end{equation}
where
$v_a=\partial_a \varepsilon_n/\hbar$  is the band velocity and $f^\prime_0(\varepsilon_n)=\partial f_0/\partial\varepsilon_n$.
This equation constitutes the central result of our work.

\textit{\textcolor{blue}{Consistency with wavepacket-based theories}--}Recent works have discussed the differences among existing theories of the intrinsic nonlinear conductivity and their physical origins~\cite{Xiao25prl,Guo26prl,Qiang26advs}.
Among these theories, our approach can be directly compared on the same footing with wavepacket-based theories (including wavepacket dynamics~\cite{Gao14prl}, wavepacket-based perturbation theory~\cite{Qiang26advs}, and the Schrieffer-Wolff transformation approach~\cite{Kaplan24prl}).  This is because they all express the current as an integral of the electron velocity weighted by the distribution function as shown in Eq.~\eqref{Eq: J}. 

We now compare our approach with wavepacket-based theories.
The key distinction among these approaches lies in how the velocity is derived.  In wavepacket dynamics~\cite{Gao14prl}, the starting point is the absence of a well-defined position for a Bloch state.
Therefore, a localized wavepacket is constructed by superposing Bloch states over $\mathbf k$, making its center $\mathbf r_c$ well-defined. Then, the wavepacket velocity $\dot{\mathbf r}_c$ is determined by the semiclassical equations of motion. On the other hand, wavepacket-based perturbation theory~\cite{Qiang26advs} and the Schrieffer-Wolff transformation approach~\cite{Kaplan24prl} also rely on the semiclassical equations of motion and obtain the velocity indirectly. Specifically, they calculate the field-corrected energy and Berry connection, and substitute them into the semiclassical equations of motion to obtain the velocity. By contrast, our approach starts from the well-defined velocity of a Bloch state as shown in Eq.~\eqref{Eq: vmatrix}. Thus, we obtain the velocity directly from the velocity expectation value over the field-dressed periodic function, without constructing a wavepacket.

Although these approaches derive the velocity through different routes, the wavepacket-based theories recover Eq.~\eqref{Eq: v} once the effective energy is properly identified. At first order in the electric field, the velocity given by the original semiclassical equations of motion is~\cite{Sundaram99prb}
\begin{equation}\label{Eq: v_(1)}
v_a^{(1)}=\frac{e}{\hbar}\Omega_n^{ab}E_b,
\end{equation}
which coincides with our first-order velocity in  Eq.~\eqref{Eq: v}. Equation~\eqref{Eq: v_(1)} is later extended to second order by incorporating second-order energy correction and first-order Berry-curvature correction~\cite{Gao14prl},
\begin{equation}
v_a^{(2)}
=
\frac{1}{\hbar}
\partial_a\bar{\varepsilon}_n^{(2)}
+
\frac{e}{\hbar}
\Omega_n^{(1),ab}E_b.
\label{Eq: v2wp}
\end{equation}
Here,  $\bar{\varepsilon}_{n}^{(2)}$ is the effective second-order energy correction,  and $\Omega_n^{(1),ab}=\partial_a\mathcal A_n^{(1),b}-\partial_b\mathcal A_n^{(1),a}$ is the first-order Berry-curvature correction. 
Wavepacket dynamics~\cite{Gao14prl}, wavepacket-based perturbation theory~\cite{Qiang26advs}, and the Schrieffer-Wolff transformation approach~\cite{Kaplan24prl} give the same first-order Berry-connection correction $\mathcal A_n^{(1),a}$ and second-order energy correction $\varepsilon_n^{(2)}$.
By subtracting $eE_a\mathcal A_n^{(1),a}$ from $\varepsilon_n^{(2)}$, all these wavepacket-based theories yield the same effective second-order energy correction $\bar\varepsilon_n^{(2)}$ and recover Eq.~\eqref{Eq: v}.

Finally, we clarify which energy should enter the distribution function $f_0$. As demonstrated in Ref.~\cite{Xiao25prl},  the intrinsic current should be evaluated with  $f_0(\tilde\varepsilon_n)$ rather than $f_0(\varepsilon_n)$. In wavepacket dynamics, this effective energy $\tilde\varepsilon_n$ is distinct from the wavepacket energy~\cite{Gao14prl},
$\tilde{\mathcal E}_n^{\mathrm{wp}}=\varepsilon_n+e\mathbf E\cdot\mathbf r_c^{(0)}+\varepsilon_n^{(2)}$,
where $\mathbf r_c^{(0)}$ is the zeroth-order wavepacket center. The potential energy evaluated at the wavepacket center
[$\mathbf r_c=\mathbf r_c^{(0)}+\mathbfcal A_n^{(1)}$] must be subtracted. This subtraction gives the effective energy
$\tilde\varepsilon_n=\tilde{\mathcal E}_n^{\mathrm{wp}}-e\mathbf E\cdot\mathbf r_c
=\varepsilon_n+\bar\varepsilon_n^{(2)}$.
The same treatment should also be applied in both perturbation theory and the Schrieffer-Wolff transformation approach, as shown in Eq.~\eqref{Eq: ee}. Thus, all these approaches lead to the same distribution function $f_0(\tilde\varepsilon_n)$, which is evaluated at the effective energy $\tilde\varepsilon_n$.
As a result, all these wavepacket-based theories become consistent once the effective  energy entering semiclassical equations of motion and the distribution function is properly identified. They then reproduce the same velocity, distribution function, and intrinsic nonlinear conductivity as those obtained in our theory.

\textit{\textcolor{blue}{Model calculations}--}
To demonstrate the intrinsic nonlinear Hall conductivity in Eq.~\eqref{Eq: sigma_int},
we consider a tight-binding model of antiferromagnetic bilayer CrSBr~\cite{Dong26natmat}
\begin{equation}
\begin{aligned}
\hat{\mathcal{H}}_0
={}&\sum_{\mathbf{k},l,\sigma}
\Bigl\{
[E_d(\mathbf{k})-\sigma m_lJ]
\hat{d}^\dagger_{\mathbf{k}l\sigma}
\hat{d}_{\mathbf{k}l\sigma}\\
&+[E_p(\mathbf{k})
+\sigma\nu_l\lambda_R\sin(k_xa)]
\hat{p}^\dagger_{\mathbf{k}l\sigma}
\hat{p}_{\mathbf{k}l\sigma}\\
&+2\nu_l\Bigl[
(w+w''\cos(k_yb))\cos(k_xa/2)\\
&\qquad+w'\cos(3k_xa/2)
\Bigr]
\bigl(
\hat{d}^\dagger_{\mathbf{k}l\sigma}
\hat{p}_{\mathbf{k}l\sigma}
+\mathrm{h.c.}
\bigr)
\Bigr\}\\
&+4v\sum_{\mathbf{k},\sigma}
\cos(k_xa/2)\cos(k_yb/2)
\bigl(
\hat{p}^\dagger_{\mathbf{k}1\sigma}
\hat{p}_{\mathbf{k}2\sigma}
+\mathrm{h.c.}
\bigr).
\end{aligned}
\label{Eq: CrSBr-model}
\end{equation}
Here,
$\hat{d}_{\mathbf{k}l\sigma}$ and
$\hat{p}_{\mathbf{k}l\sigma}$ annihilate electrons in the Cr
$3d_{3y^2-r^2}$ and Br $4p_z$ orbitals, respectively,
$l=1,2$ labels the layers, $\sigma=\pm1$ labels spin,
$E_d(\mathbf{k})=\Delta-2t_d\cos(k_yb)$,
$E_p(\mathbf{k})=-2t_p\cos(k_xa)-2t_p'\cos(2k_xa)$,
$a$ and $b$ are the lattice constants along $x$ and $y$ directions, respectively, $\Delta$ is the $d$--$p$ on-site energy difference,
$t_d,t_p,t_p',w,w',w'',v$ are the hopping amplitudes,
$J$ represents exchange coupling,  $\lambda_R$ represents the Rashba spin--orbit coupling,
and $\nu_1=-\nu_2=1$ with $m_l=\nu_l$ specifies the  antiferromagnetic 
order.

\begin{figure}[t]
\centering
\includegraphics[width=\columnwidth]{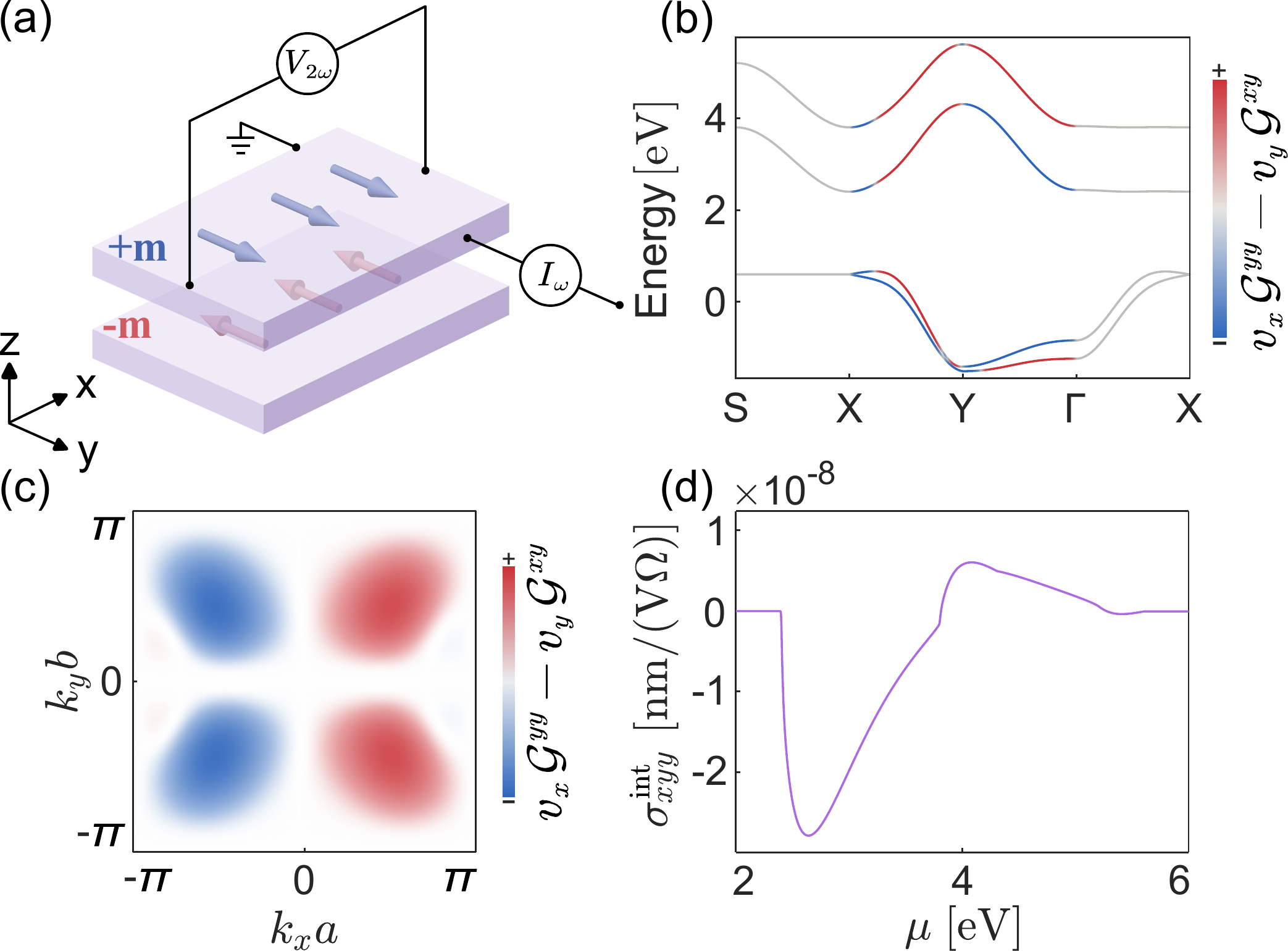}
\caption{
Model calculations for the bilayer CrSBr described by
Eq.~\eqref{Eq: CrSBr-model}.
(a) Antiferromagnetic order and measurement of $\sigma_{xyy}^{\mathrm{int}}$ with
a current $I_\omega$ applied along $y$ and a 
second-harmonic voltage $V_{2\omega}$ measured along $x$.
(b) Band dispersions along S--X--Y--$\Gamma$--X,
colored by  $v_x\mathcal G_n^{yy}-v_y\mathcal G_n^{xy}$.
(c) Momentum-space distribution of
$v_x\mathcal G_n^{yy}-v_y\mathcal G_n^{xy}$ for the highest conduction band.
(d) Intrinsic nonlinear Hall conductivity
$\sigma^{\mathrm{int}}_{xyy}$ as a function of the chemical potential $\mu$ at $T=20$ K.
The model parameters are
$\Delta=3.8$ eV, $t_d=0.35$ eV, $t_p=0.4$ eV,
$t_p'=0.1$ eV, $w=0.4$ eV, $w'=0.1$ eV,
$w''=-0.32$ eV, $J=0.7$ eV, $\lambda_R=0.1$ eV,
and $v=0.05$ eV~\cite{Dong26natmat}.
The lattice constants are $a=3.512$ \AA\ and
$b=4.762$ \AA~\cite{Boix22advmater}.
}
\label{Fig: CrSBr}
\end{figure}

As illustrated in Fig.~\ref{Fig: CrSBr}(a), the Hamiltonian in Eq.~\eqref{Eq: CrSBr-model} depicts a bilayer CrSBr, which has
parallel magnetic moments within each layer and antiparallel
moments between the layers.
This material breaks inversion, time reversal, and the $n$-fold rotation ,
while preserving combined inversion and time reversal, thereby permitting an intrinsic nonlinear Hall response~\cite{Gao21prl,YangSY21prl}.
For the specific response $\sigma_{xyy}^{\mathrm{int}}$,
it can be measured experimentally  as shown in Fig.~\ref{Fig: CrSBr}(a).
Figure~\ref{Fig: CrSBr}(b) shows the band dispersions of Eq.~\eqref{Eq: CrSBr-model} colored by
the quantum-geometric quantity $v_x\mathcal G_n^{yy}-v_y\mathcal G_n^{xy}$
governing $\sigma_{xyy}^{\mathrm{int}}$.
Figure~\ref{Fig: CrSBr}(c) maps this quantity in momentum space
for the highest conduction band.
Figure~\ref{Fig: CrSBr}(d)
shows the chemical potential dependence of intrinsic nonlinear Hall conductivity $\sigma^{\mathrm{int}}_{xyy}$, which is suppressed
in the band gap and peaks upon light electron doping.

\textit{\textcolor{blue}{Note added}--}During the  refereeing process of the manuscript, we became aware of a related work~\cite{Tang26arXiv}, which also uses perturbation theory to calculate the intrinsic nonlinear conductivity.  Nevertheless, there are several differences. First, in Ref.~\cite{Tang26arXiv}, the position matrix element, velocity matrix element, and velocity expectation value are all evaluated in the Bloch representation $|\psi_n(\mathbf k)\rangle$. By contrast, our derivation is formulated entirely in the periodic function representation $|n(\mathbf{k})\rangle$. Second, Ref.~\cite{Tang26arXiv} directly applies the standard perturbation formula after replacing the discrete band label $n$ by the extended label $\{n,\mathbf k\}$. Instead, we start from the perturbative equations and solve them order by order.
Third, we systematically compare our formulation with the wavepacket-based theories and demonstrate their consistency. Together, our work and Ref.~\cite{Tang26arXiv} complement each other, establishing a comprehensive perturbative framework for intrinsic nonlinear transport.

\begin{acknowledgments}
This work is supported by the National Key R$\&$D Program of China (2022YFA1403700), the National Natural Science Foundation of China (12525401 and 12350402), the Quantum Science and Technology--National Science and Technology Major Project (Grant No. 2021ZD0302400), the Guangdong Provincial Key Laboratory of Topological Matter (2026B1212010004), the Guangdong Basic and Applied Basic Research Foundation (2023B0303000011), the Guangdong Provincial Quantum Science Strategic Initiative (GDZX2201001 and GDZX2401001), the Science, Technology and Innovation Commission of Shenzhen Municipality (ZDSYS20190902092905285), High-level Special Funds (G03050K004), the New Cornerstone Science Foundation through the XPLORER PRIZE, the China Scholarship Council Program (Grant No. 202508440272), and the Center for Computational Science and Engineering of SUSTech.
\end{acknowledgments}

\clearpage
\onecolumngrid
\setcounter{page}{1}
\setcounter{section}{0}
\setcounter{equation}{0}
\setcounter{figure}{0}
\setcounter{table}{0}
\setcounter{secnumdepth}{3}
\renewcommand{\thepage}{S\arabic{page}}
\renewcommand{\thesection}{S\Roman{section}}
\renewcommand{\thetable}{S\arabic{table}}
\renewcommand{\theequation}{S\arabic{equation}}
\renewcommand{\thefigure}{S\arabic{figure}}
\renewcommand{\theHequation}{supp.\arabic{equation}}
\renewcommand{\theHsection}{supp.\arabic{section}}
\renewcommand{\theHfigure}{supp.\arabic{figure}}
\renewcommand{\theHtable}{supp.\arabic{table}}
\pdfbookmark[0]{Supplemental Material}{supplement}
\phantomsection\label{supp:title}
\makeatletter
\newcommand{\suppsection}[2]{%
  \section{#1}\def\@currentlabelname{#1}\label{#2}}
\makeatother

\begin{center}
{\large\bfseries Supplemental Material for\\
		``Intrinsic second-order nonlinear Hall effect from perturbation theory''\par}
\medskip
Zhi-Hao Wei\ensuremath{^{1,\ast}}, Xiao-Bin Qiang\ensuremath{^{1,2,\ast}}, Renjie Tong\ensuremath{^{1}}, Hai-Zhou Lu\ensuremath{^{1,3,\dagger}}, and X. C. Xie\ensuremath{^{4,5,6}}\par
\smallskip
{\small\itshape
\textsuperscript{1}State Key Laboratory of Quantum Functional Materials, Department of Physics, Guangdong Provincial Key Laboratory of Topological Matter, and Guangdong Basic Research Center of Excellence for Quantum Science, Southern University of Science and Technology (SUSTech), Shenzhen 518055, China\par
\textsuperscript{2}Division of Physics and Applied Physics, School of Physical and Mathematical Sciences, Nanyang Technological University, 21 Nanyang Link, 637371, Singapore\par
\textsuperscript{3}Quantum Science Center of Guangdong-Hong Kong-Macao Greater Bay Area (Guangdong), Shenzhen 518045, China\par
\textsuperscript{4}International Center for Quantum Materials, School of Physics, Peking University, Beijing 100871, China\par
\textsuperscript{5}Interdisciplinary Center for Theoretical Physics and Information
Sciences (ICTPIS), Fudan University, Shanghai 200433, China\par
\textsuperscript{6}Hefei National Laboratory, Hefei 230088, China\par
}
\smallskip
(Dated: \today)
\end{center}
\begin{center}\begin{minipage}{0.78\textwidth}\small
This Supplemental Material provides detailed derivations and supporting calculations, which includes \mbox{Refs.~\cite{supp-Ashcroft76book,supp-Gao14prl,supp-Griffiths18book,supp-Sakurai20book}}
\end{minipage}\end{center}
\begin{center}\bfseries CONTENTS\end{center}
\newcommand{\suppcontentsline}[1]{%
  \noindent\hyperref[#1]{\ref*{#1}.\quad\nameref*{#1}}\dotfill\pageref{#1}\par}
\suppcontentsline{supp:bare-velocity}
\suppcontentsline{supp:perturbation}
\suppcontentsline{supp:perturbed-velocity}
\noindent\hyperref[supp:references]{References}\dotfill\pageref{supp:references}\par
\vfill
{\footnotesize
\noindent\(\!^{\ast}\)\,These authors contributed equally to this work.\par
\noindent\(\!^{\dagger}\)\,Corresponding author: luhz@sustech.edu.cn\par
}
\clearpage
\suppsection{Bare velocity of the Bloch state}{supp:bare-velocity}
In this Supplemental Material, we derive the velocity of a Bloch electron in a uniform electric field up to second order using perturbation theory.
We start from the unperturbed problem in a periodic crystal. 
The single-electron eigenstate in the crystal is Bloch state,
\begin{equation}
	\hat{\mathcal{H}}_0|\psi _n\left( \mathbf{k} \right) \rangle
	=
	\varepsilon _n\left( \mathbf{k} \right)
	|\psi _n\left( \mathbf{k} \right) \rangle ,
	\label{supp:Eq: bs_eg}
\end{equation}
where $\hat{\mathcal{H}}_0$ is the unperturbed crystal Hamiltonian, $|\psi _n\left( \mathbf{k} \right) \rangle$ is the Bloch state labeled by the band index $n$ and the crystal momentum $\mathbf{k}$, and $\varepsilon _n\left( \mathbf{k} \right)$ is the band dispersion. 
Each Bloch state can be written as a plane-wave factor multiplying a cell-periodic function,
\begin{equation}
	|\psi _n\left( \mathbf{k} \right) \rangle
	=
	e^{i\mathbf{k}\cdot \hat{\mathbf{r}}}
	|n\left( \mathbf{k} \right) \rangle .
	\label{supp:Eq: bs_decom}
\end{equation}
Substituting Eq.~\eqref{supp:Eq: bs_decom} into Eq.~\eqref{supp:Eq: bs_eg}, we obtain
\begin{equation}
	\hat{\mathcal H}_0(\mathbf{k})|n(\mathbf{k})\rangle
	=
	\varepsilon_n(\mathbf{k})|n(\mathbf{k})\rangle ,
\end{equation}
where 
$	\hat{\mathcal H}_0(\mathbf{k})
	=
	e^{-i\mathbf{k}\cdot\hat{\mathbf r}}
	\hat{\mathcal H}_0
	e^{i\mathbf{k}\cdot\hat{\mathbf r}}$
is the Bloch Hamiltonian.

In the presence of a uniform electric field $\mathbf E$,
the Hamiltonian becomes
\begin{equation}
	\hat{\mathcal H}
	=
	\hat{\mathcal H}_0
	+
	e\mathbf E\cdot\hat{\mathbf r},
\end{equation}
where $-e$ is the electron charge and $\hat{\mathbf r}$ is the position operator. 
In the periodic function representation, it becomes
\begin{equation}
	\hat{\mathcal H}({\mathbf k})
	=
	e^{-i\mathbf k\cdot\hat{\mathbf r}}
	\hat{\mathcal H}
	e^{i\mathbf k\cdot\hat{\mathbf r}}
	=
	\hat{\mathcal H}_0(\mathbf k)
	+
	e\mathbf E\cdot\hat{\mathbf r}.
	\label{supp:Eq: Hk}
\end{equation}
To obtain the velocity of a Bloch electron in an electric field, we treat the electric-field term $e\mathbf E\cdot\hat{\mathbf r}$ as a perturbation and construct the field-dressed periodic function $|\tilde n(\mathbf k)\rangle$. 
The velocity is then obtained by evaluating the expectation value of the $\mathbf k$-space velocity operator,
\begin{equation}
    \tilde{\mathbf v}
	=
	\langle \tilde n(\mathbf k)|
	\hat{\mathbf v}(\mathbf k)
	|\tilde n(\mathbf k)\rangle ,
\end{equation}
and expanding the result order by order in $\mathbf E$.

First, we clarify the velocity operator matrix element in the periodic function representation.
Starting from the Heisenberg equation, the
corresponding $\mathbf{k}$-space velocity operator is
\begin{equation}
	\hat{\mathbf v}(\mathbf{k})
	=	\frac{i}{\hbar}
	\left[
	\hat{\mathcal H}(\mathbf{k}),
	\hat{\mathbf r}
	\right]=
	\frac{i}{\hbar}
	\left[
	\hat{\mathcal H}_0(\mathbf{k}),
	\hat{\mathbf r}
	\right],
\end{equation}
where $\hbar$ is the reduced Planck constant.
Therefore, its matrix element can be expressed in terms of the position operator matrix element as
\begin{equation}
		\langle n(\mathbf{k})|\hat{\mathbf v}(\mathbf{k})|m(\mathbf{k})\rangle
		=
		\frac{i}{\hbar}
		\langle n(\mathbf{k})|
		\left[
		\hat{\mathcal H}_0(\mathbf{k}),
		\hat{\mathbf r}
		\right]
		|m(\mathbf{k})\rangle
		=
		\frac{i}{\hbar}
		\left(
		\varepsilon_n-\varepsilon_m
		\right)
		\langle n(\mathbf{k})|\hat{\mathbf r}|m(\mathbf{k})\rangle .
	\label{supp:Eq: velement_view1}
\end{equation}
Equation~\eqref{supp:Eq: velement_view1} does not by itself provide well-defined velocity operator matrix element.  In particular, the diagonal position matrix element $\langle n(\mathbf{k})|\hat{\mathbf r}|n(\mathbf{k})\rangle$ is not  well-defined, so the diagonal velocity cannot be determined from this relation.
To obtain the full velocity operator matrix element, we differentiate the Bloch Hamiltonian,
\begin{equation}
	\begin{aligned}
		\partial_{\mathbf{k}}\hat{\mathcal H}_0(\mathbf{k})
		=
		\partial_{\mathbf{k}}
		\big(
		e^{-i\mathbf{k}\cdot\hat{\mathbf r}}
		\hat{\mathcal H}_0
		e^{i\mathbf{k}\cdot\hat{\mathbf r}}
		\big)
		=
		-i\hat{\mathbf r}\hat{\mathcal H}_0(\mathbf{k})
		+
		i\hat{\mathcal H}_0(\mathbf{k})\hat{\mathbf r}
		=
		i
		\left[
		\hat{\mathcal H}_0(\mathbf{k}),
		\hat{\mathbf r}
		\right]
		=
		\hbar\hat{\mathbf v}(\mathbf{k}),
	\end{aligned}
\end{equation}
where $\partial_{\mathbf k}$ represent $\partial/(\partial\mathbf k) $.
As a result, the  velocity operator can be rewritten as
\begin{equation}
	\hat{\mathbf v}(\mathbf{k})
	=
	\frac{1}{\hbar}
	\partial_{\mathbf{k}}\hat{\mathcal H}_0(\mathbf{k}).
\end{equation}
Using this expression, we find
\begin{equation}
	\begin{aligned}
		\langle n(\mathbf{k})|\hat{\mathbf v}(\mathbf{k})|m(\mathbf{k})\rangle
		&=
		\frac{1}{\hbar}
		\langle n(\mathbf{k})|
		\partial_{\mathbf{k}}\hat{\mathcal H}_0(\mathbf{k})
		|m(\mathbf{k})\rangle
		\\
		&=
		\frac{1}{\hbar}
		\langle n|
		\partial_{\mathbf{k}}
\textit{}		\big[
		\hat{\mathcal H}_0(\mathbf{k})|m\rangle
		\big]
		-
		\hat{\mathcal H}_0(\mathbf{k})
		|\partial_{\mathbf{k}}m\rangle
		\\
		&=
		\frac{1}{\hbar}
		\partial_{\mathbf{k}}\varepsilon_m(\mathbf{k})\delta_{mn}
		+
		\frac{i}{\hbar}
		\left[
		\varepsilon_n(\mathbf{k})-\varepsilon_m(\mathbf{k})
		\right]
		\mathbfcal{A}_{nm},
	\end{aligned}
	\label{supp:Eq: velement_view2}
\end{equation}
where $\mathbfcal A_{nm}=\langle n(\mathbf{k})|i\partial_{\mathbf k}|m(\mathbf{k})\rangle$ is the interband Berry connection.
Equation~\eqref{supp:Eq: velement_view2} provides the complete velocity operator matrix element,
\begin{equation}\left\{
\begin{aligned}
&\langle n|\hat{\mathbf v}|n\rangle=
\frac{1}{\hbar}
\partial_{\mathbf{k}}\varepsilon_n ,\\
&\langle n|\hat{\mathbf v}|m\rangle
=
\frac{i}{\hbar}
\left(
\varepsilon_n-\varepsilon_m
\right)
\mathbfcal{A}_{nm},
\qquad m\neq n .
\end{aligned}\right.
\label{supp:Eq: vmatrix}
\end{equation}

We can also derive the position operator matrix element in periodic function representation by comparing Eqs.~\eqref{supp:Eq: velement_view1} and
\eqref{supp:Eq: velement_view2}. For $m\neq n$, we find that the off-diagonal position operator matrix element is indeed represented by the interband Berry connection 	$\mathbfcal A_{nm}$. For $m=n$, 
Eq.~\eqref{supp:Eq: velement_view2} gives $\partial_{\mathbf k}\varepsilon_n(\mathbf k)/\hbar$,
whereas Eq.~\eqref{supp:Eq: velement_view1} would give zero for any ordinary diagonal
matrix element. This mismatch shows that the diagonal
matrix element of the position operator is not well-defined within the periodic function representation and cannot be identified as intraband Berry connection  $\mathbfcal A_{n}=\langle n(\mathbf{k})|i\partial_{\mathbf k}|n(\mathbf{k})\rangle$,
\begin{equation}\left\{
	\begin{aligned}
	&\langle n|\hat{\mathbf r}|n\rangle
	\ne
	\mathbfcal A_{n},\\
	&\langle n|\hat{\mathbf r}|m\rangle
	=
	\mathbfcal A_{nm},
	\qquad m\neq n.
	\end{aligned}\right.
	\label{supp:Eq: rmatrix}
\end{equation}

\suppsection{Bloch perturbation theory}{supp:perturbation}
\label{supp:Sec: bloch_PT}
Second, we treat the electric field term $\hat{V}=e\mathbf E\cdot \hat{\mathbf r}$ as the perturbation and expand the periodic function $|n(\mathbf k)\rangle$ into its field-dressed form $|\tilde n(\mathbf k)\rangle$.
To account for the velocity up to second order, we write its expansion up to second order in the electric field as
\begin{equation}
	|\tilde n(\mathbf{k})\rangle
	=
	|n(\mathbf{k})\rangle
	+
	|n(\mathbf{k})^{(1)}\rangle
	+
	|n(\mathbf{k})^{(2')}\rangle
	+
	|n(\mathbf{k})^{(2)}\rangle,
	\label{supp:Eq: nexpansion}
\end{equation}
where $|n(\mathbf{k})^{(1)}\rangle=\sum_{m\neq n}\mathcal C_{mn}|m(\mathbf{k})\rangle$ and $|n(\mathbf{k})^{(2)}\rangle=\sum_{m\neq n}\mathcal D_{mn}|m(\mathbf{k})\rangle$ are the first- and second-order interband corrections, respectively, and $|n(\mathbf{k})^{(2')}\rangle=\delta_2|n(\mathbf{k})\rangle$ is the second-order normalization correction.

In standard perturbation theory, the second-order interband correction takes the form
\begin{equation}
	|n^{(2)}\rangle
	=
	\sum_{m\neq n}
	\left(
	\sum_{l\neq n}
	\frac{V_{ml}V_{ln}}{\varepsilon_{nm}\varepsilon_{nl}}
	-
	\frac{V_{nn}V_{mn}}{\varepsilon_{nm}^2}
	\right)
	|m\rangle ,\label{supp:Eq: n2sd}
\end{equation}
where $V_{nm}=\langle n|\hat V|m\rangle$ and $\varepsilon_{nm}=\varepsilon_n-\varepsilon_m$. 
However, this formula contains the diagonal matrix element $V_{nn}=e\mathbf E\cdot\langle n|\hat{\mathbf r}|n\rangle$
, which is not well-defined as shown in Eq.~\eqref{supp:Eq: rmatrix}. 
Therefore, standard perturbation theory cannot be directly applied to obtain the second-order interband correction.

To avoid the ill-defined diagonal matrix elements of $\hat{\mathbf r}$ in the periodic function representation, we instead expand the full Bloch state and use the matrix elements of the position operator in the Bloch representation,
\begin{equation}
	\langle \psi _n\left( \mathbf{k} \right) |\hat{\mathbf{r}}|\psi _m\left( \mathbf{k}^{\prime} \right) \rangle
	=
	i\partial _{\mathbf{k}}\delta \left( \mathbf{k}-\mathbf{k}^{\prime} \right) \delta _{nm}
	+\delta \left( \mathbf{k}-\mathbf{k}^{\prime} \right) \mathbfcal{A} _{nm}\left( \mathbf{k}^{\prime} \right).
	\label{supp:Eq: rinbs}
\end{equation}
The field-dressed Bloch state is written as
\begin{equation}
	|\tilde\psi_n(\mathbf k)\rangle
    =
	|\psi_n(\mathbf k)\rangle
	+
	|\psi_n(\mathbf k)^{(1)}\rangle
	+
	|\psi_n(\mathbf k)^{(2')}\rangle
	+
	|\psi_n(\mathbf k)^{(2)}\rangle ,
	\label{supp:Eq: psi_expansion}
\end{equation}
where $|\psi_n(\mathbf k)^{(1)}\rangle=\sum_{m\neq n}\mathcal C_{mn}|\psi_m(\mathbf k)\rangle$ and $|\psi_n(\mathbf k)^{(2)}\rangle=\sum_{m\neq n}\mathcal D_{mn}|\psi_m(\mathbf k)\rangle$ are the first- and second-order interband corrections, respectively, and $|\psi_n(\mathbf k)^{(2')}\rangle=\delta_2|\psi_n(\mathbf k)\rangle$ is the second-order normalization correction.
After this expansion is obtained, we factor out the plane-wave part $|\tilde\psi_n(\mathbf k)\rangle=e^{i\mathbf k\cdot\hat{\mathbf r}}|\tilde n(\mathbf k)\rangle$ to obtain the expansion of the periodic function in Eq.~\eqref{supp:Eq: nexpansion}.

We emphasize that this procedure is not obtained by simply replacing the discrete band label $n$ in standard perturbation theory with the continuous label $\{n,\mathbf k\}$ in Eq.~\eqref{supp:Eq: n2sd}. 
Such a replacement leads to ill-defined distributions involving the position operator,
\begin{equation}
	\begin{aligned}
		&V_{m\mathbf k',l\mathbf k''}V_{l\mathbf k'',n\mathbf k}
		\sim
		\partial_{\mathbf k'}\delta(\mathbf k'-\mathbf k'')
		\,\delta(\mathbf k''-\mathbf k),
		\\
		&V_{n\mathbf k,n\mathbf k}V_{m\mathbf k',n\mathbf k}
		\sim
		\left[
		\partial_{\mathbf k}\delta(\mathbf 0)
		+
		\delta(\mathbf 0)
		\right]
		\delta(\mathbf k'-\mathbf k),
	\end{aligned}
\end{equation}
which are mathematically ill-defined.

Instead, we use Bloch perturbation theory to determine the coefficients in Eq.~\eqref{supp:Eq: psi_expansion}, thereby obtaining the expansion in Eq.~\eqref{supp:Eq: nexpansion}.
We start from the perturbed eigenvalue equation,
\begin{equation}
	\hat{\mathcal H}|\tilde\psi_n(\mathbf k)\rangle
	=
	\tilde\varepsilon_n(\mathbf k)|\tilde\psi_n(\mathbf k)\rangle ,
	\label{supp:Eq: BPT_eigen}
\end{equation}
where $\tilde\varepsilon_n
=
\varepsilon_n
+
\varepsilon_n^{(1)}
+
\varepsilon_n^{(2)}
+\cdots $
is the field-corrected energy  with the superscript $\nu$ representing the $\nu$-th order energy correction.
We then collect terms order by order in $\mathbf E$ and project them onto the unperturbed Bloch states.
At first order, Eq.~\eqref{supp:Eq: BPT_eigen} gives
\begin{equation}
	\hat{\mathcal H}_0|\psi_n(\mathbf k)^{(1)}\rangle
	+
	e\mathbf E\cdot\hat{\mathbf r}|\psi_n(\mathbf k)\rangle
	=
	\varepsilon_n|\psi_n(\mathbf k)^{(1)}\rangle
	+
	\varepsilon_n^{(1)}|\psi_n(\mathbf k)\rangle .
	\label{supp:Eq: BPT_O1}
\end{equation}
Projecting Eq.~\eqref{supp:Eq: BPT_O1} onto $\langle\psi_n(\mathbf k')|$ and using  Eq.~\eqref{supp:Eq: rinbs} give the first-order energy correction,
\begin{equation}
	\varepsilon_n^{(1)}
	=
	e\mathbf{E}\cdot\mathbfcal A_n.
	\label{supp:Eq: BPT_e1}
\end{equation}
Projecting Eq.~\eqref{supp:Eq: BPT_O1} onto $\langle\psi_m(\mathbf k')|$ with $m\neq n$ gives the first-order interband coefficient,
\begin{equation}
	\mathcal C_{mn}
	=
	\frac{e\mathbf E\cdot\mathbfcal A_{mn}}{\varepsilon_{nm}}.
\label{supp:Eq: Cmn}
\end{equation} 
At second order, Eq.~\eqref{supp:Eq: BPT_eigen} gives
\begin{equation}
	\hat{\mathcal H}_0|\psi_n(\mathbf k)^{(2)}\rangle
	+
	e\mathbf E\cdot\hat{\mathbf r}|\psi_n(\mathbf k)^{(1)}\rangle
	=
	\varepsilon_n|\psi_n(\mathbf k)^{(2)}\rangle
	+
	\varepsilon_n^{(1)}|\psi_n(\mathbf k)^{(1)}\rangle
	+
	\varepsilon_n^{(2)}|\psi_n(\mathbf k)\rangle .
	\label{supp:Eq: BPT_O2}
\end{equation}
Projecting Eq.~\eqref{supp:Eq: BPT_O2} onto $\langle\psi_n(\mathbf k')|$ gives the second-order energy correction,
\begin{equation}
	\varepsilon_n^{(2)}
	=
	e\mathbf E\cdot
	\sum_{m\neq n}
	\mathcal C_{mn}\mathbfcal A_{nm}
	=
	\frac{e^2}{2}\mathcal{G}_n^{ab}E_aE_b ,
	\label{supp:Eq: BPT_e2}
\end{equation}
where
\begin{equation}
	\mathcal{G}_n^{ab}
	=
	2\,\mathrm{Re}
	\sum_{m\neq n}
	\frac{
		\mathcal A_{nm}^{a}\mathcal A_{mn}^{b}
	}
	{\varepsilon_n-\varepsilon_m}
	\label{supp:Eq: qmetric_def}
\end{equation} is the band-normalized quantum metric and repeated indices are implicitly summed over. We remark that the second-order energy correction $\varepsilon_n^{(2)}$ was first derived in Ref.~\cite{supp-Gao14prl}.
Projecting Eq.~\eqref{supp:Eq: BPT_O2} onto $\langle\psi_m(\mathbf k')|$ with $m\neq n$ gives the second-order interband coefficient,
\begin{equation} 
	\mathcal D_{mn}
	=
	\frac{e\mathbf E}{\varepsilon_{nm}}\cdot
	\big(
	i\partial_{\mathbf{k}}\mathcal C_{mn}
	+
	\sum_{l\neq n}\mathcal C_{ln}\mathbfcal{A}_{ml}
	-
	\mathbfcal{A}_n\mathcal C_{mn}
 	\big).
\label{supp:Eq: Dmn}
\end{equation}
The second-order normalization correction is fixed by
$\langle\tilde n(\mathbf k)|\tilde n(\mathbf k)\rangle=1+\mathcal O(E^3)$,  thus we have
\begin{equation}
	\delta_2
	=
	\sum_{m\neq n}
	\frac{\mathcal C_{nm}\mathcal C_{mn}}{2}.
\label{supp:Eq: delta2}
\end{equation}

\suppsection{Perturbed velocity up to second order in electric field}{supp:perturbed-velocity}
Third, we evaluate the velocity of a Bloch electron in an electric field using the velocity operator matrix element in Eq.~\eqref{supp:Eq: vmatrix} and the field-dressed periodic function in Eq.~\eqref{supp:Eq: nexpansion}, 
\begin{equation}
\langle \tilde n|\hat{\mathbf v}|\tilde n\rangle
=
\big[
\langle n|
+
\langle n^{(1)}|
+
\langle n^{(2)}|
+
\langle n^{(2')}|
\big]
\hat{\mathbf v}
\big[
|n\rangle
+
|n^{(1)}\rangle
+
|n^{(2)}\rangle
+
|n^{(2')}\rangle
\big].
\end{equation}
At zeroth order, we obtain
\begin{equation}
	\mathbf v
	=
	\langle n|\hat{\mathbf v}|n\rangle
	=
	\frac{1}{\hbar}\partial_{\mathbf{k}}\varepsilon_n.
	\label{supp:Eq: v0}
\end{equation}
At first order, we obtain
\begin{equation}
	\begin{aligned}
		\mathbf v^{(1)}
		&=
		\langle n|\hat{\mathbf v}|n^{(1)}\rangle
		+\text{c.c.}
		\\
		&=
		\sum_{m\neq n}
		\mathcal C_{mn}
		\big[
		\frac{1}{\hbar}\partial_{\mathbf{k}}\varepsilon_m\,\delta_{mn}
		+
		\frac{i}{\hbar}
		\left(
		\varepsilon_n-\varepsilon_m
		\right)
		\mathbfcal{A}_{nm}
		\big]
		+\text{c.c.}
		\\
		&=
		\sum_{m\neq n}
		e\mathbf E\cdot\mathbfcal{ A}_{mn}
		\frac{i}{\hbar}
		\mathbfcal{A}_{nm}
		+\text{c.c.}
		\\
		&=
		\mathrm{Re}
		\Big\{
		\sum_{m\neq n}
		\frac{2i}{\hbar}
		e\mathbf E\cdot\mathbfcal{A}_{mn}
		\mathbfcal{A}_{nm}
		\Big\}.
	\end{aligned}
\end{equation}
Therefore, the $a$-th component is
\begin{equation}
	v_a^{(1)}
	=
	\mathrm{Im}
	\Big\{
	-\sum_{m\neq n}
	\frac{2}{\hbar}
	eE_b\mathcal A_{mn}^b\mathcal A_{nm}^a
	\Big\}
	=
	\frac{e}{\hbar}\Omega_n^{ab}E_b ,
	\label{supp:Eq: v1}
\end{equation}
where $\Omega_n^{ab}
=
-2\,\mathrm{Im}
\sum_{m\neq n}
\mathcal A_{nm}^{a}\mathcal A_{mn}^{b}$ is the Berry curvature. 

Next, we compute the second-order velocity,
\begin{equation}
	\mathbf v^{(2)}
	=
	\langle n^{(1)}|\hat{\mathbf v}|n^{(1)}\rangle
	+
	\langle n|\hat{\mathbf v}|n^{(2)}\rangle
	+\text{c.c.}
	+
	\langle n|\hat{\mathbf v}|n^{(2')}\rangle
	+\text{c.c.} .
	\label{supp:Eq: v2_decom}
\end{equation}
The first term in Eq.~\eqref{supp:Eq: v2_decom} gives
\begin{equation}
	\begin{aligned}
		\langle n^{(1)}|\hat{\mathbf v}|n^{(1)}\rangle
		&=
		\sum_{m,l\neq n}
		\mathcal C_{ln}^*\mathcal C_{mn}
		\langle l|\hat{\mathbf v}|m\rangle
		\\
		&=
		\sum_{m,l\neq n}
		\mathcal C_{ln}^*\mathcal C_{mn}
		\big[
		\frac{1}{\hbar}\partial_{\mathbf{k}}\varepsilon_m\,\delta_{lm}
		+
		\frac{i}{\hbar}
		\left(
		\varepsilon_l-\varepsilon_m
		\right)
		\mathbfcal{A}_{lm}
		\big]
		\\
		&=
		-\sum_{m\neq n}
		\mathcal C_{nm}\mathcal C_{mn}
		\frac{1}{\hbar}
		\partial_{\mathbf{k}}\varepsilon_m
		-
		\frac{i}{\hbar}
		\sum_{m,l\neq n}
		\mathcal C_{nl}\mathcal C_{mn}
		\left(
		\varepsilon_l-\varepsilon_m
		\right)
		\mathbfcal{A}_{lm}.
	\end{aligned}
	\label{supp:Eq: v11}
\end{equation}
The contribution involving $|n^{(2)}\rangle$ in Eq.~\eqref{supp:Eq: v2_decom}  is
\begin{equation}
	\begin{aligned}
		\langle n|\hat{\mathbf v}|n^{(2)}\rangle+\text{c.c.}
		&=
		\sum_{m\neq n}
		\mathcal D_{mn}
		\big[
		\frac{1}{\hbar}\partial_{\mathbf{k}}\varepsilon_m\,\delta_{mn}
		+
		\frac{i}{\hbar}
		\left(
		\varepsilon_n-\varepsilon_m
		\right)
		\mathbfcal {A}_{nm}
		\big]
		+\text{c.c.}
		\\
		&=
		\frac{i}{\hbar}
		\sum_{m\neq n}
		e\mathbf E\cdot
		\big(
		i\partial_{\mathbf{k}}\mathcal C_{mn}
		+
		\sum_{l\neq n}\mathcal C_{ln}\mathbfcal{A}_{ml}
		-
		\mathbfcal {A}_n\mathcal C_{mn}
		\big)
		\mathbfcal {A}_{nm}
		+\text{c.c.} .
	\end{aligned}
	\label{supp:Eq: v02}
\end{equation}
The normalization contribution in Eq.~\eqref{supp:Eq: v2_decom}  is
\begin{equation}
	\begin{aligned}
		\langle n|\hat{\mathbf v}|n^{(2')}\rangle+\text{c.c.}
		&=
		\sum_{m\neq n}
		\frac{\mathcal C_{nm}\mathcal C_{mn}}{2}
		\frac{1}{\hbar}\partial_{\mathbf{k}}\varepsilon_n
		+\text{c.c.}=
		\sum_{m\neq n}
		\mathcal C_{nm}\mathcal C_{mn}
		\frac{1}{\hbar}\partial_{\mathbf{k}}\varepsilon_n .
	\end{aligned}
	\label{supp:Eq: v02p}
\end{equation}
Combining Eqs.~\eqref{supp:Eq: v11}--\eqref{supp:Eq: v02p}, we obtain
\begin{equation}
	\begin{aligned}
		v_a^{(2)}
		=&
		\frac{i}{\hbar}
		\sum_{m\neq n}
		e\mathbf E\cdot
		\big(
		i\partial_{\mathbf{k}}\mathcal C_{mn}
		+
		\sum_{l\neq n}\mathcal C_{ln}\mathbfcal{ A}_{ml}
		-
		\mathbfcal{A}_n\mathcal C_{mn}
		\big)
		\mathcal A_{nm}^a
		+\text{c.c.}
		\\
		&+
		\sum_{m\neq n}
		\mathcal C_{nm}\mathcal C_{mn}
		\frac{1}{\hbar}
		\partial_a
		\left(
		\varepsilon_n-\varepsilon_m
		\right)
		-
		\frac{i}{\hbar}
		\sum_{m,l\neq n}
		\mathcal C_{nl}\mathcal C_{mn}
		\left(
		\varepsilon_l-\varepsilon_m
		\right)
		\mathcal A_{lm}^a .
	\end{aligned}
\end{equation}
We then decompose this expression as
\begin{equation}
	v_a^{(2)}
	=    
	v_a^{(2),1}+
	v_a^{(2),2}
	+
	v_a^{(2),3}
	+
	v_a^{(2),4},
\end{equation}
where
\begin{equation}
	\begin{aligned}
    v_a^{(2),1}
		&=
		\sum_{m\neq n}
		\mathcal C_{nm}\mathcal C_{mn}
		\frac{1}{\hbar}
		\partial_a\varepsilon_{nm},\\
		v_a^{(2),2}
		&=
		\frac{i}{\hbar}
		\sum_{m\neq n}
		e\mathbf E\cdot
		\left(
		i\partial_{\mathbf{k}}\mathcal C_{mn}
		\right)
		\mathcal A_{nm}^a
		+\text{c.c.},
		\\
		v_a^{(2),3}
		&=
		\frac{i}{\hbar}
		\sum_{m\neq n}
		e\mathbf E\cdot
		\big(
		\sum_{l\neq n}
		\mathcal C_{ln}\mathbfcal A_{ml}
		-
		\mathbfcal A_n\mathcal C_{mn}
		\big)
		\mathcal A_{nm}^a
		+\text{c.c.},
		\\
		v_a^{(2),4}
		&=-
		\frac{i}{\hbar}
		\sum_{m,l\neq n}
		\mathcal C_{nl}\mathcal C_{mn}
		\left(
		\varepsilon_l-\varepsilon_m
		\right)
		\mathcal A_{lm}^a .
	\end{aligned}
\end{equation}
First, the contribution $v_a^{(2),1}$ is evaluated as
\begin{equation}
	\begin{aligned}
		v_a^{(2),1}
		&=
		\sum_{m\neq n}
		\mathcal C_{nm}\mathcal C_{mn}
		\frac{1}{\hbar}\partial_a\varepsilon_{nm}=
		-\frac{e^2}{\hbar}E_bE_c
		\,\mathrm{Re}
		\sum_{m\neq n}
		\frac{
			\mathcal A_{nm}^b\mathcal A_{mn}^c
		}
		{\varepsilon_{nm}^2}
		\partial_a\varepsilon_{nm}.
	\end{aligned}
	\label{supp:Eq: v_partial_delta}
\end{equation}
Then, the contribution $v_a^{(2),2}$  is evaluated as
\begin{equation}
	\begin{aligned}
		v_a^{(2),2}
		&=
		\frac{i}{\hbar}
		\sum_{m\neq n}
		\left(
		e\mathbf E\cdot i\partial_{\mathbf{k}}\mathcal C_{mn}
		\right)
		\mathcal A_{nm}^a
		+\text{c.c.}
		\\
		&=
		-\frac{2e^2}{\hbar}E_bE_c
		\,\mathrm{Re}
		\sum_{m\neq n}
		\partial_b
		\left(
		\frac{\mathcal A_{mn}^c}{\varepsilon_{nm}}
		\right)
		\mathcal A_{nm}^a
		\\
		&=
		-\frac{2e^2}{\hbar}E_bE_c
		\,\mathrm{Re}
		\sum_{m\neq n}
		\frac{
			\mathcal A_{nm}^a\partial_b\mathcal A_{mn}^c
		}
		{\varepsilon_{nm}}
		+
		\frac{2e^2}{\hbar}E_bE_c
		\,\mathrm{Re}
		\sum_{m\neq n}
		\frac{
			\mathcal A_{nm}^a\mathcal A_{mn}^c
		}
		{\varepsilon_{nm}^2}
		\partial_b\varepsilon_{nm}.
	\end{aligned}
	\label{supp:Eq: v_partial_C}
\end{equation}
We next treat the $v_a^{(2),3}$ and $v_a^{(2),4}$. The $v_a^{(2),3}$ is evaluated as
\begin{equation}
	\begin{aligned}
		v_a^{(2),3}
		&=
		\frac{i}{\hbar}
		\sum_{m\neq n}
		e\mathbf E\cdot
		\Big(
		\sum_{l\neq n}\mathcal C_{ln}\mathbfcal{A}_{ml}
		-
		\mathbfcal{A}_n\mathcal C_{mn}
		\Big)
		\mathcal A_{nm}^a
		+\text{c.c.}
		\\
		&=
		\frac{i e^2}{\hbar}E_bE_c
		\sum_{m\neq n}
		\Big(
		\sum_{l\neq n}
		\frac{\mathcal A_{ln}^c}{\varepsilon_{nl}}
		\mathcal A_{ml}^b
		-
		\mathcal A_n^b
		\frac{\mathcal A_{mn}^c}{\varepsilon_{nm}}
		\Big)
		\mathcal A_{nm}^a
		+\text{c.c.}
		\\
		&=
		\frac{i e^2}{\hbar}E_bE_c
		\sum_{m\neq n}
		\frac{1}{\varepsilon_{nm}}
		\Big(
		\sum_{l\neq n}
		\mathcal A_{nl}^a\mathcal A_{lm}^b\mathcal A_{mn}^c
		-
		\mathcal A_{nm}^a\mathcal A_n^b\mathcal A_{mn}^c
		\Big)
		+\text{c.c.} .
	\end{aligned}
	\label{supp:Eq: v_conn}
\end{equation}
The $v_a^{(2),4}$ is evaluated as
\begin{equation}
	\begin{aligned}
		v_a^{(2),4}
		&=
		-\frac{i}{\hbar}
		\sum_{m,l\neq n}
		\mathcal C_{nl}\mathcal C_{mn}
		\left(
		\varepsilon_l-\varepsilon_m
		\right)
		\mathcal A_{lm}^a
		\\
		&=
		-\frac{i e^2}{\hbar}E_bE_c
		\sum_{m,l\neq n}
		\frac{\mathcal A_{nl}^b}{\varepsilon_{ln}}
		\frac{\mathcal A_{mn}^c}{\varepsilon_{nm}}
		\left(
		\varepsilon_{nm}-\varepsilon_{nl}
		\right)
		\mathcal A_{lm}^a
		\\
		&=
		\frac{i e^2}{\hbar}E_bE_c
		\sum_{m,l\neq n}
		\frac{
			\mathcal A_{lm}^a\mathcal A_{nl}^b\mathcal A_{mn}^c
		}
		{\varepsilon_{nl}}
		-
		\frac{i e^2}{\hbar}E_bE_c
		\sum_{m,l\neq n}
		\frac{
			\mathcal A_{lm}^a\mathcal A_{nl}^b\mathcal A_{mn}^c
		}
		{\varepsilon_{nm}}
		\\
		&=
		\frac{i e^2}{\hbar}E_bE_c
		\sum_{m,l\neq n}
		\frac{
			\mathcal A_{ml}^a\mathcal A_{nm}^b\mathcal A_{ln}^c
		}
		{\varepsilon_{nm}}
		-
		\frac{i e^2}{\hbar}E_bE_c
		\sum_{m,l\neq n}
		\frac{
			\mathcal A_{lm}^a\mathcal A_{nl}^b\mathcal A_{mn}^c
		}
		{\varepsilon_{nm}} .
	\end{aligned}
	\label{supp:Eq: v_off}
\end{equation}
Adding Eqs.~\eqref{supp:Eq: v_conn} and \eqref{supp:Eq: v_off}, we obtain
\begin{equation}
	\begin{aligned}
		v_a^{(2),3}+v_a^{(2),4}
		=&
		\frac{i e^2}{\hbar}E_bE_c
		\sum_{m\neq n}
		\frac{1}{\varepsilon_{nm}}
		\Big[
		\sum_{l\neq n}
		\left(
		\mathcal A_{nl}^a\mathcal A_{lm}^b
		-
		\mathcal A_{lm}^a\mathcal A_{nl}^b
		\right)
		\mathcal A_{mn}^c
		-
		\mathcal A_{nm}^a\mathcal A_n^b\mathcal A_{mn}^c
		\Big]
	+\text{c.c.}.
	\end{aligned}
	\label{supp:Eq: v_conn_off_before_flatness}
\end{equation}
Using the flatness identity
\begin{equation}
	\partial_a\mathcal A_{nm}^b-\partial_b\mathcal A_{nm}^a
	=
	i\sum_l
	\left(
	\mathcal A_{nl}^a\mathcal A_{lm}^b
	-
	\mathcal A_{nl}^b\mathcal A_{lm}^a
	\right),
	\label{supp:Eq: flatness_identity}
\end{equation}
and removing $l=n$ part, we obtain
\begin{equation}
	\partial_a\mathcal A_{nm}^b-\partial_b\mathcal A_{nm}^a
	-
	i
	\left(
	\mathcal A_n^a\mathcal A_{nm}^b
	-
	\mathcal A_n^b\mathcal A_{nm}^a
	\right)
	=
	i\sum_{l\neq n}
	\left(
	\mathcal A_{nl}^a\mathcal A_{lm}^b
	-
	\mathcal A_{nl}^b\mathcal A_{lm}^a
	\right).
	\label{supp:Eq: flatness_excluding_n}
\end{equation}
Then, using Eq.~\eqref{supp:Eq: flatness_excluding_n} in
Eq.~\eqref{supp:Eq: v_conn_off_before_flatness}, we find
\begin{equation}
	\begin{aligned}
		v_a^{(2),3}+v_a^{(2),4}
		&=
		\frac{i e^2}{\hbar}E_bE_c
		\sum_{m\neq n}
		\frac{1}{\varepsilon_{nm}}
		\Big[
		\frac{
			\left(
			\partial_a\mathcal A_{nm}^b
			-
			\partial_b\mathcal A_{nm}^a
			\right)
			-
			i
			\left(
			\mathcal A_n^a\mathcal A_{nm}^b
			-
			\mathcal A_n^b\mathcal A_{nm}^a
			\right)
		}{i}
		\mathcal A_{mn}^c
		-
		\mathcal A_{nm}^a\mathcal A_n^b\mathcal A_{mn}^c
		\Big]
		+\text{c.c.}
		\\
		&=
		\frac{e^2}{\hbar}E_bE_c
		\sum_{m\neq n}
		\frac{
			\left(
			\partial_a\mathcal A_{nm}^b
			-
			\partial_b\mathcal A_{nm}^a
			\right)
			\mathcal A_{mn}^c
		}
		{\varepsilon_{nm}}
		-
		\frac{i e^2}{\hbar}E_bE_c
		\sum_{m\neq n}
		\frac{
			\mathcal A_n^a\mathcal A_{nm}^b\mathcal A_{mn}^c
		}
		{\varepsilon_{nm}}
		+\text{c.c.}
		\\
		&=
		\frac{2e^2}{\hbar}E_bE_c
		\,\mathrm{Re}
		\sum_{m\neq n}
		\frac{
			\left(
			\partial_a\mathcal A_{nm}^b
			-
			\partial_b\mathcal A_{nm}^a
			\right)
			\mathcal A_{mn}^c
		}
		{\varepsilon_{nm}}
		\\
		&=
		\frac{e^2}{\hbar}E_bE_c
		\,\mathrm{Re}
		\sum_{m\neq n}
		\frac{
			\partial_a
			\left(
			\mathcal A_{nm}^b\mathcal A_{mn}^c
			\right)
		}
		{\varepsilon_{nm}}
		-
		\frac{2e^2}{\hbar}E_bE_c
		\,\mathrm{Re}
		\sum_{m\neq n}
		\frac{
			\left(
			\partial_b\mathcal A_{nm}^a
			\right)
			\mathcal A_{mn}^c
		}
		{\varepsilon_{nm}} .
	\end{aligned}
	\label{supp:Eq: v_offconn}
\end{equation}
By combining
Eqs.~\eqref{supp:Eq: v_partial_delta}, \eqref{supp:Eq: v_partial_C}, and
\eqref{supp:Eq: v_offconn}, we obtain
\begin{equation}
	\begin{aligned}
		v_a^{(2)}
		=&
		\frac{e^2}{\hbar}E_bE_c
		\,\mathrm{Re}
		\sum_{m\neq n}
		\frac{
			\partial_a
			\left(
			\mathcal A_{nm}^b\mathcal A_{mn}^c
			\right)
		}
		{\varepsilon_{nm}}	-
		\frac{e^2}{\hbar}E_bE_c
		\,\mathrm{Re}
		\sum_{m\neq n}
		\frac{
			\mathcal A_{nm}^b\mathcal A_{mn}^c
		}
		{\varepsilon_{nm}^2}
		\partial_a\varepsilon_{nm}
		\\
		&-
		\frac{2e^2}{\hbar}E_bE_c
		\,\mathrm{Re}
		\sum_{m\neq n}
		\frac{
			\left(
			\partial_b\mathcal A_{nm}^a
			\right)
			\mathcal A_{mn}^c
		}
		{\varepsilon_{nm}}
	     -
		\frac{2e^2}{\hbar}E_bE_c
		\,\mathrm{Re}
		\sum_{m\neq n}
		\frac{
			\mathcal A_{nm}^a\partial_b\mathcal A_{mn}^c
		}
		{\varepsilon_{nm}}
		\\
		&+
		\frac{2e^2}{\hbar}E_bE_c
		\,\mathrm{Re}
		\sum_{m\neq n}
		\frac{
			\mathcal A_{nm}^a\mathcal A_{mn}^c
		}
		{\varepsilon_{nm}^2}
		\partial_b\varepsilon_{nm}.
	\end{aligned}
	\label{supp:Eq: v2_before_metric}
\end{equation}
According to the definition of the band-normalized quantum metric in Eq.~\eqref{supp:Eq: qmetric_def}, we can rewrite the result within a compact form.
The  first two terms of Eq.~\eqref{supp:Eq: v2_before_metric} give
\begin{equation}
	\begin{aligned}
		\frac{e^2}{2\hbar}
		\left(
		\partial_a\mathcal{G}_n^{bc}
		\right)
		E_bE_c
		&=
		\frac{e^2}{\hbar}E_bE_c
		\,\mathrm{Re}
		\sum_{m\neq n}
		\frac{
			\partial_a
			\left(
			\mathcal A_{nm}^b\mathcal A_{mn}^c
			\right)
		}
		{\varepsilon_{nm}}
		-
		\frac{e^2}{\hbar}E_bE_c
		\,\mathrm{Re}
		\sum_{m\neq n}
		\mathcal A_{nm}^b\mathcal A_{mn}^c
		\frac{\partial_a\varepsilon_{nm}}
		{\varepsilon_{nm}^2}.
	\end{aligned}
\end{equation}
The remaining terms of Eq.~\eqref{supp:Eq: v2_before_metric} give
\begin{equation}
	\begin{aligned}
		-\frac{e^2}{\hbar}
		\left(
		\partial_b\mathcal{G}_n^{ac}
		\right)
		E_bE_c
		&=
		-\frac{2e^2}{\hbar}E_bE_c
		\,\mathrm{Re}
		\sum_{m\neq n}
		\frac{
			\partial_b
			\left(
			\mathcal A_{nm}^a\mathcal A_{mn}^c
			\right)
		}
		{\varepsilon_{nm}}
		+
		\frac{2e^2}{\hbar}E_bE_c
		\,\mathrm{Re}
		\sum_{m\neq n}
		\mathcal A_{nm}^a\mathcal A_{mn}^c
		\frac{\partial_b\varepsilon_{nm}}
		{\varepsilon_{nm}^2}
		\\
		&=
		-\frac{2e^2}{\hbar}E_bE_c
		\,\mathrm{Re}
		\sum_{m\neq n}
		\frac{
			\left(
			\partial_b\mathcal A_{nm}^a
			\right)
			\mathcal A_{mn}^c
			+
			\mathcal A_{nm}^a
			\left(
			\partial_b\mathcal A_{mn}^c
			\right)
		}
		{\varepsilon_{nm}}
		+
		\frac{2e^2}{\hbar}E_bE_c
		\,\mathrm{Re}
		\sum_{m\neq n}
		\mathcal A_{nm}^a\mathcal A_{mn}^c
		\frac{\partial_b\varepsilon_{nm}}
		{\varepsilon_{nm}^2}.
	\end{aligned}
\end{equation}
Consequently, we obtain the final expression of the second-order velocity as
\begin{equation}
	v_a^{(2)}
	=
	\frac{e^2}{2\hbar}
	\left(
	\partial_a\mathcal{G}_n^{bc}
	\right)
	E_bE_c
	-
	\frac{e^2}{\hbar}
	\left(
	\partial_b\mathcal{G}_n^{ac}
	\right)
	E_bE_c .
\label{supp:Eq: v2}
\end{equation}
This is the formal result of this work.

\makeatletter
\renewcommand{\bibsection}{\section*{References}\phantomsection\label{supp:references}}
\begin{NAT@thebibliography}{4}
	\makeatletter
	\providecommand \@ifxundefined [1]{%
		\@ifx{#1\undefined}
	}%
	\providecommand \@ifnum [1]{%
		\ifnum #1\expandafter \@firstoftwo
		\else \expandafter \@secondoftwo
		\fi
	}%
	\providecommand \@ifx [1]{%
		\ifx #1\expandafter \@firstoftwo
		\else \expandafter \@secondoftwo
		\fi
	}%
	\providecommand \natexlab [1]{#1}%
	\providecommand \enquote  [1]{``#1''}%
	\providecommand \bibnamefont  [1]{#1}%
	\providecommand \bibfnamefont [1]{#1}%
	\providecommand \citenamefont [1]{#1}%
	\providecommand \href@noop [0]{\@secondoftwo}%
	\providecommand \href [0]{\begingroup \@sanitize@url \@href}%
	\providecommand \@href[1]{\@@startlink{#1}\@@href}%
	\providecommand \@@href[1]{\endgroup#1\@@endlink}%
	\providecommand \@sanitize@url [0]{\catcode `\\12\catcode `\$12\catcode
		`\&12\catcode `\#12\catcode `\^12\catcode `\_12\catcode `\%12\relax}%
	\providecommand \@@startlink[1]{}%
	\providecommand \@@endlink[0]{}%
	\providecommand \url  [0]{\begingroup\@sanitize@url \@url }%
	\providecommand \@url [1]{\endgroup\@href {#1}{\urlprefix }}%
	\providecommand \urlprefix  [0]{URL }%
	\providecommand \Eprint [0]{\href }%
	\providecommand \doibase [0]{https://doi.org/}%
	\providecommand \selectlanguage [0]{\@gobble}%
	\providecommand \bibinfo  [0]{\@secondoftwo}%
	\providecommand \bibfield  [0]{\@secondoftwo}%
	\providecommand \translation [1]{[#1]}%
	\providecommand \BibitemOpen [0]{}%
	\providecommand \bibitemStop [0]{}%
	\providecommand \bibitemNoStop [0]{.\EOS\space}%
	\providecommand \EOS [0]{\spacefactor3000\relax}%
	\providecommand \BibitemShut  [1]{\csname bibitem#1\endcsname}%
	\let\auto@bib@innerbib\@empty
	\bibitem [{\citenamefont {Ashcroft}\ and\ \citenamefont
		{Mermin}(1976)}]{supp-Ashcroft76book}%
	\BibitemOpen
	\bibfield  {author} {\bibinfo {author} {\bibfnamefont {N.~W.}\ \bibnamefont
			{Ashcroft}}\ and\ \bibinfo {author} {\bibfnamefont {N.~D.}\ \bibnamefont
			{Mermin}},\ }\href@noop {} {\emph {\bibinfo {title} {{Solid State
					Physics}}}}\ (\bibinfo  {publisher} {Saunders College Publishing},\ \bibinfo
	{address} {Philadelphia},\ \bibinfo {year} {1976})\BibitemShut {NoStop}%
	\bibitem [{\citenamefont {Gao}\ \emph {et~al.}(2014)\citenamefont {Gao},
		\citenamefont {Yang},\ and\ \citenamefont {Niu}}]{supp-Gao14prl}%
	\BibitemOpen
	\bibfield  {author} {\bibinfo {author} {\bibfnamefont {Y.}~\bibnamefont
			{Gao}}, \bibinfo {author} {\bibfnamefont {S.~A.}\ \bibnamefont {Yang}},\ and\
		\bibinfo {author} {\bibfnamefont {Q.}~\bibnamefont {Niu}},\ }\bibfield
	{title} {\bibinfo {title} {Field induced positional shift of {Bloch}
			electrons and its dynamical implications},\ }\href
	{https://doi.org/10.1103/PhysRevLett.112.166601} {\bibfield  {journal}
		{\bibinfo  {journal} {Phys. Rev. Lett.}\ }\textbf {\bibinfo {volume} {112}},\
		\bibinfo {pages} {166601} (\bibinfo {year} {2014})}\BibitemShut {NoStop}%
	\bibitem [{\citenamefont {Griffiths}\ and\ \citenamefont
		{Schroeter}(2018)}]{supp-Griffiths18book}%
	\BibitemOpen
	\bibfield  {author} {\bibinfo {author} {\bibfnamefont {D.~J.}\ \bibnamefont
			{Griffiths}}\ and\ \bibinfo {author} {\bibfnamefont {D.~F.}\ \bibnamefont
			{Schroeter}},\ }\href@noop {} {\emph {\bibinfo {title} {{Introduction to
					Quantum Mechanics}}}},\ \bibinfo {edition} {3rd}\ ed.\ (\bibinfo  {publisher}
	{Cambridge University Press},\ \bibinfo {address} {Cambridge},\ \bibinfo
	{year} {2018})\BibitemShut {NoStop}%
	\bibitem [{\citenamefont {Sakurai}\ and\ \citenamefont
		{Napolitano}(2020)}]{supp-Sakurai20book}%
	\BibitemOpen
	\bibfield  {author} {\bibinfo {author} {\bibfnamefont {J.~J.}\ \bibnamefont
			{Sakurai}}\ and\ \bibinfo {author} {\bibfnamefont {J.}~\bibnamefont
			{Napolitano}},\ }\href@noop {} {\emph {\bibinfo {title} {{Modern Quantum
					Mechanics}}}},\ \bibinfo {edition} {3rd}\ ed.\ (\bibinfo  {publisher}
	{Cambridge University Press},\ \bibinfo {address} {Cambridge},\ \bibinfo
	{year} {2020})\BibitemShut {NoStop}%
\end{NAT@thebibliography}
\makeatother

\begin{thebibliography}{81}%
	\makeatletter
	\providecommand \@ifxundefined [1]{%
		\@ifx{#1\undefined}
	}%
	\providecommand \@ifnum [1]{%
		\ifnum #1\expandafter \@firstoftwo
		\else \expandafter \@secondoftwo
		\fi
	}%
	\providecommand \@ifx [1]{%
		\ifx #1\expandafter \@firstoftwo
		\else \expandafter \@secondoftwo
		\fi
	}%
	\providecommand \natexlab [1]{#1}%
	\providecommand \enquote  [1]{``#1''}%
	\providecommand \bibnamefont  [1]{#1}%
	\providecommand \bibfnamefont [1]{#1}%
	\providecommand \citenamefont [1]{#1}%
	\providecommand \href@noop [0]{\@secondoftwo}%
	\providecommand \href [0]{\begingroup \@sanitize@url \@href}%
	\providecommand \@href[1]{\@@startlink{#1}\@@href}%
	\providecommand \@@href[1]{\endgroup#1\@@endlink}%
	\providecommand \@sanitize@url [0]{\catcode `\\12\catcode `\$12\catcode
		`\&12\catcode `\#12\catcode `\^12\catcode `\_12\catcode `\%12\relax}%
	\providecommand \@@startlink[1]{}%
	\providecommand \@@endlink[0]{}%
	\providecommand \url  [0]{\begingroup\@sanitize@url \@url }%
	\providecommand \@url [1]{\endgroup\@href {#1}{\urlprefix }}%
	\providecommand \urlprefix  [0]{URL }%
	\providecommand \Eprint [0]{\href }%
	\providecommand \doibase [0]{https://doi.org/}%
	\providecommand \selectlanguage [0]{\@gobble}%
	\providecommand \bibinfo  [0]{\@secondoftwo}%
	\providecommand \bibfield  [0]{\@secondoftwo}%
	\providecommand \translation [1]{[#1]}%
	\providecommand \BibitemOpen [0]{}%
	\providecommand \bibitemStop [0]{}%
	\providecommand \bibitemNoStop [0]{.\EOS\space}%
	\providecommand \EOS [0]{\spacefactor3000\relax}%
	\providecommand \BibitemShut  [1]{\csname bibitem#1\endcsname}%
	\let\auto@bib@innerbib\@empty
	\bibitem [{\citenamefont {Hall}(1880)}]{Hall1880}%
	\BibitemOpen
	\bibfield  {author} {\bibinfo {author} {\bibfnamefont {E.~H.}\ \bibnamefont
			{Hall}},\ }\bibfield  {title} {\bibinfo {title} {On a new action of the
			magnet on electric currents},\ }\href
	{https://doi.org/10.2475/ajs.s3-19.111.200} {\bibfield  {journal} {\bibinfo
			{journal} {Am. J. Sci.}\ }\textbf {\bibinfo {volume} {19}},\ \bibinfo {pages}
		{200} (\bibinfo {year} {1880})}\BibitemShut {NoStop}%
	\bibitem [{\citenamefont {Karplus}\ and\ \citenamefont
		{Luttinger}(1954)}]{Karplus54pr}%
	\BibitemOpen
	\bibfield  {author} {\bibinfo {author} {\bibfnamefont {R.}~\bibnamefont
			{Karplus}}\ and\ \bibinfo {author} {\bibfnamefont {J.~M.}\ \bibnamefont
			{Luttinger}},\ }\bibfield  {title} {\bibinfo {title} {{Hall} effect in
			ferromagnetics},\ }\href {https://doi.org/10.1103/PhysRev.95.1154} {\bibfield
		{journal} {\bibinfo  {journal} {Phys. Rev.}\ }\textbf {\bibinfo {volume}
			{95}},\ \bibinfo {pages} {1154} (\bibinfo {year} {1954})}\BibitemShut
	{NoStop}%
	\bibitem [{\citenamefont {Haldane}(1988)}]{Haldane88prl}%
	\BibitemOpen
	\bibfield  {author} {\bibinfo {author} {\bibfnamefont {F.~D.~M.}\
			\bibnamefont {Haldane}},\ }\bibfield  {title} {\bibinfo {title} {Model for a
			quantum {Hall} effect without {Landau} levels: Condensed-matter realization
			of the ``parity anomaly''},\ }\href
	{https://doi.org/10.1103/PhysRevLett.61.2015} {\bibfield  {journal} {\bibinfo
			{journal} {Phys. Rev. Lett.}\ }\textbf {\bibinfo {volume} {61}},\ \bibinfo
		{pages} {2015} (\bibinfo {year} {1988})}\BibitemShut {NoStop}%
	\bibitem [{\citenamefont {Jungwirth}\ \emph {et~al.}(2002)\citenamefont
		{Jungwirth}, \citenamefont {Niu},\ and\ \citenamefont
		{MacDonald}}]{Jungwirth02prl}%
	\BibitemOpen
	\bibfield  {author} {\bibinfo {author} {\bibfnamefont {T.}~\bibnamefont
			{Jungwirth}}, \bibinfo {author} {\bibfnamefont {Q.}~\bibnamefont {Niu}},\
		and\ \bibinfo {author} {\bibfnamefont {A.~H.}\ \bibnamefont {MacDonald}},\
	}\bibfield  {title} {\bibinfo {title} {Anomalous {Hall} effect in
			ferromagnetic semiconductors},\ }\href
	{https://doi.org/10.1103/PhysRevLett.88.207208} {\bibfield  {journal}
		{\bibinfo  {journal} {Phys. Rev. Lett.}\ }\textbf {\bibinfo {volume} {88}},\
		\bibinfo {pages} {207208} (\bibinfo {year} {2002})}\BibitemShut {NoStop}%
	\bibitem [{\citenamefont {Fang}\ \emph {et~al.}(2003)\citenamefont {Fang},
		\citenamefont {Nagaosa}, \citenamefont {Takahashi} \emph
		{et~al.}}]{Fang03science}%
	\BibitemOpen
	\bibfield  {author} {\bibinfo {author} {\bibfnamefont {Z.}~\bibnamefont
			{Fang}}, \bibinfo {author} {\bibfnamefont {N.}~\bibnamefont {Nagaosa}},
		\bibinfo {author} {\bibfnamefont {K.~S.}\ \bibnamefont {Takahashi}}, \emph
		{et~al.},\ }\bibfield  {title} {\bibinfo {title} {The anomalous {Hall} effect
			and magnetic monopoles in momentum space},\ }\href
	{https://doi.org/10.1126/science.1089408} {\bibfield  {journal} {\bibinfo
			{journal} {Science}\ }\textbf {\bibinfo {volume} {302}},\ \bibinfo {pages}
		{92} (\bibinfo {year} {2003})}\BibitemShut {NoStop}%
	\bibitem [{\citenamefont {Yao}\ \emph {et~al.}(2004)\citenamefont {Yao},
		\citenamefont {Kleinman}, \citenamefont {MacDonald}, \citenamefont {Sinova},
		\citenamefont {Jungwirth}, \citenamefont {Wang}, \citenamefont {Wang},\ and\
		\citenamefont {Niu}}]{Yao04prl}%
	\BibitemOpen
	\bibfield  {author} {\bibinfo {author} {\bibfnamefont {Y.}~\bibnamefont
			{Yao}}, \bibinfo {author} {\bibfnamefont {L.}~\bibnamefont {Kleinman}},
		\bibinfo {author} {\bibfnamefont {A.~H.}\ \bibnamefont {MacDonald}}, \bibinfo
		{author} {\bibfnamefont {J.}~\bibnamefont {Sinova}}, \bibinfo {author}
		{\bibfnamefont {T.}~\bibnamefont {Jungwirth}}, \bibinfo {author}
		{\bibfnamefont {D.-s.}\ \bibnamefont {Wang}}, \bibinfo {author}
		{\bibfnamefont {E.}~\bibnamefont {Wang}},\ and\ \bibinfo {author}
		{\bibfnamefont {Q.}~\bibnamefont {Niu}},\ }\bibfield  {title} {\bibinfo
		{title} {First principles calculation of anomalous {Hall} conductivity in
			ferromagnetic bcc {Fe}},\ }\href
	{https://doi.org/10.1103/PhysRevLett.92.037204} {\bibfield  {journal}
		{\bibinfo  {journal} {Phys. Rev. Lett.}\ }\textbf {\bibinfo {volume} {92}},\
		\bibinfo {pages} {037204} (\bibinfo {year} {2004})}\BibitemShut {NoStop}%
	\bibitem [{\citenamefont {Nagaosa}\ \emph {et~al.}(2010)\citenamefont
		{Nagaosa}, \citenamefont {Sinova}, \citenamefont {Onoda}, \citenamefont
		{MacDonald},\ and\ \citenamefont {Ong}}]{Nagaosa10rmp}%
	\BibitemOpen
	\bibfield  {author} {\bibinfo {author} {\bibfnamefont {N.}~\bibnamefont
			{Nagaosa}}, \bibinfo {author} {\bibfnamefont {J.}~\bibnamefont {Sinova}},
		\bibinfo {author} {\bibfnamefont {S.}~\bibnamefont {Onoda}}, \bibinfo
		{author} {\bibfnamefont {A.~H.}\ \bibnamefont {MacDonald}},\ and\ \bibinfo
		{author} {\bibfnamefont {N.~P.}\ \bibnamefont {Ong}},\ }\bibfield  {title}
	{\bibinfo {title} {Anomalous {Hall} effect},\ }\href
	{https://doi.org/10.1103/RevModPhys.82.1539} {\bibfield  {journal} {\bibinfo
			{journal} {Rev. Mod. Phys.}\ }\textbf {\bibinfo {volume} {82}},\ \bibinfo
		{pages} {1539} (\bibinfo {year} {2010})}\BibitemShut {NoStop}%
	\bibitem [{\citenamefont {Yu}\ \emph {et~al.}(2010)\citenamefont {Yu},
		\citenamefont {Zhang}, \citenamefont {Zhang}, \citenamefont {Zhang},
		\citenamefont {Dai},\ and\ \citenamefont {Fang}}]{YuR10science}%
	\BibitemOpen
	\bibfield  {author} {\bibinfo {author} {\bibfnamefont {R.}~\bibnamefont
			{Yu}}, \bibinfo {author} {\bibfnamefont {W.}~\bibnamefont {Zhang}}, \bibinfo
		{author} {\bibfnamefont {H.-J.}\ \bibnamefont {Zhang}}, \bibinfo {author}
		{\bibfnamefont {S.-C.}\ \bibnamefont {Zhang}}, \bibinfo {author}
		{\bibfnamefont {X.}~\bibnamefont {Dai}},\ and\ \bibinfo {author}
		{\bibfnamefont {Z.}~\bibnamefont {Fang}},\ }\bibfield  {title} {\bibinfo
		{title} {Quantized anomalous {Hall} effect in magnetic topological
			insulators},\ }\href {https://doi.org/10.1126/science.1187485} {\bibfield
		{journal} {\bibinfo  {journal} {Science}\ }\textbf {\bibinfo {volume}
			{329}},\ \bibinfo {pages} {61} (\bibinfo {year} {2010})}\BibitemShut
	{NoStop}%
	\bibitem [{\citenamefont {Chang}\ \emph {et~al.}(2013)\citenamefont {Chang},
		\citenamefont {Zhang}, \citenamefont {Feng}, \citenamefont {Shen},
		\citenamefont {Zhang}, \citenamefont {Guo}, \citenamefont {Li}, \citenamefont
		{Ou}, \citenamefont {Wei}, \citenamefont {Wang} \emph
		{et~al.}}]{XueQK13science}%
	\BibitemOpen
	\bibfield  {author} {\bibinfo {author} {\bibfnamefont {C.-Z.}\ \bibnamefont
			{Chang}}, \bibinfo {author} {\bibfnamefont {J.}~\bibnamefont {Zhang}},
		\bibinfo {author} {\bibfnamefont {X.}~\bibnamefont {Feng}}, \bibinfo {author}
		{\bibfnamefont {J.}~\bibnamefont {Shen}}, \bibinfo {author} {\bibfnamefont
			{Z.}~\bibnamefont {Zhang}}, \bibinfo {author} {\bibfnamefont
			{M.}~\bibnamefont {Guo}}, \bibinfo {author} {\bibfnamefont {K.}~\bibnamefont
			{Li}}, \bibinfo {author} {\bibfnamefont {Y.}~\bibnamefont {Ou}}, \bibinfo
		{author} {\bibfnamefont {P.}~\bibnamefont {Wei}}, \bibinfo {author}
		{\bibfnamefont {L.-L.}\ \bibnamefont {Wang}}, \emph {et~al.},\ }\bibfield
	{title} {\bibinfo {title} {Experimental observation of the quantum anomalous
			{Hall} effect in a magnetic topological insulator},\ }\href
	{https://doi.org/10.1126/science.1234414} {\bibfield  {journal} {\bibinfo
			{journal} {Science}\ }\textbf {\bibinfo {volume} {340}},\ \bibinfo {pages}
		{167} (\bibinfo {year} {2013})}\BibitemShut {NoStop}%
	\bibitem [{\citenamefont {Burkov}(2014)}]{Burkov14prl}%
	\BibitemOpen
	\bibfield  {author} {\bibinfo {author} {\bibfnamefont {A.~A.}\ \bibnamefont
			{Burkov}},\ }\bibfield  {title} {\bibinfo {title} {Anomalous {Hall} effect in
			{Weyl} metals},\ }\href {https://doi.org/10.1103/PhysRevLett.113.187202}
	{\bibfield  {journal} {\bibinfo  {journal} {Phys. Rev. Lett.}\ }\textbf
		{\bibinfo {volume} {113}},\ \bibinfo {pages} {187202} (\bibinfo {year}
		{2014})}\BibitemShut {NoStop}%
	\bibitem [{\citenamefont {Chang}\ \emph {et~al.}(2023)\citenamefont {Chang},
		\citenamefont {Liu},\ and\ \citenamefont {MacDonald}}]{ChangCZ23rmp}%
	\BibitemOpen
	\bibfield  {author} {\bibinfo {author} {\bibfnamefont {C.-Z.}\ \bibnamefont
			{Chang}}, \bibinfo {author} {\bibfnamefont {C.-X.}\ \bibnamefont {Liu}},\
		and\ \bibinfo {author} {\bibfnamefont {A.~H.}\ \bibnamefont {MacDonald}},\
	}\bibfield  {title} {\bibinfo {title} {Colloquium: Quantum anomalous {Hall}
			effect},\ }\href {https://doi.org/10.1103/RevModPhys.95.011002} {\bibfield
		{journal} {\bibinfo  {journal} {Rev. Mod. Phys.}\ }\textbf {\bibinfo {volume}
			{95}},\ \bibinfo {pages} {011002} (\bibinfo {year} {2023})}\BibitemShut
	{NoStop}%
	\bibitem [{\citenamefont {Du}\ \emph {et~al.}(2021{\natexlab{a}})\citenamefont
		{Du}, \citenamefont {Lu},\ and\ \citenamefont {Xie}}]{Du21natrevphys}%
	\BibitemOpen
	\bibfield  {author} {\bibinfo {author} {\bibfnamefont {Z.~Z.}\ \bibnamefont
			{Du}}, \bibinfo {author} {\bibfnamefont {H.-Z.}\ \bibnamefont {Lu}},\ and\
		\bibinfo {author} {\bibfnamefont {X.~C.}\ \bibnamefont {Xie}},\ }\bibfield
	{title} {\bibinfo {title} {Nonlinear {Hall} effects},\ }\href
	{https://doi.org/10.1038/s42254-021-00359-5} {\bibfield  {journal} {\bibinfo
			{journal} {Nat. Rev. Phys.}\ }\textbf {\bibinfo {volume} {3}},\ \bibinfo
		{pages} {744} (\bibinfo {year} {2021}{\natexlab{a}})}\BibitemShut {NoStop}%
	\bibitem [{\citenamefont {Liu}\ \emph {et~al.}(2025)\citenamefont {Liu},
		\citenamefont {Qiang}, \citenamefont {Lu},\ and\ \citenamefont
		{Xie}}]{Liu24nsr}%
	\BibitemOpen
	\bibfield  {author} {\bibinfo {author} {\bibfnamefont {T.}~\bibnamefont
			{Liu}}, \bibinfo {author} {\bibfnamefont {X.-B.}\ \bibnamefont {Qiang}},
		\bibinfo {author} {\bibfnamefont {H.-Z.}\ \bibnamefont {Lu}},\ and\ \bibinfo
		{author} {\bibfnamefont {X.~C.}\ \bibnamefont {Xie}},\ }\bibfield  {title}
	{\bibinfo {title} {Quantum geometry in condensed matter},\ }\href
	{https://doi.org/10.1093/nsr/nwae334} {\bibfield  {journal} {\bibinfo
			{journal} {Natl. Sci. Rev.}\ }\textbf {\bibinfo {volume} {12}},\ \bibinfo
		{pages} {nwae334} (\bibinfo {year} {2025})}\BibitemShut {NoStop}%
	\bibitem [{\citenamefont {Jiang}\ \emph
		{et~al.}(2025{\natexlab{a}})\citenamefont {Jiang}, \citenamefont {Holder},\
		and\ \citenamefont {Yan}}]{YanBH25rppp}%
	\BibitemOpen
	\bibfield  {author} {\bibinfo {author} {\bibfnamefont {Y.}~\bibnamefont
			{Jiang}}, \bibinfo {author} {\bibfnamefont {T.}~\bibnamefont {Holder}},\ and\
		\bibinfo {author} {\bibfnamefont {B.}~\bibnamefont {Yan}},\ }\bibfield
	{title} {\bibinfo {title} {Revealing quantum geometry in nonlinear quantum
			materials},\ }\href {https://doi.org/10.1088/1361-6633/ade454} {\bibfield
		{journal} {\bibinfo  {journal} {Rep. Prog. Phys.}\ }\textbf {\bibinfo
			{volume} {88}},\ \bibinfo {pages} {076502} (\bibinfo {year}
		{2025}{\natexlab{a}})}\BibitemShut {NoStop}%
	\bibitem [{\citenamefont {Ma}\ \emph {et~al.}(2019)\citenamefont {Ma},
		\citenamefont {Xu}, \citenamefont {Shen}, \citenamefont {MacNeill},
		\citenamefont {Fatemi}, \citenamefont {Mier~Valdivia}, \citenamefont {Wu},
		\citenamefont {Chang}, \citenamefont {Du}, \citenamefont {Hsu}, \citenamefont
		{Gibson}, \citenamefont {Fang}, \citenamefont {Kaxiras}, \citenamefont
		{Watanabe}, \citenamefont {Taniguchi}, \citenamefont {Cava}, \citenamefont
		{Lu}, \citenamefont {Lin}, \citenamefont {Fu}, \citenamefont {Gedik},\ and\
		\citenamefont {Jarillo-Herrero}}]{Ma19nature}%
	\BibitemOpen
	\bibfield  {author} {\bibinfo {author} {\bibfnamefont {Q.}~\bibnamefont
			{Ma}}, \bibinfo {author} {\bibfnamefont {S.-Y.}\ \bibnamefont {Xu}}, \bibinfo
		{author} {\bibfnamefont {H.}~\bibnamefont {Shen}}, \bibinfo {author}
		{\bibfnamefont {D.}~\bibnamefont {MacNeill}}, \bibinfo {author}
		{\bibfnamefont {V.}~\bibnamefont {Fatemi}}, \bibinfo {author} {\bibfnamefont
			{A.~M.}\ \bibnamefont {Mier~Valdivia}}, \bibinfo {author} {\bibfnamefont
			{S.}~\bibnamefont {Wu}}, \bibinfo {author} {\bibfnamefont {T.-R.}\
			\bibnamefont {Chang}}, \bibinfo {author} {\bibfnamefont {Z.}~\bibnamefont
			{Du}}, \bibinfo {author} {\bibfnamefont {C.-H.}\ \bibnamefont {Hsu}},
		\bibinfo {author} {\bibfnamefont {Q.~D.}\ \bibnamefont {Gibson}}, \bibinfo
		{author} {\bibfnamefont {S.}~\bibnamefont {Fang}}, \bibinfo {author}
		{\bibfnamefont {E.}~\bibnamefont {Kaxiras}}, \bibinfo {author} {\bibfnamefont
			{K.}~\bibnamefont {Watanabe}}, \bibinfo {author} {\bibfnamefont
			{T.}~\bibnamefont {Taniguchi}}, \bibinfo {author} {\bibfnamefont {R.~J.}\
			\bibnamefont {Cava}}, \bibinfo {author} {\bibfnamefont {H.-Z.}\ \bibnamefont
			{Lu}}, \bibinfo {author} {\bibfnamefont {H.}~\bibnamefont {Lin}}, \bibinfo
		{author} {\bibfnamefont {L.}~\bibnamefont {Fu}}, \bibinfo {author}
		{\bibfnamefont {N.}~\bibnamefont {Gedik}},\ and\ \bibinfo {author}
		{\bibfnamefont {P.}~\bibnamefont {Jarillo-Herrero}},\ }\bibfield  {title}
	{\bibinfo {title} {Observation of the nonlinear {Hall} effect under
			time-reversal-symmetric conditions},\ }\href
	{https://doi.org/10.1038/s41586-018-0807-6} {\bibfield  {journal} {\bibinfo
			{journal} {Nature}\ }\textbf {\bibinfo {volume} {565}},\ \bibinfo {pages}
		{337} (\bibinfo {year} {2019})}\BibitemShut {NoStop}%
	\bibitem [{\citenamefont {Kang}\ \emph {et~al.}(2019)\citenamefont {Kang},
		\citenamefont {Li}, \citenamefont {Sohn}, \citenamefont {Shan},\ and\
		\citenamefont {Mak}}]{Kang19natmat}%
	\BibitemOpen
	\bibfield  {author} {\bibinfo {author} {\bibfnamefont {K.}~\bibnamefont
			{Kang}}, \bibinfo {author} {\bibfnamefont {T.}~\bibnamefont {Li}}, \bibinfo
		{author} {\bibfnamefont {E.}~\bibnamefont {Sohn}}, \bibinfo {author}
		{\bibfnamefont {J.}~\bibnamefont {Shan}},\ and\ \bibinfo {author}
		{\bibfnamefont {K.~F.}\ \bibnamefont {Mak}},\ }\bibfield  {title} {\bibinfo
		{title} {Nonlinear anomalous {Hall} effect in few-layer {WTe}$_2$},\ }\href
	{https://doi.org/10.1038/s41563-019-0294-7} {\bibfield  {journal} {\bibinfo
			{journal} {Nat. Mater.}\ }\textbf {\bibinfo {volume} {18}},\ \bibinfo {pages}
		{324} (\bibinfo {year} {2019})}\BibitemShut {NoStop}%
	\bibitem [{\citenamefont {Tiwari}\ \emph {et~al.}(2021)\citenamefont {Tiwari},
		\citenamefont {Chen}, \citenamefont {Zhong}, \citenamefont {Drueke},
		\citenamefont {Koo}, \citenamefont {Kaczmarek}, \citenamefont {Xiao},
		\citenamefont {Gao}, \citenamefont {Luo}, \citenamefont {Niu}, \citenamefont
		{Sun}, \citenamefont {Yan}, \citenamefont {Zhao},\ and\ \citenamefont
		{Tsen}}]{Tiwari21natcomm}%
	\BibitemOpen
	\bibfield  {author} {\bibinfo {author} {\bibfnamefont {A.}~\bibnamefont
			{Tiwari}}, \bibinfo {author} {\bibfnamefont {F.}~\bibnamefont {Chen}},
		\bibinfo {author} {\bibfnamefont {S.}~\bibnamefont {Zhong}}, \bibinfo
		{author} {\bibfnamefont {E.}~\bibnamefont {Drueke}}, \bibinfo {author}
		{\bibfnamefont {J.}~\bibnamefont {Koo}}, \bibinfo {author} {\bibfnamefont
			{A.}~\bibnamefont {Kaczmarek}}, \bibinfo {author} {\bibfnamefont
			{C.}~\bibnamefont {Xiao}}, \bibinfo {author} {\bibfnamefont {J.}~\bibnamefont
			{Gao}}, \bibinfo {author} {\bibfnamefont {X.}~\bibnamefont {Luo}}, \bibinfo
		{author} {\bibfnamefont {Q.}~\bibnamefont {Niu}}, \bibinfo {author}
		{\bibfnamefont {Y.}~\bibnamefont {Sun}}, \bibinfo {author} {\bibfnamefont
			{B.}~\bibnamefont {Yan}}, \bibinfo {author} {\bibfnamefont {L.}~\bibnamefont
			{Zhao}},\ and\ \bibinfo {author} {\bibfnamefont {A.~W.}\ \bibnamefont
			{Tsen}},\ }\bibfield  {title} {\bibinfo {title} {Giant $c$-axis nonlinear
			anomalous {Hall} effect in {$T_d$}-{MoTe}$_2$ and {WTe}$_2$},\ }\href
	{https://doi.org/10.1038/s41467-021-22343-5} {\bibfield  {journal} {\bibinfo
			{journal} {Nat. Commun.}\ }\textbf {\bibinfo {volume} {12}},\ \bibinfo
		{pages} {2049} (\bibinfo {year} {2021})}\BibitemShut {NoStop}%
	\bibitem [{\citenamefont {Ma}\ \emph {et~al.}(2022)\citenamefont {Ma},
		\citenamefont {Chen}, \citenamefont {Yananose}, \citenamefont {Zhou},
		\citenamefont {Wang}, \citenamefont {Li}, \citenamefont {Zhu}, \citenamefont
		{Wu}, \citenamefont {Xu}, \citenamefont {Yu}, \citenamefont {Qiu},
		\citenamefont {Stroppa},\ and\ \citenamefont {Loh}}]{Ma22natcomm}%
	\BibitemOpen
	\bibfield  {author} {\bibinfo {author} {\bibfnamefont {T.}~\bibnamefont
			{Ma}}, \bibinfo {author} {\bibfnamefont {H.}~\bibnamefont {Chen}}, \bibinfo
		{author} {\bibfnamefont {K.}~\bibnamefont {Yananose}}, \bibinfo {author}
		{\bibfnamefont {X.}~\bibnamefont {Zhou}}, \bibinfo {author} {\bibfnamefont
			{L.}~\bibnamefont {Wang}}, \bibinfo {author} {\bibfnamefont {R.}~\bibnamefont
			{Li}}, \bibinfo {author} {\bibfnamefont {Z.}~\bibnamefont {Zhu}}, \bibinfo
		{author} {\bibfnamefont {Z.}~\bibnamefont {Wu}}, \bibinfo {author}
		{\bibfnamefont {Q.-H.}\ \bibnamefont {Xu}}, \bibinfo {author} {\bibfnamefont
			{J.}~\bibnamefont {Yu}}, \bibinfo {author} {\bibfnamefont {C.-W.}\
			\bibnamefont {Qiu}}, \bibinfo {author} {\bibfnamefont {A.}~\bibnamefont
			{Stroppa}},\ and\ \bibinfo {author} {\bibfnamefont {K.~P.}\ \bibnamefont
			{Loh}},\ }\bibfield  {title} {\bibinfo {title} {Growth of bilayer {MoTe}$_2$
			single crystals with strong non-linear {Hall} effect},\ }\href
	{https://doi.org/10.1038/s41467-022-33201-3} {\bibfield  {journal} {\bibinfo
			{journal} {Nat. Commun.}\ }\textbf {\bibinfo {volume} {13}},\ \bibinfo
		{pages} {5465} (\bibinfo {year} {2022})}\BibitemShut {NoStop}%
	\bibitem [{\citenamefont {Ye}\ \emph {et~al.}(2023)\citenamefont {Ye},
		\citenamefont {Liu}, \citenamefont {Zhu}, \citenamefont {Xu}, \citenamefont
		{Yang}, \citenamefont {Shang}, \citenamefont {Liu},\ and\ \citenamefont
		{Liao}}]{Ye23prl}%
	\BibitemOpen
	\bibfield  {author} {\bibinfo {author} {\bibfnamefont {X.-G.}\ \bibnamefont
			{Ye}}, \bibinfo {author} {\bibfnamefont {H.}~\bibnamefont {Liu}}, \bibinfo
		{author} {\bibfnamefont {P.-F.}\ \bibnamefont {Zhu}}, \bibinfo {author}
		{\bibfnamefont {W.-Z.}\ \bibnamefont {Xu}}, \bibinfo {author} {\bibfnamefont
			{S.~A.}\ \bibnamefont {Yang}}, \bibinfo {author} {\bibfnamefont
			{N.}~\bibnamefont {Shang}}, \bibinfo {author} {\bibfnamefont
			{K.}~\bibnamefont {Liu}},\ and\ \bibinfo {author} {\bibfnamefont {Z.-M.}\
			\bibnamefont {Liao}},\ }\bibfield  {title} {\bibinfo {title} {Control over
			{Berry} curvature dipole with electric field in {WTe$_2$}},\ }\href
	{https://doi.org/10.1103/PhysRevLett.130.016301} {\bibfield  {journal}
		{\bibinfo  {journal} {Phys. Rev. Lett.}\ }\textbf {\bibinfo {volume} {130}},\
		\bibinfo {pages} {016301} (\bibinfo {year} {2023})}\BibitemShut {NoStop}%
	\bibitem [{\citenamefont {Huang}\ \emph
		{et~al.}(2023{\natexlab{a}})\citenamefont {Huang}, \citenamefont {Wu},
		\citenamefont {Hu}, \citenamefont {Cai}, \citenamefont {Li}, \citenamefont
		{An}, \citenamefont {Feng}, \citenamefont {Ye}, \citenamefont {Lin},
		\citenamefont {Law},\ and\ \citenamefont {Wang}}]{Huang23nsr}%
	\BibitemOpen
	\bibfield  {author} {\bibinfo {author} {\bibfnamefont {M.}~\bibnamefont
			{Huang}}, \bibinfo {author} {\bibfnamefont {Z.}~\bibnamefont {Wu}}, \bibinfo
		{author} {\bibfnamefont {J.}~\bibnamefont {Hu}}, \bibinfo {author}
		{\bibfnamefont {X.}~\bibnamefont {Cai}}, \bibinfo {author} {\bibfnamefont
			{E.}~\bibnamefont {Li}}, \bibinfo {author} {\bibfnamefont {L.}~\bibnamefont
			{An}}, \bibinfo {author} {\bibfnamefont {X.}~\bibnamefont {Feng}}, \bibinfo
		{author} {\bibfnamefont {Z.}~\bibnamefont {Ye}}, \bibinfo {author}
		{\bibfnamefont {N.}~\bibnamefont {Lin}}, \bibinfo {author} {\bibfnamefont
			{K.~T.}\ \bibnamefont {Law}},\ and\ \bibinfo {author} {\bibfnamefont
			{N.}~\bibnamefont {Wang}},\ }\bibfield  {title} {\bibinfo {title} {Giant
			nonlinear {Hall} effect in twisted bilayer {WSe}$_2$},\ }\href
	{https://doi.org/10.1093/nsr/nwac232} {\bibfield  {journal} {\bibinfo
			{journal} {Natl. Sci. Rev.}\ }\textbf {\bibinfo {volume} {10}},\ \bibinfo
		{pages} {nwac232} (\bibinfo {year} {2023}{\natexlab{a}})}\BibitemShut
	{NoStop}%
	\bibitem [{\citenamefont {Shvetsov}\ \emph {et~al.}(2019)\citenamefont
		{Shvetsov}, \citenamefont {Esin}, \citenamefont {Timonina}, \citenamefont
		{Kolesnikov},\ and\ \citenamefont {Deviatov}}]{Shvetsov19jetpl}%
	\BibitemOpen
	\bibfield  {author} {\bibinfo {author} {\bibfnamefont {O.~O.}\ \bibnamefont
			{Shvetsov}}, \bibinfo {author} {\bibfnamefont {V.~D.}\ \bibnamefont {Esin}},
		\bibinfo {author} {\bibfnamefont {A.~V.}\ \bibnamefont {Timonina}}, \bibinfo
		{author} {\bibfnamefont {N.~N.}\ \bibnamefont {Kolesnikov}},\ and\ \bibinfo
		{author} {\bibfnamefont {E.~V.}\ \bibnamefont {Deviatov}},\ }\bibfield
	{title} {\bibinfo {title} {Nonlinear {Hall} effect in three-dimensional
			{Weyl} and {Dirac} semimetals},\ }\href
	{https://doi.org/10.1134/S0021364019110018} {\bibfield  {journal} {\bibinfo
			{journal} {JETP Lett.}\ }\textbf {\bibinfo {volume} {109}},\ \bibinfo {pages}
		{715} (\bibinfo {year} {2019})}\BibitemShut {NoStop}%
	\bibitem [{\citenamefont {Kumar}\ \emph {et~al.}(2021)\citenamefont {Kumar},
		\citenamefont {Hsu}, \citenamefont {Sharma}, \citenamefont {Chang},
		\citenamefont {Yu}, \citenamefont {Wang}, \citenamefont {Eda}, \citenamefont
		{Liang},\ and\ \citenamefont {Yang}}]{Kumar21natnano}%
	\BibitemOpen
	\bibfield  {author} {\bibinfo {author} {\bibfnamefont {D.}~\bibnamefont
			{Kumar}}, \bibinfo {author} {\bibfnamefont {C.-H.}\ \bibnamefont {Hsu}},
		\bibinfo {author} {\bibfnamefont {R.}~\bibnamefont {Sharma}}, \bibinfo
		{author} {\bibfnamefont {T.-R.}\ \bibnamefont {Chang}}, \bibinfo {author}
		{\bibfnamefont {P.}~\bibnamefont {Yu}}, \bibinfo {author} {\bibfnamefont
			{J.}~\bibnamefont {Wang}}, \bibinfo {author} {\bibfnamefont {G.}~\bibnamefont
			{Eda}}, \bibinfo {author} {\bibfnamefont {G.}~\bibnamefont {Liang}},\ and\
		\bibinfo {author} {\bibfnamefont {H.}~\bibnamefont {Yang}},\ }\bibfield
	{title} {\bibinfo {title} {Room-temperature nonlinear {Hall} effect and
			wireless radiofrequency rectification in {Weyl} semimetal {TaIrTe}$_4$},\
	}\href {https://doi.org/10.1038/s41565-020-00839-3} {\bibfield  {journal}
		{\bibinfo  {journal} {Nat. Nanotechnol.}\ }\textbf {\bibinfo {volume} {16}},\
		\bibinfo {pages} {421} (\bibinfo {year} {2021})}\BibitemShut {NoStop}%
	\bibitem [{\citenamefont {Min}\ \emph {et~al.}(2023)\citenamefont {Min},
		\citenamefont {Tan}, \citenamefont {Xie}, \citenamefont {Miao}, \citenamefont
		{Zhang}, \citenamefont {Lee}, \citenamefont {Gopalan}, \citenamefont {Liu},
		\citenamefont {Alem}, \citenamefont {Yan},\ and\ \citenamefont
		{Mao}}]{Min23natcomm}%
	\BibitemOpen
	\bibfield  {author} {\bibinfo {author} {\bibfnamefont {L.}~\bibnamefont
			{Min}}, \bibinfo {author} {\bibfnamefont {H.}~\bibnamefont {Tan}}, \bibinfo
		{author} {\bibfnamefont {Z.}~\bibnamefont {Xie}}, \bibinfo {author}
		{\bibfnamefont {L.}~\bibnamefont {Miao}}, \bibinfo {author} {\bibfnamefont
			{R.}~\bibnamefont {Zhang}}, \bibinfo {author} {\bibfnamefont {S.~H.}\
			\bibnamefont {Lee}}, \bibinfo {author} {\bibfnamefont {V.}~\bibnamefont
			{Gopalan}}, \bibinfo {author} {\bibfnamefont {C.-X.}\ \bibnamefont {Liu}},
		\bibinfo {author} {\bibfnamefont {N.}~\bibnamefont {Alem}}, \bibinfo {author}
		{\bibfnamefont {B.}~\bibnamefont {Yan}},\ and\ \bibinfo {author}
		{\bibfnamefont {Z.}~\bibnamefont {Mao}},\ }\bibfield  {title} {\bibinfo
		{title} {Strong room-temperature bulk nonlinear {Hall} effect in a
			spin-valley locked {Dirac} material},\ }\href
	{https://doi.org/10.1038/s41467-023-35989-0} {\bibfield  {journal} {\bibinfo
			{journal} {Nat. Commun.}\ }\textbf {\bibinfo {volume} {14}},\ \bibinfo
		{pages} {364} (\bibinfo {year} {2023})}\BibitemShut {NoStop}%
	\bibitem [{\citenamefont {Lee}\ \emph {et~al.}(2024)\citenamefont {Lee},
		\citenamefont {Wang}, \citenamefont {Chen}, \citenamefont {Kwon},
		\citenamefont {Hwang}, \citenamefont {Cho}, \citenamefont {Son},
		\citenamefont {Han}, \citenamefont {Choi}, \citenamefont {Kim}, \citenamefont
		{Mo}, \citenamefont {Petrovic}, \citenamefont {Hwang}, \citenamefont {Park},
		\citenamefont {Jang},\ and\ \citenamefont {Ryu}}]{Lee24natcomm}%
	\BibitemOpen
	\bibfield  {author} {\bibinfo {author} {\bibfnamefont {J.-E.}\ \bibnamefont
			{Lee}}, \bibinfo {author} {\bibfnamefont {A.}~\bibnamefont {Wang}}, \bibinfo
		{author} {\bibfnamefont {S.}~\bibnamefont {Chen}}, \bibinfo {author}
		{\bibfnamefont {M.}~\bibnamefont {Kwon}}, \bibinfo {author} {\bibfnamefont
			{J.}~\bibnamefont {Hwang}}, \bibinfo {author} {\bibfnamefont
			{M.}~\bibnamefont {Cho}}, \bibinfo {author} {\bibfnamefont {K.-H.}\
			\bibnamefont {Son}}, \bibinfo {author} {\bibfnamefont {D.-S.}\ \bibnamefont
			{Han}}, \bibinfo {author} {\bibfnamefont {J.~W.}\ \bibnamefont {Choi}},
		\bibinfo {author} {\bibfnamefont {Y.~D.}\ \bibnamefont {Kim}}, \bibinfo
		{author} {\bibfnamefont {S.-K.}\ \bibnamefont {Mo}}, \bibinfo {author}
		{\bibfnamefont {C.}~\bibnamefont {Petrovic}}, \bibinfo {author}
		{\bibfnamefont {C.}~\bibnamefont {Hwang}}, \bibinfo {author} {\bibfnamefont
			{S.~Y.}\ \bibnamefont {Park}}, \bibinfo {author} {\bibfnamefont
			{C.}~\bibnamefont {Jang}},\ and\ \bibinfo {author} {\bibfnamefont
			{H.}~\bibnamefont {Ryu}},\ }\bibfield  {title} {\bibinfo {title}
		{Spin-orbit-splitting-driven nonlinear {Hall} effect in {NbIrTe$_4$}},\
	}\href {https://doi.org/10.1038/s41467-024-47643-4} {\bibfield  {journal}
		{\bibinfo  {journal} {Nat. Commun.}\ }\textbf {\bibinfo {volume} {15}},\
		\bibinfo {pages} {3971} (\bibinfo {year} {2024})}\BibitemShut {NoStop}%
	\bibitem [{\citenamefont {Jiang}\ \emph
		{et~al.}(2025{\natexlab{b}})\citenamefont {Jiang}, \citenamefont {Xi},
		\citenamefont {Li}, \citenamefont {He}, \citenamefont {Ma}, \citenamefont
		{Mao}, \citenamefont {Taniguchi}, \citenamefont {Watanabe}, \citenamefont
		{Rhodes}, \citenamefont {Zhang}, \citenamefont {Xiao},\ and\ \citenamefont
		{Wang}}]{Jiang25natcomm}%
	\BibitemOpen
	\bibfield  {author} {\bibinfo {author} {\bibfnamefont {H.}~\bibnamefont
			{Jiang}}, \bibinfo {author} {\bibfnamefont {T.}~\bibnamefont {Xi}}, \bibinfo
		{author} {\bibfnamefont {J.}~\bibnamefont {Li}}, \bibinfo {author}
		{\bibfnamefont {Y.}~\bibnamefont {He}}, \bibinfo {author} {\bibfnamefont
			{H.}~\bibnamefont {Ma}}, \bibinfo {author} {\bibfnamefont {Y.}~\bibnamefont
			{Mao}}, \bibinfo {author} {\bibfnamefont {T.}~\bibnamefont {Taniguchi}},
		\bibinfo {author} {\bibfnamefont {K.}~\bibnamefont {Watanabe}}, \bibinfo
		{author} {\bibfnamefont {D.~A.}\ \bibnamefont {Rhodes}}, \bibinfo {author}
		{\bibfnamefont {Y.}~\bibnamefont {Zhang}}, \bibinfo {author} {\bibfnamefont
			{J.}~\bibnamefont {Xiao}},\ and\ \bibinfo {author} {\bibfnamefont
			{Y.}~\bibnamefont {Wang}},\ }\bibfield  {title} {\bibinfo {title} {Probing
			interplay of topological properties and electron correlation in {TaIrTe$_4$}
			via nonlinear {Hall} effect},\ }\href
	{https://doi.org/10.1038/s41467-025-61347-3} {\bibfield  {journal} {\bibinfo
			{journal} {Nat. Commun.}\ }\textbf {\bibinfo {volume} {16}},\ \bibinfo
		{pages} {6351} (\bibinfo {year} {2025}{\natexlab{b}})}\BibitemShut {NoStop}%
	\bibitem [{\citenamefont {Xi}\ \emph {et~al.}(2025)\citenamefont {Xi},
		\citenamefont {Jiang}, \citenamefont {Li}, \citenamefont {He}, \citenamefont
		{Gu}, \citenamefont {Fox}, \citenamefont {Primeau}, \citenamefont {Mao},
		\citenamefont {Rollins}, \citenamefont {Taniguchi}, \citenamefont {Watanabe},
		\citenamefont {van~der Weide}, \citenamefont {Rhodes}, \citenamefont {Zhang},
		\citenamefont {Wang},\ and\ \citenamefont {Xiao}}]{Xi25natelectron}%
	\BibitemOpen
	\bibfield  {author} {\bibinfo {author} {\bibfnamefont {T.}~\bibnamefont
			{Xi}}, \bibinfo {author} {\bibfnamefont {H.}~\bibnamefont {Jiang}}, \bibinfo
		{author} {\bibfnamefont {J.}~\bibnamefont {Li}}, \bibinfo {author}
		{\bibfnamefont {Y.}~\bibnamefont {He}}, \bibinfo {author} {\bibfnamefont
			{Y.}~\bibnamefont {Gu}}, \bibinfo {author} {\bibfnamefont {C.}~\bibnamefont
			{Fox}}, \bibinfo {author} {\bibfnamefont {L.}~\bibnamefont {Primeau}},
		\bibinfo {author} {\bibfnamefont {Y.}~\bibnamefont {Mao}}, \bibinfo {author}
		{\bibfnamefont {J.}~\bibnamefont {Rollins}}, \bibinfo {author} {\bibfnamefont
			{T.}~\bibnamefont {Taniguchi}}, \bibinfo {author} {\bibfnamefont
			{K.}~\bibnamefont {Watanabe}}, \bibinfo {author} {\bibfnamefont
			{D.}~\bibnamefont {van~der Weide}}, \bibinfo {author} {\bibfnamefont
			{D.}~\bibnamefont {Rhodes}}, \bibinfo {author} {\bibfnamefont
			{Y.}~\bibnamefont {Zhang}}, \bibinfo {author} {\bibfnamefont
			{Y.}~\bibnamefont {Wang}},\ and\ \bibinfo {author} {\bibfnamefont
			{J.}~\bibnamefont {Xiao}},\ }\bibfield  {title} {\bibinfo {title} {Terahertz
			sensing based on the nonlinear electrodynamics of the two-dimensional
			correlated topological semimetal {TaIrTe$_4$}},\ }\href
	{https://doi.org/10.1038/s41928-025-01397-z} {\bibfield  {journal} {\bibinfo
			{journal} {Nat. Electron.}\ }\textbf {\bibinfo {volume} {8}},\ \bibinfo
		{pages} {578} (\bibinfo {year} {2025})}\BibitemShut {NoStop}%
	\bibitem [{\citenamefont {Ho}\ \emph {et~al.}(2021)\citenamefont {Ho},
		\citenamefont {Chang}, \citenamefont {Hsieh}, \citenamefont {Lo},
		\citenamefont {Huang}, \citenamefont {Vu}, \citenamefont {Ortix},\ and\
		\citenamefont {Chen}}]{Ho21natelectron}%
	\BibitemOpen
	\bibfield  {author} {\bibinfo {author} {\bibfnamefont {S.-C.}\ \bibnamefont
			{Ho}}, \bibinfo {author} {\bibfnamefont {C.-H.}\ \bibnamefont {Chang}},
		\bibinfo {author} {\bibfnamefont {Y.-C.}\ \bibnamefont {Hsieh}}, \bibinfo
		{author} {\bibfnamefont {S.-T.}\ \bibnamefont {Lo}}, \bibinfo {author}
		{\bibfnamefont {B.}~\bibnamefont {Huang}}, \bibinfo {author} {\bibfnamefont
			{T.-H.-Y.}\ \bibnamefont {Vu}}, \bibinfo {author} {\bibfnamefont
			{C.}~\bibnamefont {Ortix}},\ and\ \bibinfo {author} {\bibfnamefont {T.-M.}\
			\bibnamefont {Chen}},\ }\bibfield  {title} {\bibinfo {title} {{Hall} effects
			in artificially corrugated bilayer graphene without breaking time-reversal
			symmetry},\ }\href {https://doi.org/10.1038/s41928-021-00537-5} {\bibfield
		{journal} {\bibinfo  {journal} {Nat. Electron.}\ }\textbf {\bibinfo {volume}
			{4}},\ \bibinfo {pages} {116} (\bibinfo {year} {2021})}\BibitemShut {NoStop}%
	\bibitem [{\citenamefont {Sinha}\ \emph {et~al.}(2022)\citenamefont {Sinha},
		\citenamefont {Adak}, \citenamefont {Chakraborty}, \citenamefont {Das},
		\citenamefont {Debnath}, \citenamefont {Sangani}, \citenamefont {Watanabe},
		\citenamefont {Taniguchi}, \citenamefont {Waghmare}, \citenamefont
		{Agarwal},\ and\ \citenamefont {Deshmukh}}]{Sinha22natphys}%
	\BibitemOpen
	\bibfield  {author} {\bibinfo {author} {\bibfnamefont {S.}~\bibnamefont
			{Sinha}}, \bibinfo {author} {\bibfnamefont {P.~C.}\ \bibnamefont {Adak}},
		\bibinfo {author} {\bibfnamefont {A.}~\bibnamefont {Chakraborty}}, \bibinfo
		{author} {\bibfnamefont {K.}~\bibnamefont {Das}}, \bibinfo {author}
		{\bibfnamefont {K.}~\bibnamefont {Debnath}}, \bibinfo {author} {\bibfnamefont
			{L.~D.~V.}\ \bibnamefont {Sangani}}, \bibinfo {author} {\bibfnamefont
			{K.}~\bibnamefont {Watanabe}}, \bibinfo {author} {\bibfnamefont
			{T.}~\bibnamefont {Taniguchi}}, \bibinfo {author} {\bibfnamefont {U.~V.}\
			\bibnamefont {Waghmare}}, \bibinfo {author} {\bibfnamefont {A.}~\bibnamefont
			{Agarwal}},\ and\ \bibinfo {author} {\bibfnamefont {M.~M.}\ \bibnamefont
			{Deshmukh}},\ }\bibfield  {title} {\bibinfo {title} {{Berry} curvature dipole
			senses topological transition in a moir{\'e} superlattice},\ }\href
	{https://doi.org/10.1038/s41567-022-01606-y} {\bibfield  {journal} {\bibinfo
			{journal} {Nat. Phys.}\ }\textbf {\bibinfo {volume} {18}},\ \bibinfo {pages}
		{765} (\bibinfo {year} {2022})}\BibitemShut {NoStop}%
	\bibitem [{\citenamefont {Huang}\ \emph
		{et~al.}(2023{\natexlab{b}})\citenamefont {Huang}, \citenamefont {Wu},
		\citenamefont {Zhang}, \citenamefont {Feng}, \citenamefont {Zhou},
		\citenamefont {Wang} \emph {et~al.}}]{Huang23prl}%
	\BibitemOpen
	\bibfield  {author} {\bibinfo {author} {\bibfnamefont {M.}~\bibnamefont
			{Huang}}, \bibinfo {author} {\bibfnamefont {Z.}~\bibnamefont {Wu}}, \bibinfo
		{author} {\bibfnamefont {X.}~\bibnamefont {Zhang}}, \bibinfo {author}
		{\bibfnamefont {X.}~\bibnamefont {Feng}}, \bibinfo {author} {\bibfnamefont
			{Z.}~\bibnamefont {Zhou}}, \bibinfo {author} {\bibfnamefont {S.}~\bibnamefont
			{Wang}}, \emph {et~al.},\ }\bibfield  {title} {\bibinfo {title} {Intrinsic
			nonlinear {Hall} effect and gate-switchable {Berry} curvature sliding in
			twisted bilayer graphene},\ }\href
	{https://doi.org/10.1103/PhysRevLett.131.066301} {\bibfield  {journal}
		{\bibinfo  {journal} {Phys. Rev. Lett.}\ }\textbf {\bibinfo {volume} {131}},\
		\bibinfo {pages} {066301} (\bibinfo {year} {2023}{\natexlab{b}})}\BibitemShut
	{NoStop}%
	\bibitem [{\citenamefont {Lu}\ \emph {et~al.}(2024)\citenamefont {Lu},
		\citenamefont {Zhang}, \citenamefont {Wang}, \citenamefont {Zhao},
		\citenamefont {Zhou}, \citenamefont {Gao}, \citenamefont {Chen},
		\citenamefont {Law},\ and\ \citenamefont {Loh}}]{Lu24natcomm}%
	\BibitemOpen
	\bibfield  {author} {\bibinfo {author} {\bibfnamefont {X.~F.}\ \bibnamefont
			{Lu}}, \bibinfo {author} {\bibfnamefont {C.-P.}\ \bibnamefont {Zhang}},
		\bibinfo {author} {\bibfnamefont {N.}~\bibnamefont {Wang}}, \bibinfo {author}
		{\bibfnamefont {D.}~\bibnamefont {Zhao}}, \bibinfo {author} {\bibfnamefont
			{X.}~\bibnamefont {Zhou}}, \bibinfo {author} {\bibfnamefont {W.}~\bibnamefont
			{Gao}}, \bibinfo {author} {\bibfnamefont {X.~H.}\ \bibnamefont {Chen}},
		\bibinfo {author} {\bibfnamefont {K.~T.}\ \bibnamefont {Law}},\ and\ \bibinfo
		{author} {\bibfnamefont {K.~P.}\ \bibnamefont {Loh}},\ }\bibfield  {title}
	{\bibinfo {title} {Nonlinear transport and radio frequency rectification in
			{BiTeBr} at room temperature},\ }\href
	{https://doi.org/10.1038/s41467-023-44439-w} {\bibfield  {journal} {\bibinfo
			{journal} {Nat. Commun.}\ }\textbf {\bibinfo {volume} {15}},\ \bibinfo
		{pages} {245} (\bibinfo {year} {2024})}\BibitemShut {NoStop}%
	\bibitem [{\citenamefont {Makushko}\ \emph {et~al.}(2024)\citenamefont
		{Makushko}, \citenamefont {Kovalev}, \citenamefont {Zabila}, \citenamefont
		{Ilyakov}, \citenamefont {Ponomaryov}, \citenamefont {Arshad}, \citenamefont
		{Prajapati}, \citenamefont {de~Oliveira}, \citenamefont {Deinert},
		\citenamefont {Chekhonin}, \citenamefont {Veremchuk}, \citenamefont {Kosub},
		\citenamefont {Skourski}, \citenamefont {Ganss}, \citenamefont {Makarov},\
		and\ \citenamefont {Ortix}}]{Makushko24natelectron}%
	\BibitemOpen
	\bibfield  {author} {\bibinfo {author} {\bibfnamefont {P.}~\bibnamefont
			{Makushko}}, \bibinfo {author} {\bibfnamefont {S.}~\bibnamefont {Kovalev}},
		\bibinfo {author} {\bibfnamefont {Y.}~\bibnamefont {Zabila}}, \bibinfo
		{author} {\bibfnamefont {I.}~\bibnamefont {Ilyakov}}, \bibinfo {author}
		{\bibfnamefont {A.}~\bibnamefont {Ponomaryov}}, \bibinfo {author}
		{\bibfnamefont {A.}~\bibnamefont {Arshad}}, \bibinfo {author} {\bibfnamefont
			{G.~L.}\ \bibnamefont {Prajapati}}, \bibinfo {author} {\bibfnamefont {T.~V.
				A.~G.}\ \bibnamefont {de~Oliveira}}, \bibinfo {author} {\bibfnamefont
			{J.-C.}\ \bibnamefont {Deinert}}, \bibinfo {author} {\bibfnamefont
			{P.}~\bibnamefont {Chekhonin}}, \bibinfo {author} {\bibfnamefont
			{I.}~\bibnamefont {Veremchuk}}, \bibinfo {author} {\bibfnamefont
			{T.}~\bibnamefont {Kosub}}, \bibinfo {author} {\bibfnamefont
			{Y.}~\bibnamefont {Skourski}}, \bibinfo {author} {\bibfnamefont
			{F.}~\bibnamefont {Ganss}}, \bibinfo {author} {\bibfnamefont
			{D.}~\bibnamefont {Makarov}},\ and\ \bibinfo {author} {\bibfnamefont
			{C.}~\bibnamefont {Ortix}},\ }\bibfield  {title} {\bibinfo {title} {A tunable
			room-temperature nonlinear {Hall} effect in elemental bismuth thin films},\
	}\href {https://doi.org/10.1038/s41928-024-01118-y} {\bibfield  {journal}
		{\bibinfo  {journal} {Nat. Electron.}\ }\textbf {\bibinfo {volume} {7}},\
		\bibinfo {pages} {207} (\bibinfo {year} {2024})}\BibitemShut {NoStop}%
	\bibitem [{\citenamefont {Cheng}\ \emph {et~al.}(2024)\citenamefont {Cheng},
		\citenamefont {Gao}, \citenamefont {Zheng}, \citenamefont {Chen},
		\citenamefont {Liu}, \citenamefont {Zhang}, \citenamefont {Zhu},
		\citenamefont {Li}, \citenamefont {Li},\ and\ \citenamefont
		{Zeng}}]{Cheng24natcomm}%
	\BibitemOpen
	\bibfield  {author} {\bibinfo {author} {\bibfnamefont {B.}~\bibnamefont
			{Cheng}}, \bibinfo {author} {\bibfnamefont {Y.}~\bibnamefont {Gao}}, \bibinfo
		{author} {\bibfnamefont {Z.}~\bibnamefont {Zheng}}, \bibinfo {author}
		{\bibfnamefont {S.}~\bibnamefont {Chen}}, \bibinfo {author} {\bibfnamefont
			{Z.}~\bibnamefont {Liu}}, \bibinfo {author} {\bibfnamefont {L.}~\bibnamefont
			{Zhang}}, \bibinfo {author} {\bibfnamefont {Q.}~\bibnamefont {Zhu}}, \bibinfo
		{author} {\bibfnamefont {H.}~\bibnamefont {Li}}, \bibinfo {author}
		{\bibfnamefont {L.}~\bibnamefont {Li}},\ and\ \bibinfo {author}
		{\bibfnamefont {C.}~\bibnamefont {Zeng}},\ }\bibfield  {title} {\bibinfo
		{title} {Giant nonlinear {Hall} and wireless rectification effects at room
			temperature in the elemental semiconductor tellurium},\ }\href
	{https://doi.org/10.1038/s41467-024-49706-y} {\bibfield  {journal} {\bibinfo
			{journal} {Nat. Commun.}\ }\textbf {\bibinfo {volume} {15}},\ \bibinfo
		{pages} {5513} (\bibinfo {year} {2024})}\BibitemShut {NoStop}%
	\bibitem [{\citenamefont {Hu}\ \emph {et~al.}(2025)\citenamefont {Hu},
		\citenamefont {Zhao}, \citenamefont {Yang}, \citenamefont {Zhao},
		\citenamefont {Lei}, \citenamefont {Li}, \citenamefont {Lai}, \citenamefont
		{Yu}, \citenamefont {Park}, \citenamefont {Wong}, \citenamefont {Sharma},
		\citenamefont {Eda}, \citenamefont {Yang}, \citenamefont {Xu}, \citenamefont
		{Wang},\ and\ \citenamefont {Yang}}]{Hu25natnano}%
	\BibitemOpen
	\bibfield  {author} {\bibinfo {author} {\bibfnamefont {F.}~\bibnamefont
			{Hu}}, \bibinfo {author} {\bibfnamefont {P.}~\bibnamefont {Zhao}}, \bibinfo
		{author} {\bibfnamefont {L.}~\bibnamefont {Yang}}, \bibinfo {author}
		{\bibfnamefont {S.}~\bibnamefont {Zhao}}, \bibinfo {author} {\bibfnamefont
			{J.}~\bibnamefont {Lei}}, \bibinfo {author} {\bibfnamefont {W.-J.}\
			\bibnamefont {Li}}, \bibinfo {author} {\bibfnamefont {J.}~\bibnamefont
			{Lai}}, \bibinfo {author} {\bibfnamefont {Z.}~\bibnamefont {Yu}}, \bibinfo
		{author} {\bibfnamefont {H.}~\bibnamefont {Park}}, \bibinfo {author}
		{\bibfnamefont {C.~Q.}\ \bibnamefont {Wong}}, \bibinfo {author}
		{\bibfnamefont {R.}~\bibnamefont {Sharma}}, \bibinfo {author} {\bibfnamefont
			{G.}~\bibnamefont {Eda}}, \bibinfo {author} {\bibfnamefont {S.~A.}\
			\bibnamefont {Yang}}, \bibinfo {author} {\bibfnamefont {X.}~\bibnamefont
			{Xu}}, \bibinfo {author} {\bibfnamefont {F.}~\bibnamefont {Wang}},\ and\
		\bibinfo {author} {\bibfnamefont {H.}~\bibnamefont {Yang}},\ }\bibfield
	{title} {\bibinfo {title} {Ultrabroadband nonlinear {Hall} rectifier using
			{SnTe}},\ }\href {https://doi.org/10.1038/s41565-025-01993-2} {\bibfield
		{journal} {\bibinfo  {journal} {Nat. Nanotechnol.}\ }\textbf {\bibinfo
			{volume} {20}},\ \bibinfo {pages} {1588} (\bibinfo {year}
		{2025})}\BibitemShut {NoStop}%
	\bibitem [{\citenamefont {Zhang}\ \emph
		{et~al.}(2026{\natexlab{a}})\citenamefont {Zhang}, \citenamefont {Ding},
		\citenamefont {Xie}, \citenamefont {Li}, \citenamefont {Moynihan},
		\citenamefont {Sanchez}, \citenamefont {Zhu}, \citenamefont {Gao},
		\citenamefont {Iwasa}, \citenamefont {Alexe},\ and\ \citenamefont
		{Yang}}]{ZhangHB26natcomm}%
	\BibitemOpen
	\bibfield  {author} {\bibinfo {author} {\bibfnamefont {H.-B.}\ \bibnamefont
			{Zhang}}, \bibinfo {author} {\bibfnamefont {Z.-Y.}\ \bibnamefont {Ding}},
		\bibinfo {author} {\bibfnamefont {Y.-N.}\ \bibnamefont {Xie}}, \bibinfo
		{author} {\bibfnamefont {Z.-H.}\ \bibnamefont {Li}}, \bibinfo {author}
		{\bibfnamefont {E.}~\bibnamefont {Moynihan}}, \bibinfo {author}
		{\bibfnamefont {A.~M.}\ \bibnamefont {Sanchez}}, \bibinfo {author}
		{\bibfnamefont {W.}~\bibnamefont {Zhu}}, \bibinfo {author} {\bibfnamefont
			{Y.}~\bibnamefont {Gao}}, \bibinfo {author} {\bibfnamefont {Y.}~\bibnamefont
			{Iwasa}}, \bibinfo {author} {\bibfnamefont {M.}~\bibnamefont {Alexe}},\ and\
		\bibinfo {author} {\bibfnamefont {M.-M.}\ \bibnamefont {Yang}},\ }\bibfield
	{title} {\bibinfo {title} {Magnetization generation and giant nonlinear
			transport at symmetry-engineered interfaces},\ }\href
	{https://doi.org/10.1038/s41467-025-66149-1} {\bibfield  {journal} {\bibinfo
			{journal} {Nat. Commun.}\ }\textbf {\bibinfo {volume} {17}},\ \bibinfo
		{pages} {361} (\bibinfo {year} {2026}{\natexlab{a}})}\BibitemShut {NoStop}%
	\bibitem [{\citenamefont {Godinho}\ \emph {et~al.}(2018)\citenamefont
		{Godinho}, \citenamefont {Reichlov{\'a}}, \citenamefont {Kriegner},
		\citenamefont {Nov{\'a}k}, \citenamefont {Olejn{\'\i}k}, \citenamefont
		{Ka{\v{s}}par}, \citenamefont {{\v{S}}ob{\'a}{\v{n}}}, \citenamefont
		{Wadley}, \citenamefont {Campion}, \citenamefont {Otxoa} \emph
		{et~al.}}]{Jungwirth18nc}%
	\BibitemOpen
	\bibfield  {author} {\bibinfo {author} {\bibfnamefont {J.}~\bibnamefont
			{Godinho}}, \bibinfo {author} {\bibfnamefont {H.}~\bibnamefont
			{Reichlov{\'a}}}, \bibinfo {author} {\bibfnamefont {D.}~\bibnamefont
			{Kriegner}}, \bibinfo {author} {\bibfnamefont {V.}~\bibnamefont {Nov{\'a}k}},
		\bibinfo {author} {\bibfnamefont {K.}~\bibnamefont {Olejn{\'\i}k}}, \bibinfo
		{author} {\bibfnamefont {Z.}~\bibnamefont {Ka{\v{s}}par}}, \bibinfo {author}
		{\bibfnamefont {Z.}~\bibnamefont {{\v{S}}ob{\'a}{\v{n}}}}, \bibinfo {author}
		{\bibfnamefont {P.}~\bibnamefont {Wadley}}, \bibinfo {author} {\bibfnamefont
			{R.~P.}\ \bibnamefont {Campion}}, \bibinfo {author} {\bibfnamefont {R.~M.}\
			\bibnamefont {Otxoa}}, \emph {et~al.},\ }\bibfield  {title} {\bibinfo {title}
		{Electrically induced and detected {N{\'e}el} vector reversal in a collinear
			antiferromagnet},\ }\href
	{https://www.nature.com/articles/s41467-018-07092-2} {\bibfield  {journal}
		{\bibinfo  {journal} {Nat. Commun.}\ }\textbf {\bibinfo {volume} {9}},\
		\bibinfo {pages} {4686} (\bibinfo {year} {2018})}\BibitemShut {NoStop}%
	\bibitem [{\citenamefont {Gao}\ \emph {et~al.}(2023)\citenamefont {Gao},
		\citenamefont {Liu}, \citenamefont {Qiu}, \citenamefont {Ghosh},
		\citenamefont {Trevisan}, \citenamefont {Onishi}, \citenamefont {Hu},
		\citenamefont {Qian}, \citenamefont {Tien}, \citenamefont {Chen},
		\citenamefont {Huang}, \citenamefont {B{\'e}rub{\'e}}, \citenamefont {Li},
		\citenamefont {Tzschaschel}, \citenamefont {Dinh}, \citenamefont {Sun},
		\citenamefont {Lien}, \citenamefont {Singh}, \citenamefont {Watanabe},
		\citenamefont {Taniguchi}, \citenamefont {Bell}, \citenamefont {Lin},
		\citenamefont {Chang}, \citenamefont {Du}, \citenamefont {Bansil},
		\citenamefont {Fu}, \citenamefont {Ni}, \citenamefont {Orth}, \citenamefont
		{Ma},\ and\ \citenamefont {Xu}}]{Gao23science}%
	\BibitemOpen
	\bibfield  {author} {\bibinfo {author} {\bibfnamefont {A.}~\bibnamefont
			{Gao}}, \bibinfo {author} {\bibfnamefont {Y.-F.}\ \bibnamefont {Liu}},
		\bibinfo {author} {\bibfnamefont {J.-X.}\ \bibnamefont {Qiu}}, \bibinfo
		{author} {\bibfnamefont {B.}~\bibnamefont {Ghosh}}, \bibinfo {author}
		{\bibfnamefont {T.~V.}\ \bibnamefont {Trevisan}}, \bibinfo {author}
		{\bibfnamefont {Y.}~\bibnamefont {Onishi}}, \bibinfo {author} {\bibfnamefont
			{C.}~\bibnamefont {Hu}}, \bibinfo {author} {\bibfnamefont {T.}~\bibnamefont
			{Qian}}, \bibinfo {author} {\bibfnamefont {H.-J.}\ \bibnamefont {Tien}},
		\bibinfo {author} {\bibfnamefont {S.-W.}\ \bibnamefont {Chen}}, \bibinfo
		{author} {\bibfnamefont {M.}~\bibnamefont {Huang}}, \bibinfo {author}
		{\bibfnamefont {D.}~\bibnamefont {B{\'e}rub{\'e}}}, \bibinfo {author}
		{\bibfnamefont {H.}~\bibnamefont {Li}}, \bibinfo {author} {\bibfnamefont
			{C.}~\bibnamefont {Tzschaschel}}, \bibinfo {author} {\bibfnamefont
			{T.}~\bibnamefont {Dinh}}, \bibinfo {author} {\bibfnamefont {Z.}~\bibnamefont
			{Sun}}, \bibinfo {author} {\bibfnamefont {S.-W.}\ \bibnamefont {Lien}},
		\bibinfo {author} {\bibfnamefont {B.}~\bibnamefont {Singh}}, \bibinfo
		{author} {\bibfnamefont {K.}~\bibnamefont {Watanabe}}, \bibinfo {author}
		{\bibfnamefont {T.}~\bibnamefont {Taniguchi}}, \bibinfo {author}
		{\bibfnamefont {D.~C.}\ \bibnamefont {Bell}}, \bibinfo {author}
		{\bibfnamefont {H.}~\bibnamefont {Lin}}, \bibinfo {author} {\bibfnamefont
			{T.-R.}\ \bibnamefont {Chang}}, \bibinfo {author} {\bibfnamefont {C.~R.}\
			\bibnamefont {Du}}, \bibinfo {author} {\bibfnamefont {A.}~\bibnamefont
			{Bansil}}, \bibinfo {author} {\bibfnamefont {L.}~\bibnamefont {Fu}}, \bibinfo
		{author} {\bibfnamefont {N.}~\bibnamefont {Ni}}, \bibinfo {author}
		{\bibfnamefont {P.~P.}\ \bibnamefont {Orth}}, \bibinfo {author}
		{\bibfnamefont {Q.}~\bibnamefont {Ma}},\ and\ \bibinfo {author}
		{\bibfnamefont {S.-Y.}\ \bibnamefont {Xu}},\ }\bibfield  {title} {\bibinfo
		{title} {Quantum metric nonlinear {Hall} effect in a topological
			antiferromagnetic heterostructure},\ }\href
	{https://doi.org/10.1126/science.adf1506} {\bibfield  {journal} {\bibinfo
			{journal} {Science}\ }\textbf {\bibinfo {volume} {381}},\ \bibinfo {pages}
		{181} (\bibinfo {year} {2023})}\BibitemShut {NoStop}%
	\bibitem [{\citenamefont {Wang}\ \emph {et~al.}(2023)\citenamefont {Wang},
		\citenamefont {Kaplan}, \citenamefont {Zhang} \emph {et~al.}}]{Wang23nature}%
	\BibitemOpen
	\bibfield  {author} {\bibinfo {author} {\bibfnamefont {N.}~\bibnamefont
			{Wang}}, \bibinfo {author} {\bibfnamefont {D.}~\bibnamefont {Kaplan}},
		\bibinfo {author} {\bibfnamefont {Z.}~\bibnamefont {Zhang}}, \emph {et~al.},\
	}\bibfield  {title} {\bibinfo {title} {Quantum-metric-induced nonlinear
			transport in a topological antiferromagnet},\ }\href
	{https://doi.org/10.1038/s41586-023-06363-3} {\bibfield  {journal} {\bibinfo
			{journal} {Nature}\ }\textbf {\bibinfo {volume} {621}},\ \bibinfo {pages}
		{487} (\bibinfo {year} {2023})}\BibitemShut {NoStop}%
	\bibitem [{\citenamefont {Zhao}\ \emph {et~al.}(2025)\citenamefont {Zhao},
		\citenamefont {Xing}, \citenamefont {Zhao}, \citenamefont {Chen},
		\citenamefont {Hong}, \citenamefont {Yin}, \citenamefont {Liu}, \citenamefont
		{Le}, \citenamefont {Gayles}, \citenamefont {Tang}, \citenamefont {Fang},
		\citenamefont {Yan},\ and\ \citenamefont {Karel}}]{ZhaoW25natcomm}%
	\BibitemOpen
	\bibfield  {author} {\bibinfo {author} {\bibfnamefont {W.}~\bibnamefont
			{Zhao}}, \bibinfo {author} {\bibfnamefont {K.}~\bibnamefont {Xing}}, \bibinfo
		{author} {\bibfnamefont {Y.}~\bibnamefont {Zhao}}, \bibinfo {author}
		{\bibfnamefont {L.}~\bibnamefont {Chen}}, \bibinfo {author} {\bibfnamefont
			{M.}~\bibnamefont {Hong}}, \bibinfo {author} {\bibfnamefont {Y.}~\bibnamefont
			{Yin}}, \bibinfo {author} {\bibfnamefont {Y.}~\bibnamefont {Liu}}, \bibinfo
		{author} {\bibfnamefont {D.~K.}\ \bibnamefont {Le}}, \bibinfo {author}
		{\bibfnamefont {J.}~\bibnamefont {Gayles}}, \bibinfo {author} {\bibfnamefont
			{F.}~\bibnamefont {Tang}}, \bibinfo {author} {\bibfnamefont {Y.}~\bibnamefont
			{Fang}}, \bibinfo {author} {\bibfnamefont {B.}~\bibnamefont {Yan}},\ and\
		\bibinfo {author} {\bibfnamefont {J.}~\bibnamefont {Karel}},\ }\bibfield
	{title} {\bibinfo {title} {Room temperature quantum metric effect in
			{TbMn$_6$Sn$_6$}},\ }\href {https://doi.org/10.1038/s41467-025-62096-z}
	{\bibfield  {journal} {\bibinfo  {journal} {Nat. Commun.}\ }\textbf {\bibinfo
			{volume} {16}},\ \bibinfo {pages} {6881} (\bibinfo {year}
		{2025})}\BibitemShut {NoStop}%
	\bibitem [{\citenamefont {Mali}\ \emph {et~al.}(2026)\citenamefont {Mali},
		\citenamefont {Zhao}, \citenamefont {Wang}, \citenamefont {Sarker},
		\citenamefont {Chen}, \citenamefont {Li}, \citenamefont {Zhu}, \citenamefont
		{Liu}, \citenamefont {Gopalan}, \citenamefont {Yan},\ and\ \citenamefont
		{Mao}}]{Mali26natcomm}%
	\BibitemOpen
	\bibfield  {author} {\bibinfo {author} {\bibfnamefont {S.}~\bibnamefont
			{Mali}}, \bibinfo {author} {\bibfnamefont {Y.}~\bibnamefont {Zhao}}, \bibinfo
		{author} {\bibfnamefont {Y.}~\bibnamefont {Wang}}, \bibinfo {author}
		{\bibfnamefont {S.}~\bibnamefont {Sarker}}, \bibinfo {author} {\bibfnamefont
			{Y.}~\bibnamefont {Chen}}, \bibinfo {author} {\bibfnamefont {Z.}~\bibnamefont
			{Li}}, \bibinfo {author} {\bibfnamefont {J.}~\bibnamefont {Zhu}}, \bibinfo
		{author} {\bibfnamefont {Y.}~\bibnamefont {Liu}}, \bibinfo {author}
		{\bibfnamefont {V.}~\bibnamefont {Gopalan}}, \bibinfo {author} {\bibfnamefont
			{B.}~\bibnamefont {Yan}},\ and\ \bibinfo {author} {\bibfnamefont
			{Z.}~\bibnamefont {Mao}},\ }\bibfield  {title} {\bibinfo {title} {Probing
			hidden symmetry via nonlinear transport in an altermagnet candidate
			{Ca}$_3${Ru}$_2${O}$_7$},\ }\href
	{https://doi.org/10.1038/s41467-026-69739-9} {\bibfield  {journal} {\bibinfo
			{journal} {Nat. Commun.}\ }\textbf {\bibinfo {volume} {17}},\ \bibinfo
		{pages} {3074} (\bibinfo {year} {2026})}\BibitemShut {NoStop}%
	\bibitem [{\citenamefont {Li}\ \emph {et~al.}(2026)\citenamefont {Li},
		\citenamefont {Dai}, \citenamefont {Zhang}, \citenamefont {Zhang},
		\citenamefont {Li}, \citenamefont {Wei}, \citenamefont {Guo}, \citenamefont
		{Li}, \citenamefont {Fei}, \citenamefont {Zhang}, \citenamefont {Wang},
		\citenamefont {Wang},\ and\ \citenamefont {Song}}]{LiX26natcomm}%
	\BibitemOpen
	\bibfield  {author} {\bibinfo {author} {\bibfnamefont {X.}~\bibnamefont
			{Li}}, \bibinfo {author} {\bibfnamefont {Z.}~\bibnamefont {Dai}}, \bibinfo
		{author} {\bibfnamefont {S.}~\bibnamefont {Zhang}}, \bibinfo {author}
		{\bibfnamefont {H.}~\bibnamefont {Zhang}}, \bibinfo {author} {\bibfnamefont
			{C.}~\bibnamefont {Li}}, \bibinfo {author} {\bibfnamefont {B.}~\bibnamefont
			{Wei}}, \bibinfo {author} {\bibfnamefont {F.}~\bibnamefont {Guo}}, \bibinfo
		{author} {\bibfnamefont {C.}~\bibnamefont {Li}}, \bibinfo {author}
		{\bibfnamefont {F.}~\bibnamefont {Fei}}, \bibinfo {author} {\bibfnamefont
			{M.}~\bibnamefont {Zhang}}, \bibinfo {author} {\bibfnamefont
			{X.}~\bibnamefont {Wang}}, \bibinfo {author} {\bibfnamefont {H.}~\bibnamefont
			{Wang}},\ and\ \bibinfo {author} {\bibfnamefont {F.}~\bibnamefont {Song}},\
	}\bibfield  {title} {\bibinfo {title} {Higher odd-order nonlinear {Hall}
			effect in magnetic topological insulator {Mn(Bi$_{1-x}$Sb$_x$)$_2$Te$_4$}},\
	}\href {https://doi.org/10.1038/s41467-026-73418-0} {\bibfield  {journal}
		{\bibinfo  {journal} {Nat. Commun.}\ }\textbf {\bibinfo {volume} {17}},\
		\bibinfo {pages} {6652} (\bibinfo {year} {2026})}\BibitemShut {NoStop}%
	\bibitem [{\citenamefont {Gao}\ \emph {et~al.}(2014)\citenamefont {Gao},
		\citenamefont {Yang},\ and\ \citenamefont {Niu}}]{Gao14prl}%
	\BibitemOpen
	\bibfield  {author} {\bibinfo {author} {\bibfnamefont {Y.}~\bibnamefont
			{Gao}}, \bibinfo {author} {\bibfnamefont {S.~A.}\ \bibnamefont {Yang}},\ and\
		\bibinfo {author} {\bibfnamefont {Q.}~\bibnamefont {Niu}},\ }\bibfield
	{title} {\bibinfo {title} {Field induced positional shift of {Bloch}
			electrons and its dynamical implications},\ }\href
	{https://doi.org/10.1103/PhysRevLett.112.166601} {\bibfield  {journal}
		{\bibinfo  {journal} {Phys. Rev. Lett.}\ }\textbf {\bibinfo {volume} {112}},\
		\bibinfo {pages} {166601} (\bibinfo {year} {2014})}\BibitemShut {NoStop}%
	\bibitem [{\citenamefont {Gao}(2019)}]{Gao19fop}%
	\BibitemOpen
	\bibfield  {author} {\bibinfo {author} {\bibfnamefont {Y.}~\bibnamefont
			{Gao}},\ }\bibfield  {title} {\bibinfo {title} {Semiclassical dynamics and
			nonlinear charge current},\ }\href
	{https://doi.org/10.1007/s11467-019-0887-2} {\bibfield  {journal} {\bibinfo
			{journal} {Front. Phys.}\ }\textbf {\bibinfo {volume} {14}},\ \bibinfo
		{pages} {33404} (\bibinfo {year} {2019})}\BibitemShut {NoStop}%
	\bibitem [{\citenamefont {Wang}\ \emph {et~al.}(2021)\citenamefont {Wang},
		\citenamefont {Gao},\ and\ \citenamefont {Xiao}}]{Gao21prl}%
	\BibitemOpen
	\bibfield  {author} {\bibinfo {author} {\bibfnamefont {C.}~\bibnamefont
			{Wang}}, \bibinfo {author} {\bibfnamefont {Y.}~\bibnamefont {Gao}},\ and\
		\bibinfo {author} {\bibfnamefont {D.}~\bibnamefont {Xiao}},\ }\bibfield
	{title} {\bibinfo {title} {Intrinsic nonlinear {Hall} effect in
			antiferromagnetic tetragonal {CuMnAs}},\ }\href
	{https://doi.org/10.1103/PhysRevLett.127.277201} {\bibfield  {journal}
		{\bibinfo  {journal} {Phys. Rev. Lett.}\ }\textbf {\bibinfo {volume} {127}},\
		\bibinfo {pages} {277201} (\bibinfo {year} {2021})}\BibitemShut {NoStop}%
	\bibitem [{\citenamefont {Liu}\ \emph {et~al.}(2021)\citenamefont {Liu},
		\citenamefont {Zhao}, \citenamefont {Huang}, \citenamefont {Wu},
		\citenamefont {Sheng}, \citenamefont {Xiao},\ and\ \citenamefont
		{Yang}}]{YangSY21prl}%
	\BibitemOpen
	\bibfield  {author} {\bibinfo {author} {\bibfnamefont {H.}~\bibnamefont
			{Liu}}, \bibinfo {author} {\bibfnamefont {J.}~\bibnamefont {Zhao}}, \bibinfo
		{author} {\bibfnamefont {Y.-X.}\ \bibnamefont {Huang}}, \bibinfo {author}
		{\bibfnamefont {W.}~\bibnamefont {Wu}}, \bibinfo {author} {\bibfnamefont
			{X.-L.}\ \bibnamefont {Sheng}}, \bibinfo {author} {\bibfnamefont
			{C.}~\bibnamefont {Xiao}},\ and\ \bibinfo {author} {\bibfnamefont {S.~A.}\
			\bibnamefont {Yang}},\ }\bibfield  {title} {\bibinfo {title} {Intrinsic
			second-order anomalous {Hall} effect and its application in compensated
			antiferromagnets},\ }\href {https://doi.org/10.1103/PhysRevLett.127.277202}
	{\bibfield  {journal} {\bibinfo  {journal} {Phys. Rev. Lett.}\ }\textbf
		{\bibinfo {volume} {127}},\ \bibinfo {pages} {277202} (\bibinfo {year}
		{2021})}\BibitemShut {NoStop}%
	\bibitem [{\citenamefont {Moore}\ and\ \citenamefont
		{Orenstein}(2010)}]{Moore10prl}%
	\BibitemOpen
	\bibfield  {author} {\bibinfo {author} {\bibfnamefont {J.~E.}\ \bibnamefont
			{Moore}}\ and\ \bibinfo {author} {\bibfnamefont {J.}~\bibnamefont
			{Orenstein}},\ }\bibfield  {title} {\bibinfo {title} {Confinement-induced
			{Berry} phase and helicity-dependent photocurrents},\ }\href
	{https://doi.org/10.1103/PhysRevLett.105.026805} {\bibfield  {journal}
		{\bibinfo  {journal} {Phys. Rev. Lett.}\ }\textbf {\bibinfo {volume} {105}},\
		\bibinfo {pages} {026805} (\bibinfo {year} {2010})}\BibitemShut {NoStop}%
	\bibitem [{\citenamefont {Sodemann}\ and\ \citenamefont
		{Fu}(2015)}]{Sodemann15prl}%
	\BibitemOpen
	\bibfield  {author} {\bibinfo {author} {\bibfnamefont {I.}~\bibnamefont
			{Sodemann}}\ and\ \bibinfo {author} {\bibfnamefont {L.}~\bibnamefont {Fu}},\
	}\bibfield  {title} {\bibinfo {title} {Quantum nonlinear {Hall} effect
			induced by {Berry} curvature dipole in time-reversal invariant materials},\
	}\href {https://doi.org/10.1103/PhysRevLett.115.216806} {\bibfield  {journal}
		{\bibinfo  {journal} {Phys. Rev. Lett.}\ }\textbf {\bibinfo {volume} {115}},\
		\bibinfo {pages} {216806} (\bibinfo {year} {2015})}\BibitemShut {NoStop}%
	\bibitem [{\citenamefont {Low}\ \emph {et~al.}(2015)\citenamefont {Low},
		\citenamefont {Jiang},\ and\ \citenamefont {Guinea}}]{Low15prb}%
	\BibitemOpen
	\bibfield  {author} {\bibinfo {author} {\bibfnamefont {T.}~\bibnamefont
			{Low}}, \bibinfo {author} {\bibfnamefont {Y.}~\bibnamefont {Jiang}},\ and\
		\bibinfo {author} {\bibfnamefont {F.}~\bibnamefont {Guinea}},\ }\bibfield
	{title} {\bibinfo {title} {Topological currents in black phosphorus with
			broken inversion symmetry},\ }\href
	{https://doi.org/10.1103/PhysRevB.92.235447} {\bibfield  {journal} {\bibinfo
			{journal} {Phys. Rev. B}\ }\textbf {\bibinfo {volume} {92}},\ \bibinfo
		{pages} {235447} (\bibinfo {year} {2015})}\BibitemShut {NoStop}%
	\bibitem [{\citenamefont {Du}\ \emph {et~al.}(2018)\citenamefont {Du},
		\citenamefont {Wang}, \citenamefont {Lu},\ and\ \citenamefont
		{Xie}}]{Du18prl}%
	\BibitemOpen
	\bibfield  {author} {\bibinfo {author} {\bibfnamefont {Z.~Z.}\ \bibnamefont
			{Du}}, \bibinfo {author} {\bibfnamefont {C.~M.}\ \bibnamefont {Wang}},
		\bibinfo {author} {\bibfnamefont {H.-Z.}\ \bibnamefont {Lu}},\ and\ \bibinfo
		{author} {\bibfnamefont {X.~C.}\ \bibnamefont {Xie}},\ }\bibfield  {title}
	{\bibinfo {title} {Band signatures for strong nonlinear {Hall} effect in
			bilayer {WTe}$_2$},\ }\href {https://doi.org/10.1103/PhysRevLett.121.266601}
	{\bibfield  {journal} {\bibinfo  {journal} {Phys. Rev. Lett.}\ }\textbf
		{\bibinfo {volume} {121}},\ \bibinfo {pages} {266601} (\bibinfo {year}
		{2018})}\BibitemShut {NoStop}%
	\bibitem [{\citenamefont {Watanabe}\ and\ \citenamefont
		{Oshikawa}(2020)}]{Watanabe20prb}%
	\BibitemOpen
	\bibfield  {author} {\bibinfo {author} {\bibfnamefont {H.}~\bibnamefont
			{Watanabe}}\ and\ \bibinfo {author} {\bibfnamefont {M.}~\bibnamefont
			{Oshikawa}},\ }\bibfield  {title} {\bibinfo {title} {Generalized {$f$}-sum
			rules and {Kohn} formulas on nonlinear conductivities},\ }\href
	{https://doi.org/10.1103/PhysRevB.102.165137} {\bibfield  {journal} {\bibinfo
			{journal} {Phys. Rev. B}\ }\textbf {\bibinfo {volume} {102}},\ \bibinfo
		{pages} {165137} (\bibinfo {year} {2020})}\BibitemShut {NoStop}%
	\bibitem [{\citenamefont {Resta}(2021)}]{Resta21arXiv}%
	\BibitemOpen
	\bibfield  {author} {\bibinfo {author} {\bibfnamefont {R.}~\bibnamefont
			{Resta}},\ }\bibfield  {title} {\bibinfo {title} {Linear and nonlinear {Hall}
			conductivity in presence of interaction and disorder},\ }\href
	{https://arxiv.org/abs/2101.10949} {\bibfield  {journal} {\bibinfo  {journal}
			{arXiv:2101.10949}\ } (\bibinfo {year} {2021})}\BibitemShut {NoStop}%
	\bibitem [{\citenamefont {Hurd}(1972)}]{Hurd72book}%
	\BibitemOpen
	\bibfield  {author} {\bibinfo {author} {\bibfnamefont {C.~M.}\ \bibnamefont
			{Hurd}},\ }\href {https://doi.org/10.1007/978-1-4757-0465-5} {\emph {\bibinfo
			{title} {The {Hall} Effect in Metals and Alloys}}}\ (\bibinfo  {publisher}
	{Plenum Press},\ \bibinfo {address} {New York},\ \bibinfo {year}
	{1972})\BibitemShut {NoStop}%
	\bibitem [{\citenamefont {Du}\ \emph {et~al.}(2019)\citenamefont {Du},
		\citenamefont {Wang}, \citenamefont {Li}, \citenamefont {Lu},\ and\
		\citenamefont {Xie}}]{Du19natcomm}%
	\BibitemOpen
	\bibfield  {author} {\bibinfo {author} {\bibfnamefont {Z.~Z.}\ \bibnamefont
			{Du}}, \bibinfo {author} {\bibfnamefont {C.~M.}\ \bibnamefont {Wang}},
		\bibinfo {author} {\bibfnamefont {S.}~\bibnamefont {Li}}, \bibinfo {author}
		{\bibfnamefont {H.-Z.}\ \bibnamefont {Lu}},\ and\ \bibinfo {author}
		{\bibfnamefont {X.~C.}\ \bibnamefont {Xie}},\ }\bibfield  {title} {\bibinfo
		{title} {Disorder-induced nonlinear {Hall} effect with time-reversal
			symmetry},\ }\href {https://doi.org/10.1038/s41467-019-10941-3} {\bibfield
		{journal} {\bibinfo  {journal} {Nat. Commun.}\ }\textbf {\bibinfo {volume}
			{10}},\ \bibinfo {pages} {3047} (\bibinfo {year} {2019})}\BibitemShut
	{NoStop}%
	\bibitem [{\citenamefont {Xiao}\ \emph {et~al.}(2019)\citenamefont {Xiao},
		\citenamefont {Du},\ and\ \citenamefont {Niu}}]{Xiao19prb}%
	\BibitemOpen
	\bibfield  {author} {\bibinfo {author} {\bibfnamefont {C.}~\bibnamefont
			{Xiao}}, \bibinfo {author} {\bibfnamefont {Z.~Z.}\ \bibnamefont {Du}},\ and\
		\bibinfo {author} {\bibfnamefont {Q.}~\bibnamefont {Niu}},\ }\bibfield
	{title} {\bibinfo {title} {Theory of nonlinear {Hall} effects: Modified
			semiclassics from quantum kinetics},\ }\href
	{https://doi.org/10.1103/PhysRevB.100.165422} {\bibfield  {journal} {\bibinfo
			{journal} {Phys. Rev. B}\ }\textbf {\bibinfo {volume} {100}},\ \bibinfo
		{pages} {165422} (\bibinfo {year} {2019})}\BibitemShut {NoStop}%
	\bibitem [{\citenamefont {Du}\ \emph {et~al.}(2021{\natexlab{b}})\citenamefont
		{Du}, \citenamefont {Wang}, \citenamefont {Sun}, \citenamefont {Lu},\ and\
		\citenamefont {Xie}}]{Du21natcomm}%
	\BibitemOpen
	\bibfield  {author} {\bibinfo {author} {\bibfnamefont {Z.~Z.}\ \bibnamefont
			{Du}}, \bibinfo {author} {\bibfnamefont {C.~M.}\ \bibnamefont {Wang}},
		\bibinfo {author} {\bibfnamefont {H.-P.}\ \bibnamefont {Sun}}, \bibinfo
		{author} {\bibfnamefont {H.-Z.}\ \bibnamefont {Lu}},\ and\ \bibinfo {author}
		{\bibfnamefont {X.~C.}\ \bibnamefont {Xie}},\ }\bibfield  {title} {\bibinfo
		{title} {Quantum theory of the nonlinear {Hall} effect},\ }\href
	{https://doi.org/10.1038/s41467-021-25273-4} {\bibfield  {journal} {\bibinfo
			{journal} {Nat. Commun.}\ }\textbf {\bibinfo {volume} {12}},\ \bibinfo
		{pages} {5038} (\bibinfo {year} {2021}{\natexlab{b}})}\BibitemShut {NoStop}%
	\bibitem [{\citenamefont {Ma}\ \emph {et~al.}(2023)\citenamefont {Ma},
		\citenamefont {Arora}, \citenamefont {Vignale},\ and\ \citenamefont
		{Song}}]{Ma23prl}%
	\BibitemOpen
	\bibfield  {author} {\bibinfo {author} {\bibfnamefont {D.}~\bibnamefont
			{Ma}}, \bibinfo {author} {\bibfnamefont {A.}~\bibnamefont {Arora}}, \bibinfo
		{author} {\bibfnamefont {G.}~\bibnamefont {Vignale}},\ and\ \bibinfo {author}
		{\bibfnamefont {J.~C.~W.}\ \bibnamefont {Song}},\ }\bibfield  {title}
	{\bibinfo {title} {Anomalous skew-scattering nonlinear {Hall} effect and
			chiral photocurrents in {$\mathcal{PT}$}-symmetric antiferromagnets},\ }\href
	{https://doi.org/10.1103/PhysRevLett.131.076601} {\bibfield  {journal}
		{\bibinfo  {journal} {Phys. Rev. Lett.}\ }\textbf {\bibinfo {volume} {131}},\
		\bibinfo {pages} {076601} (\bibinfo {year} {2023})}\BibitemShut {NoStop}%
	\bibitem [{\citenamefont {Gong}\ \emph {et~al.}(2026)\citenamefont {Gong},
		\citenamefont {Du}, \citenamefont {Sun}, \citenamefont {Lu},\ and\
		\citenamefont {Xie}}]{GongZH26SB}%
	\BibitemOpen
	\bibfield  {author} {\bibinfo {author} {\bibfnamefont {Z.-H.}\ \bibnamefont
			{Gong}}, \bibinfo {author} {\bibfnamefont {Z.}~\bibnamefont {Du}}, \bibinfo
		{author} {\bibfnamefont {H.-P.}\ \bibnamefont {Sun}}, \bibinfo {author}
		{\bibfnamefont {H.-Z.}\ \bibnamefont {Lu}},\ and\ \bibinfo {author}
		{\bibfnamefont {X.}~\bibnamefont {Xie}},\ }\bibfield  {title} {\bibinfo
		{title} {Scaling analysis of quantum geometry in second-order nonlinear
			transport},\ }\href {https://doi.org/10.1016/j.scib.2026.07.062} {\bibfield
		{journal} {\bibinfo  {journal} {Science Bulletin}\ }\textbf {\bibinfo
			{volume} {71}},\ \bibinfo {pages} {4422–4428} (\bibinfo {year}
		{2026})}\BibitemShut {NoStop}%
	\bibitem [{\citenamefont {Xiao}\ \emph {et~al.}(2025)\citenamefont {Xiao},
		\citenamefont {Cao}, \citenamefont {Niu},\ and\ \citenamefont
		{Yang}}]{Xiao25prl}%
	\BibitemOpen
	\bibfield  {author} {\bibinfo {author} {\bibfnamefont {C.}~\bibnamefont
			{Xiao}}, \bibinfo {author} {\bibfnamefont {J.}~\bibnamefont {Cao}}, \bibinfo
		{author} {\bibfnamefont {Q.}~\bibnamefont {Niu}},\ and\ \bibinfo {author}
		{\bibfnamefont {S.~A.}\ \bibnamefont {Yang}},\ }\bibfield  {title} {\bibinfo
		{title} {Proper definition of intrinsic nonlinear current},\ }\href
	{https://doi.org/10.1103/g6yz-pwzh} {\bibfield  {journal} {\bibinfo
			{journal} {Phys. Rev. Lett.}\ }\textbf {\bibinfo {volume} {135}},\ \bibinfo
		{pages} {256306} (\bibinfo {year} {2025})}\BibitemShut {NoStop}%
	\bibitem [{\citenamefont {Jia}\ \emph {et~al.}(2024)\citenamefont {Jia},
		\citenamefont {Xiang}, \citenamefont {Qiao},\ and\ \citenamefont
		{Wang}}]{Jia24prb}%
	\BibitemOpen
	\bibfield  {author} {\bibinfo {author} {\bibfnamefont {J.}~\bibnamefont
			{Jia}}, \bibinfo {author} {\bibfnamefont {L.}~\bibnamefont {Xiang}}, \bibinfo
		{author} {\bibfnamefont {Z.}~\bibnamefont {Qiao}},\ and\ \bibinfo {author}
		{\bibfnamefont {J.}~\bibnamefont {Wang}},\ }\bibfield  {title} {\bibinfo
		{title} {Equivalence of semiclassical and response theories for second-order
			nonlinear ac {Hall} effects},\ }\href
	{https://doi.org/10.1103/PhysRevB.110.245406} {\bibfield  {journal} {\bibinfo
			{journal} {Phys. Rev. B}\ }\textbf {\bibinfo {volume} {110}},\ \bibinfo
		{pages} {245406} (\bibinfo {year} {2024})}\BibitemShut {NoStop}%
	\bibitem [{\citenamefont {Kaplan}\ \emph {et~al.}(2024)\citenamefont {Kaplan},
		\citenamefont {Holder},\ and\ \citenamefont {Yan}}]{Kaplan24prl}%
	\BibitemOpen
	\bibfield  {author} {\bibinfo {author} {\bibfnamefont {D.}~\bibnamefont
			{Kaplan}}, \bibinfo {author} {\bibfnamefont {T.}~\bibnamefont {Holder}},\
		and\ \bibinfo {author} {\bibfnamefont {B.}~\bibnamefont {Yan}},\ }\bibfield
	{title} {\bibinfo {title} {Unification of nonlinear anomalous {Hall} effect
			and nonreciprocal magnetoresistance in metals by quantum geometry},\ }\href
	{https://doi.org/10.1103/PhysRevLett.132.026301} {\bibfield  {journal}
		{\bibinfo  {journal} {Phys. Rev. Lett.}\ }\textbf {\bibinfo {volume} {132}},\
		\bibinfo {pages} {026301} (\bibinfo {year} {2024})}\BibitemShut {NoStop}%
	\bibitem [{\citenamefont {Zhang}\ and\ \citenamefont {Wang}(2025)}]{Wang25prb}%
	\BibitemOpen
	\bibfield  {author} {\bibinfo {author} {\bibfnamefont {A.}~\bibnamefont
			{Zhang}}\ and\ \bibinfo {author} {\bibfnamefont {C.~M.}\ \bibnamefont
			{Wang}},\ }\bibfield  {title} {\bibinfo {title} {Quantum theory of nonlinear
			electromagnetic response},\ }\href {https://doi.org/10.1103/fdqy-44yl}
	{\bibfield  {journal} {\bibinfo  {journal} {Phys. Rev. B}\ }\textbf {\bibinfo
			{volume} {112}},\ \bibinfo {pages} {075427} (\bibinfo {year}
		{2025})}\BibitemShut {NoStop}%
	\bibitem [{\citenamefont {Ulrich}\ \emph {et~al.}(2026)\citenamefont {Ulrich},
		\citenamefont {Mitscherling}, \citenamefont {Classen},\ and\ \citenamefont
		{Schnyder}}]{Ulrich26prb}%
	\BibitemOpen
	\bibfield  {author} {\bibinfo {author} {\bibfnamefont {Y.}~\bibnamefont
			{Ulrich}}, \bibinfo {author} {\bibfnamefont {J.}~\bibnamefont
			{Mitscherling}}, \bibinfo {author} {\bibfnamefont {L.}~\bibnamefont
			{Classen}},\ and\ \bibinfo {author} {\bibfnamefont {A.~P.}\ \bibnamefont
			{Schnyder}},\ }\bibfield  {title} {\bibinfo {title} {Quantum geometric origin
			of the intrinsic nonlinear {Hall} effect},\ }\href
	{https://doi.org/10.1103/4z8z-4kch} {\bibfield  {journal} {\bibinfo
			{journal} {Phys. Rev. B}\ }\textbf {\bibinfo {volume} {113}},\ \bibinfo
		{pages} {L201107} (\bibinfo {year} {2026})}\BibitemShut {NoStop}%
	\bibitem [{\citenamefont {Zhang}\ \emph
		{et~al.}(2026{\natexlab{b}})\citenamefont {Zhang}, \citenamefont {Cai},\ and\
		\citenamefont {Wang}}]{Zhang26arXiv}%
	\BibitemOpen
	\bibfield  {author} {\bibinfo {author} {\bibfnamefont {A.}~\bibnamefont
			{Zhang}}, \bibinfo {author} {\bibfnamefont {Z.}~\bibnamefont {Cai}},\ and\
		\bibinfo {author} {\bibfnamefont {C.~M.}\ \bibnamefont {Wang}},\ }\bibfield
	{title} {\bibinfo {title} {Nonlinear {Ohmic} electromagnetic response},\
	}\href {https://arxiv.org/abs/2605.16367} {\bibfield  {journal} {\bibinfo
			{journal} {arXiv:2605.16367}\ } (\bibinfo {year}
		{2026}{\natexlab{b}})}\BibitemShut {NoStop}%
	\bibitem [{\citenamefont {Shibata}(2026)}]{Junya26prb}%
    J.~Shibata, Second-order dc conductivity in the velocity-gauge Keldysh formalism:
    Gauge-invariant decomposition into nonlinear Drude, Berry curvature dipole, and
    quantum-metric responses, \href{https://doi.org/10.1103/sdhw-kmbc}
    {Phys.\ Rev.\ B (2026)}.
	\bibitem [{\citenamefont {Das}\ \emph {et~al.}(2023)\citenamefont {Das},
		\citenamefont {Lahiri}, \citenamefont {Atencia}, \citenamefont {Culcer},\
		and\ \citenamefont {Agarwal}}]{Das23prb}%
	\BibitemOpen
	\bibfield  {author} {\bibinfo {author} {\bibfnamefont {K.}~\bibnamefont
			{Das}}, \bibinfo {author} {\bibfnamefont {S.}~\bibnamefont {Lahiri}},
		\bibinfo {author} {\bibfnamefont {R.~B.}\ \bibnamefont {Atencia}}, \bibinfo
		{author} {\bibfnamefont {D.}~\bibnamefont {Culcer}},\ and\ \bibinfo {author}
		{\bibfnamefont {A.}~\bibnamefont {Agarwal}},\ }\bibfield  {title} {\bibinfo
		{title} {Intrinsic nonlinear conductivities induced by the quantum metric},\
	}\href {https://doi.org/10.1103/PhysRevB.108.L201405} {\bibfield  {journal}
		{\bibinfo  {journal} {Phys. Rev. B}\ }\textbf {\bibinfo {volume} {108}},\
		\bibinfo {pages} {L201405} (\bibinfo {year} {2023})}\BibitemShut {NoStop}%
	\bibitem [{\citenamefont {Wang}\ \emph {et~al.}(2024)\citenamefont {Wang},
		\citenamefont {Zhang}, \citenamefont {Zhu},\ and\ \citenamefont
		{Su}}]{Wang24prb}%
	\BibitemOpen
	\bibfield  {author} {\bibinfo {author} {\bibfnamefont {Y.}~\bibnamefont
			{Wang}}, \bibinfo {author} {\bibfnamefont {Z.}~\bibnamefont {Zhang}},
		\bibinfo {author} {\bibfnamefont {Z.-G.}\ \bibnamefont {Zhu}},\ and\ \bibinfo
		{author} {\bibfnamefont {G.}~\bibnamefont {Su}},\ }\bibfield  {title}
	{\bibinfo {title} {Intrinsic nonlinear {Ohmic} current},\ }\href
	{https://doi.org/10.1103/PhysRevB.109.085419} {\bibfield  {journal} {\bibinfo
			{journal} {Phys. Rev. B}\ }\textbf {\bibinfo {volume} {109}},\ \bibinfo
		{pages} {085419} (\bibinfo {year} {2024})}\BibitemShut {NoStop}%
	\bibitem [{\citenamefont {Guo}\ \emph {et~al.}(2026)\citenamefont {Guo},
		\citenamefont {Liu}, \citenamefont {Wang}, \citenamefont {Shi},\ and\
		\citenamefont {Chang}}]{Guo26prl}%
	\BibitemOpen
	\bibfield  {author} {\bibinfo {author} {\bibfnamefont {Z.}~\bibnamefont
			{Guo}}, \bibinfo {author} {\bibfnamefont {X.-Y.}\ \bibnamefont {Liu}},
		\bibinfo {author} {\bibfnamefont {H.}~\bibnamefont {Wang}}, \bibinfo {author}
		{\bibfnamefont {L.-k.}\ \bibnamefont {Shi}},\ and\ \bibinfo {author}
		{\bibfnamefont {K.}~\bibnamefont {Chang}},\ }\bibfield  {title} {\bibinfo
		{title} {Dissipation-shaped quantum geometry in nonlinear transport},\ }\href
	{https://doi.org/10.1103/rbck-mt1k} {\bibfield  {journal} {\bibinfo
			{journal} {Phys. Rev. Lett.}\ }\textbf {\bibinfo {volume} {136}},\ \bibinfo
		{pages} {206303} (\bibinfo {year} {2026})}\BibitemShut {NoStop}%
	\bibitem [{\citenamefont {Park}\ \emph {et~al.}(2026)\citenamefont {Park},
		\citenamefont {Huang}, \citenamefont {Savary},\ and\ \citenamefont
		{Balents}}]{Park26prb}%
	\BibitemOpen
	\bibfield  {author} {\bibinfo {author} {\bibfnamefont {T.}~\bibnamefont
			{Park}}, \bibinfo {author} {\bibfnamefont {X.}~\bibnamefont {Huang}},
		\bibinfo {author} {\bibfnamefont {L.}~\bibnamefont {Savary}},\ and\ \bibinfo
		{author} {\bibfnamefont {L.}~\bibnamefont {Balents}},\ }\bibfield  {title}
	{\bibinfo {title} {Quantum geometry from the {Moyal} product: Quantum kinetic
			equation and nonlinear response},\ }\href {https://doi.org/10.1103/x9pk-lb4x}
	{\bibfield  {journal} {\bibinfo  {journal} {Phys. Rev. B}\ }\textbf {\bibinfo
			{volume} {113}},\ \bibinfo {pages} {045146} (\bibinfo {year}
		{2026})}\BibitemShut {NoStop}%
	\bibitem [{\citenamefont {Ahmad}\ and\ \citenamefont
		{Xue}(2026)}]{Shakeel26prb}%
	\BibitemOpen
	\bibfield  {author} {\bibinfo {author} {\bibfnamefont {S.}~\bibnamefont
			{Ahmad}}\ and\ \bibinfo {author} {\bibfnamefont {F.}~\bibnamefont {Xue}},\
	}\bibfield  {title} {\bibinfo {title} {Length--velocity gauge equivalence of
			quantum geometric nonlinear conductivity in the adiabatic dc limit},\ }\href
	{https://doi.org/10.1103/r23j-t8kt} {\bibfield  {journal} {\bibinfo
			{journal} {Phys. Rev. B}\ }\textbf {\bibinfo {volume} {114}},\ \bibinfo
		{pages} {154428} (\bibinfo {year} {2026})}\BibitemShut {NoStop}%
	\bibitem [{\citenamefont {Resta}(2026)}]{Resta26prb}%
	\BibitemOpen
	\bibfield  {author} {\bibinfo {author} {\bibfnamefont {R.}~\bibnamefont
			{Resta}},\ }\bibfield  {title} {\bibinfo {title} {Intrinsic nonlinear {Hall}
			effect beyond {Bloch} geometry},\ }\href {https://doi.org/10.1103/rtdd-37y7}
	{\bibfield  {journal} {\bibinfo  {journal} {Phys. Rev. B}\ }\textbf {\bibinfo
			{volume} {113}},\ \bibinfo {pages} {125136} (\bibinfo {year}
		{2026})}\BibitemShut {NoStop}%
	\bibitem [{\citenamefont {Qiang}\ \emph {et~al.}(2026)\citenamefont {Qiang},
		\citenamefont {Liu}, \citenamefont {Gao}, \citenamefont {Lu},\ and\
		\citenamefont {Xie}}]{Qiang26advs}%
	\BibitemOpen
	\bibfield  {author} {\bibinfo {author} {\bibfnamefont {X.-B.}\ \bibnamefont
			{Qiang}}, \bibinfo {author} {\bibfnamefont {T.}~\bibnamefont {Liu}}, \bibinfo
		{author} {\bibfnamefont {Z.-X.}\ \bibnamefont {Gao}}, \bibinfo {author}
		{\bibfnamefont {H.-Z.}\ \bibnamefont {Lu}},\ and\ \bibinfo {author}
		{\bibfnamefont {X.~C.}\ \bibnamefont {Xie}},\ }\bibfield  {title} {\bibinfo
		{title} {A clarification on quantum-metric-induced nonlinear transport},\
	}\href {https://doi.org/10.1002/advs.202514818} {\bibfield  {journal}
		{\bibinfo  {journal} {Adv. Sci.}\ }\textbf {\bibinfo {volume} {13}},\
		\bibinfo {pages} {e14818} (\bibinfo {year} {2026})}\BibitemShut {NoStop}%
	\bibitem [{Sup()}]{Supp}%
	\BibitemOpen
	\href@noop {} {\bibinfo  {journal} {{See the \hyperref[supp:title]{Supplemental Material}
				for (Sec. SI) bare velocity of Bloch state, (Sec. SII) Bloch perturbation
				theory, and (Sec. SIII) perturbed velocity up to second order in electric
				field, which includes
				Refs.~\cite{Ashcroft76book,Gao14prl,Griffiths18book,Sakurai20book}}}\
	}\BibitemShut {NoStop}%
	\bibitem [{\citenamefont {Berry}(1984)}]{Berry84proc}%
	\BibitemOpen
	\bibfield  {journal} {  }\bibfield  {author} {\bibinfo {author} {\bibfnamefont
			{M.~V.}\ \bibnamefont {Berry}},\ }\bibfield  {title} {\bibinfo {title}
		{Quantal phase factors accompanying adiabatic changes},\ }\href
	{https://doi.org/10.1098/rspa.1984.0023} {\bibfield  {journal} {\bibinfo
			{journal} {Proc. R. Soc. Lond. A}\ }\textbf {\bibinfo {volume} {392}},\
		\bibinfo {pages} {45} (\bibinfo {year} {1984})}\BibitemShut {NoStop}%
	\bibitem [{\citenamefont {Xiao}\ \emph {et~al.}(2010)\citenamefont {Xiao},
		\citenamefont {Chang},\ and\ \citenamefont {Niu}}]{Xiao10rmp}%
	\BibitemOpen
	\bibfield  {author} {\bibinfo {author} {\bibfnamefont {D.}~\bibnamefont
			{Xiao}}, \bibinfo {author} {\bibfnamefont {M.-C.}\ \bibnamefont {Chang}},\
		and\ \bibinfo {author} {\bibfnamefont {Q.}~\bibnamefont {Niu}},\ }\bibfield
	{title} {\bibinfo {title} {{Berry} phase effects on electronic properties},\
	}\href {https://doi.org/10.1103/RevModPhys.82.1959} {\bibfield  {journal}
		{\bibinfo  {journal} {Rev. Mod. Phys.}\ }\textbf {\bibinfo {volume} {82}},\
		\bibinfo {pages} {1959} (\bibinfo {year} {2010})}\BibitemShut {NoStop}%
	\bibitem [{\citenamefont {Blount}(1962)}]{Blount62ssp}%
	\BibitemOpen
	\bibfield  {author} {\bibinfo {author} {\bibfnamefont {E.~I.}\ \bibnamefont
			{Blount}},\ }\bibfield  {title} {\bibinfo {title} {Formalisms of band
			theory},\ }in\ \href {https://doi.org/10.1016/S0081-1947(08)60459-2} {\emph
		{\bibinfo {booktitle} {Solid State Physics}}},\ Vol.~\bibinfo {volume} {13},\
	\bibinfo {editor} {edited by\ \bibinfo {editor} {\bibfnamefont
			{F.}~\bibnamefont {Seitz}}\ and\ \bibinfo {editor} {\bibfnamefont
			{D.}~\bibnamefont {Turnbull}}}\ (\bibinfo  {publisher} {Academic Press},\
	\bibinfo {address} {New York},\ \bibinfo {year} {1962})\ pp.\ \bibinfo
	{pages} {305--373}\BibitemShut {NoStop}%
	\bibitem [{\citenamefont {Sundaram}\ and\ \citenamefont
		{Niu}(1999)}]{Sundaram99prb}%
	\BibitemOpen
	\bibfield  {author} {\bibinfo {author} {\bibfnamefont {G.}~\bibnamefont
			{Sundaram}}\ and\ \bibinfo {author} {\bibfnamefont {Q.}~\bibnamefont {Niu}},\
	}\bibfield  {title} {\bibinfo {title} {Wave-packet dynamics in slowly
			perturbed crystals: gradient corrections and {Berry}-phase effects},\ }\href
	{https://doi.org/10.1103/PhysRevB.59.14915} {\bibfield  {journal} {\bibinfo
			{journal} {Phys. Rev. B}\ }\textbf {\bibinfo {volume} {59}},\ \bibinfo
		{pages} {14915} (\bibinfo {year} {1999})}\BibitemShut {NoStop}%
	\bibitem [{\citenamefont {Dong}\ \emph {et~al.}(2026)\citenamefont {Dong},
		\citenamefont {Kitamura}, \citenamefont {Itahashi}, \citenamefont {Chica},
		\citenamefont {Toyoda}, \citenamefont {Watanabe}, \citenamefont {Taniguchi},
		\citenamefont {Tanaka}, \citenamefont {Roy}, \citenamefont {Ogawa},
		\citenamefont {Morimoto}, \citenamefont {Iwasa},\ and\ \citenamefont
		{Ideue}}]{Dong26natmat}%
	\BibitemOpen
	\bibfield  {author} {\bibinfo {author} {\bibfnamefont {Y.}~\bibnamefont
			{Dong}}, \bibinfo {author} {\bibfnamefont {S.}~\bibnamefont {Kitamura}},
		\bibinfo {author} {\bibfnamefont {Y.~M.}\ \bibnamefont {Itahashi}}, \bibinfo
		{author} {\bibfnamefont {D.~G.}\ \bibnamefont {Chica}}, \bibinfo {author}
		{\bibfnamefont {S.}~\bibnamefont {Toyoda}}, \bibinfo {author} {\bibfnamefont
			{K.}~\bibnamefont {Watanabe}}, \bibinfo {author} {\bibfnamefont
			{T.}~\bibnamefont {Taniguchi}}, \bibinfo {author} {\bibfnamefont
			{M.}~\bibnamefont {Tanaka}}, \bibinfo {author} {\bibfnamefont
			{X.}~\bibnamefont {Roy}}, \bibinfo {author} {\bibfnamefont {N.}~\bibnamefont
			{Ogawa}}, \bibinfo {author} {\bibfnamefont {T.}~\bibnamefont {Morimoto}},
		\bibinfo {author} {\bibfnamefont {Y.}~\bibnamefont {Iwasa}},\ and\ \bibinfo
		{author} {\bibfnamefont {T.}~\bibnamefont {Ideue}},\ }\bibfield  {title}
	{\bibinfo {title} {Layer photovoltaic effect in a two-dimensional
			antiferromagnet with parity--time symmetry},\ }\href
	{https://doi.org/10.1038/s41563-026-02593-8} {\bibfield  {journal} {\bibinfo
			{journal} {Nat. Mater.}\ }\textbf {\bibinfo {volume} {25}},\ \bibinfo {pages}
		{1776} (\bibinfo {year} {2026})}\BibitemShut {NoStop}%
	\bibitem [{\citenamefont {Boix-Constant}\ \emph {et~al.}(2022)\citenamefont
		{Boix-Constant}, \citenamefont {Ma{\~n}as-Valero}, \citenamefont {Ruiz},
		\citenamefont {Rybakov}, \citenamefont {Konieczny}, \citenamefont {Pillet},
		\citenamefont {Baldov{\'i}},\ and\ \citenamefont
		{Coronado}}]{Boix22advmater}%
	\BibitemOpen
	\bibfield  {author} {\bibinfo {author} {\bibfnamefont {C.}~\bibnamefont
			{Boix-Constant}}, \bibinfo {author} {\bibfnamefont {S.}~\bibnamefont
			{Ma{\~n}as-Valero}}, \bibinfo {author} {\bibfnamefont {A.~M.}\ \bibnamefont
			{Ruiz}}, \bibinfo {author} {\bibfnamefont {A.}~\bibnamefont {Rybakov}},
		\bibinfo {author} {\bibfnamefont {K.~A.}\ \bibnamefont {Konieczny}}, \bibinfo
		{author} {\bibfnamefont {S.}~\bibnamefont {Pillet}}, \bibinfo {author}
		{\bibfnamefont {J.~J.}\ \bibnamefont {Baldov{\'i}}},\ and\ \bibinfo {author}
		{\bibfnamefont {E.}~\bibnamefont {Coronado}},\ }\bibfield  {title} {\bibinfo
		{title} {Probing the spin dimensionality in single-layer {CrSBr} van der
			{Waals} heterostructures by magneto-transport measurements},\ }\href
	{https://doi.org/10.1002/adma.202204940} {\bibfield  {journal} {\bibinfo
			{journal} {Adv. Mater.}\ }\textbf {\bibinfo {volume} {34}},\ \bibinfo {pages}
		{2204940} (\bibinfo {year} {2022})}\BibitemShut {NoStop}%
	\bibitem [{\citenamefont {Tang}(2026)}]{Tang26arXiv}%
	\BibitemOpen
	\bibfield  {author} {\bibinfo {author} {\bibfnamefont {P.}~\bibnamefont
			{Tang}},\ }\bibfield  {title} {\bibinfo {title} {Absence of
			quantum-metric-induced intrinsic longitudinal response},\ }\href
	{https://arxiv.org/abs/2605.02750} {\bibfield  {journal} {\bibinfo  {journal}
			{arXiv:2605.02750}\ } (\bibinfo {year} {2026})}\BibitemShut {NoStop}%
	\bibitem [{\citenamefont {Ashcroft}\ and\ \citenamefont
		{Mermin}(1976)}]{Ashcroft76book}%
	\BibitemOpen
	\bibfield  {author} {\bibinfo {author} {\bibfnamefont {N.~W.}\ \bibnamefont
			{Ashcroft}}\ and\ \bibinfo {author} {\bibfnamefont {N.~D.}\ \bibnamefont
			{Mermin}},\ }\href@noop {} {\emph {\bibinfo {title} {{Solid State
					Physics}}}}\ (\bibinfo  {publisher} {Saunders College Publishing},\ \bibinfo
	{address} {Philadelphia},\ \bibinfo {year} {1976})\BibitemShut {NoStop}%
	\bibitem [{\citenamefont {Griffiths}\ and\ \citenamefont
		{Schroeter}(2018)}]{Griffiths18book}%
	\BibitemOpen
	\bibfield  {author} {\bibinfo {author} {\bibfnamefont {D.~J.}\ \bibnamefont
			{Griffiths}}\ and\ \bibinfo {author} {\bibfnamefont {D.~F.}\ \bibnamefont
			{Schroeter}},\ }\href@noop {} {\emph {\bibinfo {title} {{Introduction to
					Quantum Mechanics}}}},\ \bibinfo {edition} {3rd}\ ed.\ (\bibinfo  {publisher}
	{Cambridge University Press},\ \bibinfo {address} {Cambridge},\ \bibinfo
	{year} {2018})\BibitemShut {NoStop}%
	\bibitem [{\citenamefont {Sakurai}\ and\ \citenamefont
		{Napolitano}(2020)}]{Sakurai20book}%
	\BibitemOpen
	\bibfield  {author} {\bibinfo {author} {\bibfnamefont {J.~J.}\ \bibnamefont
			{Sakurai}}\ and\ \bibinfo {author} {\bibfnamefont {J.}~\bibnamefont
			{Napolitano}},\ }\href@noop {} {\emph {\bibinfo {title} {{Modern Quantum
					Mechanics}}}},\ \bibinfo {edition} {3rd}\ ed.\ (\bibinfo  {publisher}
	{Cambridge University Press},\ \bibinfo {address} {Cambridge},\ \bibinfo
	{year} {2020})\BibitemShut {NoStop}%
\end{thebibliography}
\end{document}